\documentclass[fleqn,usenatbib]{mnras}

\usepackage{newtxtext,newtxmath}
\usepackage[T1]{fontenc}

\usepackage{graphicx}	
\usepackage{amsmath}	
\usepackage[dvipsnames]{xcolor}
\usepackage[flushleft]{threeparttable}
\usepackage{verbatim}
\usepackage{array}
\usepackage{tabularx}
\usepackage{colortbl}
\usepackage{hyperref}
\usepackage{float}
\usepackage{subcaption}
\usepackage{booktabs}
\usepackage{adjustbox}
\usepackage{pdflscape}
\usepackage{inputenc}
\usepackage[dvipsnames]{xcolor}
\usepackage{lscape}
\usepackage{longtable}
\usepackage{hyperref}
\usepackage{academicons}
\usepackage{xcolor}
\usepackage{orcidlink}
\usepackage{soul}

\DeclareRobustCommand{\VAN}[3]{#2}
\let\VANthebibliography\thebibliography
\def\thebibliography{\DeclareRobustCommand{\VAN}[3]{##3}\VANthebibliography}

\newcommand{\orcid}[1]{\href{https://orcid.org/#1}{\textcolor[HTML]{A6CE39}{\aiOrcid}}}

\newcommand{\kepler}{{\it Kepler}}

\newcommand{\TESS}{{\it TESS}}
\newcommand{\tess}{{\it TESS}}

\newcommand{\gaia}{{\it Gaia}}

\newcommand{\ngts}{{\it NGTS}}

\newcommand{\HARPS}{{\it HARPS}}
\newcommand{\harps}{{\it HARPS}}

\newcommand{\coralie}{{\it CORALIE}}
\newcommand{\CORALIE}{{\it CORALIE}}

\newcommand{\kms}{km\,s$^{-1}$}
\newcommand{\ms}{m\,s$^{-1}$}

\newcommand{\rstar}{\mbox{R$_{\star}$}}

\newcommand{\mjup}{\mbox{M\textsubscript{J}}}
\newcommand{\rjup}{\mbox{R\textsubscript{J}}}
\newcommand{\msun}{\mbox{M$_{\odot}$}}
\newcommand{\rsun}{\mbox{R$_{\odot}$}}

\newcommand{\rearth}{R$_{\oplus}$}

\newcommand{\teff}{T\textsubscript{eff}}
\newcommand{\tmag}{T\textsubscript{mag}}

\newcommand{\logg}{$\log g$}

\newcommand{\Ncand}{85}

\title{TESS Duotransit Candidates from the Southern Ecliptic Hemisphere}

\author[F.\ Hawthorn, S.\ Gill, D.\ Bayliss et al.]{\parbox{\textwidth}{\Large
Faith~Hawthorn*$^{1,2}$ \orcidlink{0000-0002-8675-182X},
Sam~Gill$^\dag$$^{1,2}$ \orcidlink{0000-0002-4259-0155},
Daniel~Bayliss$^\ddag$$^{1,2}$ \orcidlink{0000-0001-6023-1335},
Hugh~P.~Osborn$^{3}$ \orcidlink{0000-0002-4047-4724},
Ingrid~Pelisoli$^{1}$ \orcidlink{0000-0003-4615-6556},
Toby~Rodel$^{1,2,4}$ \orcidlink{0009-0009-2175-7284},
Kaylen~Smith~Darnbrook$^{1}$,
Peter~J.~Wheatley$^{1,2}$ \orcidlink{0000-0003-1452-2240},
David~R.~Anderson$^{1,2}$ \orcidlink{0000-0001-7416-7522},
Ioannis~Apergis$^{1,2}$ \orcidlink{0009-0004-7473-4573},
Matthew~P.~Battley$^{5}$ \orcidlink{0000-0002-1357-9774},
Matthew~R.~Burleigh$^{6}$ \orcidlink{0000-0003-0684-7803},
Sarah~L.~Casewell$^{6}$ \orcidlink{0000-0003-2478-0120},
Philipp~Eigm\"uller$^{7}$ \orcidlink{0000-0003-4096-0594},
Maximilian~N.~G{\"u}nther$^{8}$ \orcidlink{0000-0002-3164-9086},
James~S.~Jenkins$^{9,10}$ \orcidlink{0000-0003-2733-8725},
Monika~Lendl$^{5}$ \orcidlink{0000-0001-9699-1459},
Maximiliano~Moyano$^{11}$ \orcidlink{0000-0002-7927-9555},
Ares~Osborn$^{1,2}$ \orcidlink{0000-0002-5899-7750},
Gavin~Ramsay$^{12}$ \orcidlink{0000-0001-8722-9710},
Sol\`ene Ulmer-Moll$^{3}$ \orcidlink{0000-0003-2417-7006},
Jose~I.~Vines$^{11}$ \orcidlink{0000-0002-1896-2377},
Richard~West$^{1,2}$
}
\vspace{0.2cm}
\\
\parbox{\textwidth}{
The authors' affiliations are shown in Appendix \ref{sec:affiliations}.\\
E-mails: *faith.hawthorn@warwick.ac.uk, $^\dag$samuel.gill@warwick.ac.uk, $^\ddag$d.bayliss@warwick.ac.uk}\vspace{-0.3cm}}

\date{Accepted XXX. Received YYY; in original form ZZZ}

\pubyear{2022}

\begin{document}
\label{firstpage}
\pagerange{\pageref{firstpage}--\pageref{lastpage}}
\maketitle


\begin{abstract}
Discovering transiting exoplanets with long orbital periods allows us to study warm and cool planetary systems with temperatures similar to the planets in our own Solar system.  The \TESS\ mission has photometrically surveyed the entire Southern Ecliptic Hemisphere in Cycle 1 (August 2018 - July 2019), Cycle 3 (July 2020 - June 2021) and Cycle 5 (September 2022 - September 2023).  We use the observations from Cycle 1 and Cycle 3 to search for exoplanet systems that show a single transit event in each year - which we call \textit{duotransits}. The periods of these planet candidates are typically in excess of 20~days, with the lower limit determined by the duration of individual \tess\ observations. We find \Ncand\ duotransit candidates, which span a range of host star brightnesses between 8\,$<$\,\tmag\,$<$\,14, transit depths between 0.1 per cent and 1.8 per cent, and transit durations between 2 and 10~hours with the upper limit determined by our normalisation function. Of these candidates, 25 are already known, and 60 are new. We present these candidates along with the status of photometric and spectroscopic follow-up.
\end{abstract}

\begin{keywords}
techniques: photometric --- planets and satellites: detection --- planets and satellites: general
\end{keywords}


\section{Introduction} \label{sec:intro}
Longer period transiting exoplanets allow us to measure the densities of warm and cool planets, assisting with studies of atmospheric composition from transit spectroscopy \citep[e.g.][]{2012ApJ...757...18A, wang2021}.
Planetary obliquity is an indicator of planetary migration mechanisms \citep[e.g.][]{2012ApJ...757...18A}, and long period planets are amenable to studies of the Rossiter-McLaughlin effect \citep[e.g.][]{ulmermoll2023} to enable determination of the spin-orbit alignment of the system. When compared to short period hot Jupiter planets, the few well-studied long period planets discovered to date appear to be more aligned \citep{rice2022}, indicating possible migration mechanisms operating within the protoplanetary disk \citep[e.g.][]{2017MNRAS.469.4102M}. Since the launch of \TESS\ \citep[Transiting Exoplanet Survey Satellite;][]{Ricker:2015}, a total of 392\footnote{\url{https://exoplanetarchive.ipac.caltech.edu} (2023 Oct 09)} exoplanets have been confirmed using \tess\ data \citep[NASA Exoplanet Archive;][]{Akeson2013}. Of these systems, only 56 have orbital periods longer than the typical \tess\ Sector time span of 27\,days.

The \kepler\ mission \citep{kepler2010} monitored a single 105 square degree field for approximately four years, and was therefore able to discover long period transiting exoplanets.  However, most of the host stars are fainter than the capabilities of modern spectrographs to measure planetary masses and the planets cannot be confirmed and fully characterised. \citet{hsu2019} placed a limit on the occurrence rate of $\lesssim$0.247 planets per FGK star on orbital periods from 237-500\,days. The \kepler\ exoplanet sample includes 42 planets with determined periods longer than 20 days with a robust mass determination from radial velocity confirmation.

The \TESS\ spacecraft has completed surveys of the Southern Ecliptic Hemisphere on two occasions -- in Cycle 1 (July 2018 to July 2019) and in Cycle 3 (July 2020 to July 2021). \TESS\ has also re-observed the Southern Ecliptic Hemisphere in Cycle 5 (September 2022 to September 2023), but many of these observations are unavailable at the time of writing and thus are not included in this work. During these surveys a number of stars have shown a single transit (a \textit{monotransit}), whereby one transit like feature is found in the lightcurve but is not seen to repeat.  With only one transit we are not able to determine the orbital period of the planet, although with prior knowledge of the star we can calculate the probability distribution for the orbital period, e.g. \cite{osborn2022}.  A small number of these TESS monotransit candidates have been confirmed, e.g. \citet{gill2020b,gill2020a} and \citet{lendl2020a}. 

In addition to the \TESS\ monotransit candidates, there exist candidates that transit exactly once in Cycle 1 and once in Cycle 3.  These candidates are \textit{biennial duotransits}, as opposed to candidates that may have two transits in a single \tess\ cycle.  For the remainder of this paper we use the term \textit{duotransit} to refer only to such biennial duotransits. These duotransits confer two main advantages over monotransit candidates. Firstly, the true period of a duotransit candidate is limited to a discrete set of possible periods, $P_{n}$, given by
\begin{equation}
    P_{n} \in \left( \frac{\Delta T}{n} \right), n=1, ..., n_{max},
\end{equation}
where $\Delta T$ is the time difference between the Cycle 1 and the Cycle 3 transit event and $n$ is a whole number ranging from n=1 for the longest possible period to $n_{max}$ for the shortest possible period.  Since the duration of a TESS sector is $\approx$27\,days, we find $P_{n_{max}}$ is typically in the region of 20\,days unless there is more than one Sector of data in the TESS Cycle 1 or Cycle 3 observations.  The limited set of possible periods means it is 
possible to check specific orbital periods via photometric monitoring of the duotransit candidate at specific times.   This is much more efficient than the continuous monitoring required to find the periods of monotransit candidates.

Secondly, duotransits are more robust against false positive signals.  We can check that the two transit signals for a duotransit candidate match in depth, total duration, ingress and egress duration, limb darkening parameters, and impact parameter.  
By contrast, we are far less certain that a single monotransit event is not caused by some systematic or non-planetary astrophysical event. 

\citet{cooke2018} predicted that hundreds of long-period planets would be detectable as monotransit events in the \TESS\ sample, with similar results found by \citet{2019AJ....157...84V}. Furthermore, \citet{cooke2021} found that many of these will have a second detectable transit in Cycle 3 of \TESS\ data, making the true orbital period one of an average of 38 period aliases.

Both mono- and duotransits require extensive follow-up with both photometric and radial velocity observations in order to confirm the transit signal, orbital period, radius and mass. Constraining the period of a candidate can greatly reduce the number of radial velocity measurements needed to confirm the exoplanet and measure its mass \citep[e.g.][]{gill2020a}.

In this paper we set out our search for duotransits from the Cycle 1 and Cycle 3 \tess\ FFI (Full-Frame Image) lightcurves.  In Section~\ref{sec:obs} we describe the \tess\ data that we use to search for duotransit candidates.  In Section~\ref{sec:methods} we outline our search algorithm and vetting procedure.  In Section~\ref{sec:results} we set out the results of our search, including the details of the \Ncand\ duotransit candidates.  Finally in Section~\ref{sec:disc} we discuss the prospects for confirming these duotransit candidates via follow-up photometric and spectroscopic programmes.

\section{TESS observations} \label{sec:obs}
\tess\ is a space-based NASA mission focusing on the discovery of transiting exoplanets around bright stars \citep{Ricker:2015}.  \tess\ has four cameras, each equipped with a grid of four CCDs for a total combined field-of-view of TESS of 24\textdegree\ $\times$ 96\textdegree.  In Cycle 1 (2018 Jul 25 - 2019 Jul 17), \tess\ observed almost the entire Southern Ecliptic Hemisphere in 13 Sectors, each observed for a period of approximately 27\,days. Each \tess\ sector has a gap of approximately 2.4\,days, during which time the satellite downlinks data. There is a similar gap in between Sectors. Other gaps in the data can be due to technical problems with the spacecraft or scattered light (mostly from the Earth) making some image frames unusable \citep[e.g.][]{dalba2020}.  Select stars were observed with 2\,minute cadence, while the full frame images (FFIs) were observed with a cadence of 30\,minutes. 

\tess\ reobserved the Southern Ecliptic Hemisphere in Cycle 3 (2020 Jul 5 - 2021 Jun 24), with the FFIs collected with a shorter cadence of 10\,minutes. During Cycle 3, \tess\ slightly shifted its 
survey fields
in order to cover spatial gaps in between the Cycle 1 Sectors.

 For our search for duotransit candidates, we use the \tess\ full-frame image (FFI) light curves produced by the Science Processing Operations Center pipeline  \citep[SPOC; ][]{jenkinsSPOC2016}, which are publicly available from the Mikulski Archive for Space Telescopes (MAST)\footnote{Accessible at \url{https://mast.stsci.edu/}}. The SPOC pipeline produces FFI light curves for approximately 160,000 stars per sector, based on selection criteria set out in \citet{caldwell2020}.
 
We focus solely on the Southern Ecliptic Hemisphere, as at the time of writing \tess\ had completed its campaign of the Southern Ecliptic in two cycles. We use Sector 1-13 data from Cycle 1, and Sector 27-38 data from Cycle 3. In total, this results in a sample of 1,422,473 stars. We search light curves from the Presearch Data Conditioning Simple Aperture Photometry \citep[PDCSAP;][]{Stumpe2012, Stumpe2014, Smith2012} lightcurves, which have instrumental systematic trends removed from the Simple Aperture Photometry (SAP) flux, but should retain stellar variability including any transit events.
 
 In addition to using flux measurements from the FFI PDCSAP lightcurves, we also make use of the sky background and momentum centroiding measurements provided in the PDCSAP lightcurve files via the keywords SAPBKG and MOMCENTR1/MOMCENTR2 respectively.  These data allow us to rule out a large number of false candidates as described in Section~\ref{sec:vetting}.

\section{Duotransit Search} \label{sec:methods}
Our search for \tess\ duotransit candidates arises out of a more general search for \tess\ monotransits.  We therefore begin by searching the entire set of 1,422,473 SPOC PDCSAP FFI light-curves for monotransit events, and then we match up events to detect duotransit candidates.  

\subsection{Searching for Monotransit Events} \label{sec:monofind}

We begin by downloading the SAP and PDCSAP SPOC FFI light-curve products from MAST. To remove bad data points from the light-curves we discarded data points with \textsc{quality} bit values above 0.  We read the meta information for each light-curve, including the TIC ID, effective temperature, stellar radius, and metallicity if available. We assume a solar metallicity of 0~dex for stars with no metallicity value available. We split each lightcurve into segments defined by gaps in excess of 2.4 hours and normalise each separately. This accounts for gaps in individual orbits due to spacecraft data download times and regions of data flagged by the SPOC algorithm. For each segment, we smooth the lightcurve using an iterative Savitzky–Golay filter (5 iterations using a threshold of 3-$\sigma$) with a window size 48 hours \citep[e.g.][]{2022AJ....163..284H}. This window size strikes a balance of removing stellar activity but ultimately sets an upper limit on the width of transit events our algorithm can detect. Our tests show that transit widths below $\sim2.3$\,days are easily detectable using this filter width. For events wider than this, we find that the Savitzky–Golay filter begins to confuse transit signals with stellar activity and thus the event is not recovered. In addition, our tests showed that the change in cadence between Cycle 1 (30~minutes) and Cycle 3 (10~minutes) had a negligible effect on the filter. For a warm Jupiter in a circular orbit around a Solar type star, we estimate an upper limit to the possible recovered orbital periods of in excess of 1000 days.

\begin{figure}
    \centering
    \includegraphics[width=0.5\textwidth]{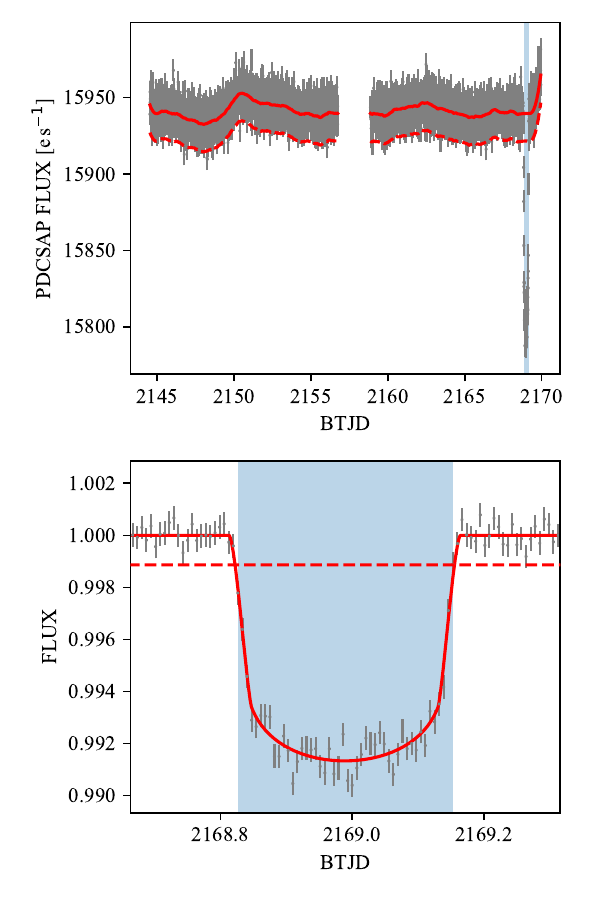}
    \caption{\textbf{Upper panel:} The \tess\ Sector 32 lightcurve of TOI-2447 (black) with detrending model (red solid line) and the 3-MAD detection threshold (red dashed line). A single event was detected highlighted in blue. \textbf{Lower panel:} An inspection of the transit event with best fitting transit model (red solid line).}
    \label{fig:monofind_example}
\end{figure}

Once the lightcurves are normalised and flattened, we pass them to \texttt{monofind}, our custom monotransit detection algorithm.  \texttt{monofind} searches for monotransit events in the lightcurve where three consecutive 30-minute data points are 3 median absolute deviations below the normalised median of the lightcurve. An example of this method is shown in Figure~\ref{fig:monofind_example}, as applied to TOI-2447 (Gill et al., submitted). For the Cycle 1 light-curves we use the native 30 minute data points, while for the Cycle 3 lightcurves we bin the native 10-minute cadence by three to a 30-minute cadence to achieve a consistent detection threshold for transit events.



\subsection{Selecting Duotransit Candidates} \label{sec:vetting}

After searching all the TESS Cycle 1 and Cycle 3 SPOC FFI lightcurves for monotransit events, we cross match detections to search for stars which show exactly one monotransit event in Cycle 1 and one monotransit event in Cycle 3. This yielded a list of 9718 duotransit candidates.

\begin{figure}
    \centering
    \includegraphics[width=\columnwidth]{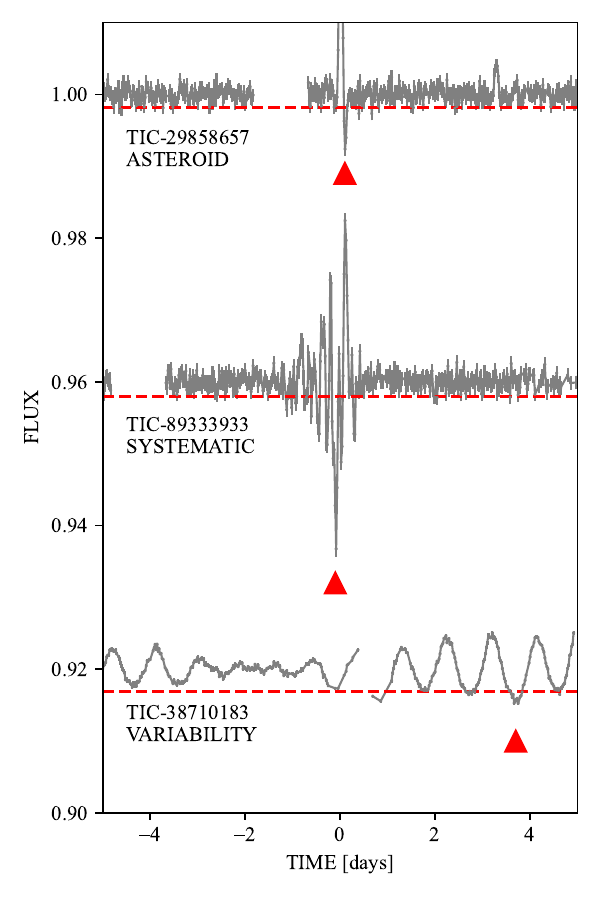}
    \caption{Three example duotransit candidates ruled out during the visual inspection of the transit events in the \tess\ light curves, with an arbitrary flux offset for clarity.  The dotted red line indicates the 3 median absolute deviations detection threshold used in \texttt{monofind}.  The red triangles indicate the event that triggered the detection by \texttt{monofind}.  From top to bottom, these example events are caused by an asteroid passing through the target and background aperture, systematic noise caused by the spacecraft, and changes in amplitude of a variable stars.}
    \label{fig:prevetted}
\end{figure}

We then performed a visual inspection of these 9718 lightcurves to check which of these showed two events that could possibly be transit events.  This allowed us to remove obvious false positives that had triggered the \texttt{monofind} algorithm but did not have two transit-shaped events in the lightcurve.  In most instances these were some kind of variable star. This quick visual inspection took us down to a total of 736 reasonable duotransit candidates. The majority of candidates that did not pass this vetting step were caused by asteroids passing through the target and/or background pixel apertures, systematics noise events due to the spacecraft, or changes in the amplitude of variable stars (see examples presented in Figure~\ref{fig:prevetted}).

We examined each of these duotransit candidates in order to ascertain which appeared to be good transiting exoplanet candidates, and which were false-positives. Examples of false-positive scenarios included those causing transit-like events such as eclipsing binary stars, \tess\ spacecraft systematics (in particular momentum dumps), stellar variability, and Solar System asteroids passing through the background and target photometric aperture.

We are able to identify events caused by asteroids crossing the photometric aperture by checking for variations in the sky background (SAPBKG) that correlate with the monotransit event. We also noticed that such events were typically very asymmetrical in shape, and therefore unlike a bona-fide monotransit event. In total, we identified 170 of our initial duotransit candidates as being due to asteroids crossing the photometric aperture in Cycle 1 or Cycle 3.

We identified monotransit events that were likely caused by eclipsing binaries by examining the results from the transit/eclipse fit from \texttt{monofind}. We designated any event with a modeled companion radius $>2\,$\rjup\ ($\sim22\,$\rearth) to be an eclipsing binary, and removed it from our duotransit candidate list. Such well-separated planets are not expected to be inflated like hot Jupiters \citep[e.g. see figure 2 from ][]{lopez2016} and thus this limit is justified.  We also designated a candidate as an eclipsing binary if there was evidence of a secondary eclipse in the lightcurve. Furthermore, we excluded events which had significant depth differences between Cycles 1 and 3. In total we identified 384 eclipsing binaries from our our initial duotransit candidate list.

We searched for blended eclipsing binaries by inspecting the lightcurves of nearby stars, by checking for centroiding offsets during the monotransit events, and by examining the Target Pixel File (TPF) using our custom \texttt{spoctpf} tool.  \texttt{spoctpf} allows us to make a lightcurve for any pixel or set of pixel in the TPF, and helps determine if the monotransit event is on the star of interest, or is on a nearby star.  In total we identified 58 blended eclipsing binaries from our our initial duotransit candidate list.

We also examined the pre-normalised and pre-flattened PDCSAP lightcurves, as well as the SAP lightcurves. This was to ensure that neither the PDCSAP algorithm nor our own normalisation and flattening had significantly altered the shape of the detected monotransit events. We also inspected by-eye the full unflagged SPOC lightcurve for additonal transit events that may have been excluded from the PDCSAP data.

Finally we checked if any of the duotransit candidates had an associated Data Validation Report \citep[DVR,][]{twicken2019}. DVRs are created for potential \TESS\ planets candidates processed with the SPOC pipeline and contain initial findings about the system including a detailed model fit and analysis of nearby stars to exclude blend scenarios. If a DVR determined a candidate to be a false-positive, we investigated the target further to assess if it warranted exclusion from our list.

Some candidates displayed depth differences between the Cycle 1 and Cycle 3 events. SPOC lightcurves use unique masks for each individual Sector which may lead to different amounts of dilution from neighboring stars depending on telescope pointing and orientation for each sector \citep{bryant2023}. This should be accounted for in the detrended SPOC data (PDCSAP) but might not be perfect, and thus small sector-dependent depth differences are expected ($\lessapprox$2 ppt), particularly in crowded fields. Some candidates had larger depth and width differences, which could be multi-planet systems or blended eclipsing binaries. However, since they do not meet our criteria for selection as duotransit candidates, they were excluded from our list. 
We also found 39 duotransit candidates that 
we associate with spacecraft systematics (stray light, momentum dumps etc.\footnote{\tess\ Instrument Handbook; \url{https://archive.stsci.edu/tess/tess_drn.html}}). This number also includes a  few events resulting from stellar variability that was poorly matched by our detrending model. 

\begin{figure}
    \centering
    \includegraphics[width=0.5\textwidth]{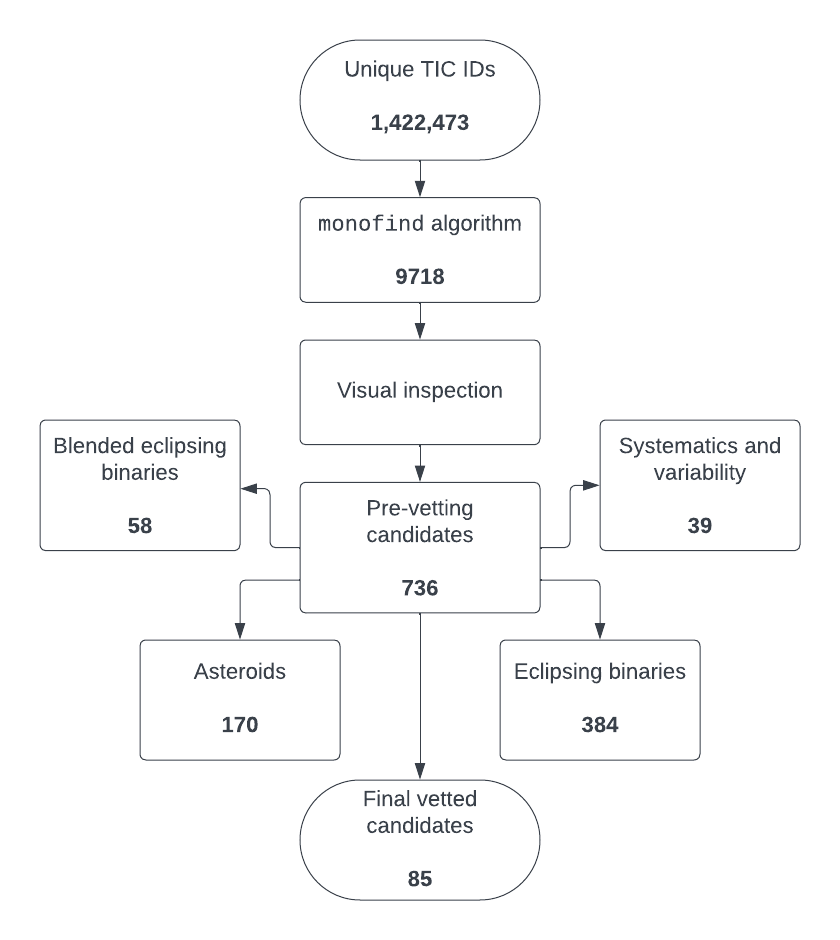}
    \caption{Flowchart of the \texttt{monofind} algorithm candidates and the vetting process that followed, with the number of candidates categorised at each stage as blended eclipsing binaries (`blends'), asteroids, eclipsing binaries, and other systematics and stellar variability features.}
    \label{fig:flowchart}
\end{figure}

A flowchart illustrating the candidate search and vetting process, and the associated numbers of candidates categorised at each stage, is shown in Figure~\ref{fig:flowchart}.

In summary, out of the 1,422,473 unique TIC IDs, 9718 duotransit candidates were identified by the \texttt{monofind} algorithm, and after visual inspection this was reduced to 736 candidates pre-vetting. As a result of the vetting process, 58 objects were classified as blended eclipsing binaries (`blends'), 170 as Solar System asteroids, 384 as eclipsing binaries, and 39 as events caused by either spacecraft systematics or poorly-detrended stellar variability. We list a sample of these events along with their designation in Table~\ref{tab:extracand}. The vetting process left us with a final number of \Ncand\ duotransit candidates that we present as likely planetary in nature in this paper.

\subsection{Modelling Candidate Duotransit Events}
\label{sec:modelling}

In order to derive planetary parameters for the candidates, as well as the most likely period aliases, we must fit the available transits.
However, most transit model fitting requires a priori knowledge of the orbital periods, and is certainly not optimised to sample the extremely narrow but widely-separated period regions found for duotransits.
We used the \texttt{MonoTools} package \citep{Osborn2022code} which fits transits in a way agnostic of orbital period and then computes the posterior density function of the planet candidate from the orbital velocity implied from the transit impact parameter, radius ratio and duration.
This has previously been used to model other duotransiting planet candidates that were subsequently confirmed \citep[e.g.][]{osborn2021,osborn2022}.

\begin{figure*}
    \centering
    \includegraphics[width=0.7975\textwidth]{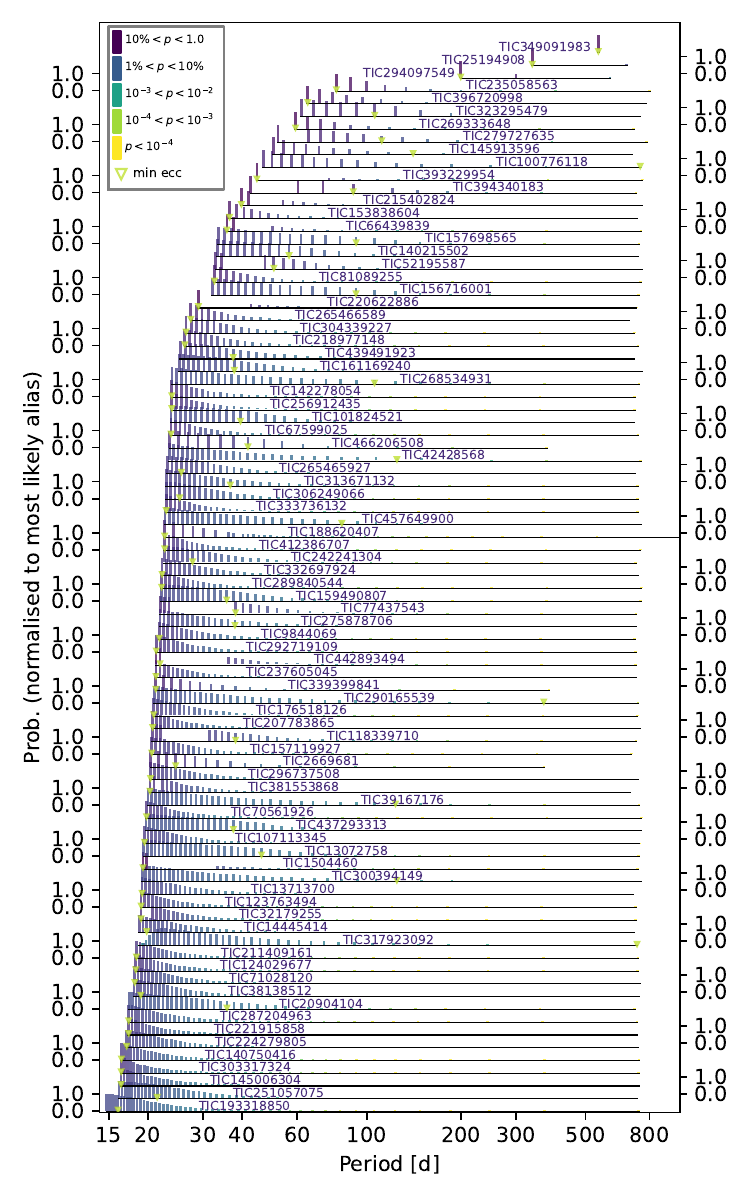}
    \caption{Probability distribution of possible periods for each duotransit candidate as computed by \texttt{MonoTools}, normalised to the most likely period alias. The vertical colourbars refer to six steps in the log probability of each alias, and the green triangles represent the minimum eccentricity aliases.}
    \label{fig:perioddist_monotools}
\end{figure*}

\texttt{MonoTools} fits transits using the \texttt{exoplanet} python library \citep{exoplanet:exoplanet}.
Stellar parameters from the \tess\ Input Catalogue \citep[TICv8;][]{Stassun2019} were used as priors for each of the fits; as was the \citet{Kipping2013_ecc} eccentricity prior, the \citet{Espinoza2016} impact parameter prior, and quadratic limb-darkening parameters constrained using theoretical predictions for the \tess\ bandpass of \citet{Claret2017}.
In the majority of cases, a cubic spline was fit to the transit-masked \tess\ PDCSAP flux in order to pre-flatten the lightcurve and a window of 5.5 transit durations was cut around the transit.
However, after modelling some of these fits appeared to poorly model the out-of-transit variability. 
In these cases we turned to a simple harmonic oscillator (SHOTerm) \texttt{celerite} Gaussian process \citep{foreman2017celerite}, which we constrained by first sampling out-of-transit data and then using the inferred hyperparameter distributions as priors for a combined model fit.
A log scatter parameter was used to encapsulate additional noise for each transit.
The transit model was then sampled using the Hamiltonian Monte Carlo of \citet{exoplanet:pymc3} using a burn-in phase of 800 steps and 1500 samples on each of four independent chains, resulting in typical effective sample sizes of 2500-3000 for each parameter for each candidate. The derived parameters are shown in full online but summarised in Table \ref{tab:candidate_planetprops}. The probabilistic planetary period distributions are shown in Table \ref{tab:candidate_planetprops} and in Figure~\ref{fig:perioddist_monotools} ordered from shortest to longest.

\section{Results} \label{sec:results}

\begin{figure}
    \centering
    \includegraphics[width=0.48\textwidth]{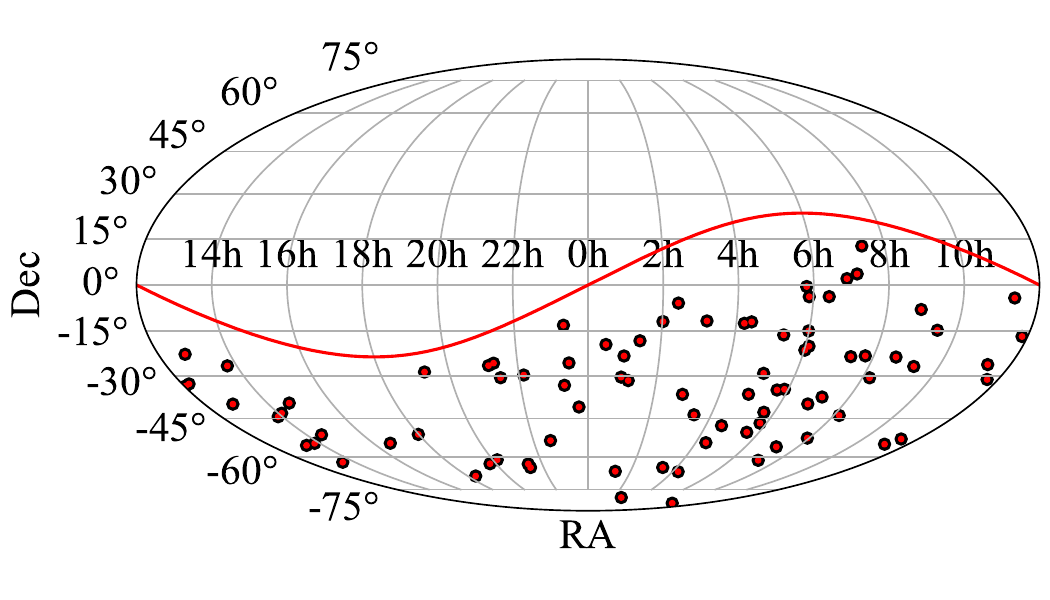}
    \caption{The ecliptic plane (solid red line) with the positions of the \Ncand\ duotransit candidates marked with red-inset black points.}
    \label{fig:SPOCall}
\end{figure}

\subsection{\Ncand\ Duotransit Planet Candidates}
\label{sec:candidates}
Following the methods described in Section\,\ref{sec:methods}, we have found a total of \Ncand\ duotransit planet candidates from our \texttt{monofind} search and vetting of the SPOC FFI lightcurves for the Southern Ecliptic \TESS\ fields (Sectors 1-13, 27-34). The stellar properties of our candidates are set out in Table~\ref{tab:candidate_stellar}, where N\textsubscript{sec} denotes the total number of \tess\ sectors in which the target was observed. The transit event parameters are set out in Tables~\ref{tab:candidate_events} \& \ref{tab:candidate_planetprops}, including the central transit times $T_{\rm c1}$ and $T_{\rm c2}$ of the Cycle 1 and Cycle 3 events respectively, and the separation between the two events in days $\Delta\,T_c$. We also include the \gaia\ flags of NSS (Non-Single Star), where 0 denotes a good single-star solution, 1 denotes an astrometric binary, 2 denotes a spectroscopic binary and 3 denotes an eclipsing binary; and the \gaia\ robust radial velocity (RV) amplitude in \kms, calculated using the standard deviations of individual RV measurements as set out in \gaia\ Data Release 3 \citep[DR3,][]{gaiadr3}. We do not remove candidates from our target list based on these values, however, since an NSS designation is often consistent with the presence of a planet and it is not clear how robust the \gaia\ DR3 RV amplitudes are \citep[e.g.][]{seabroke2021}. In Table~\ref{tab:candidate_planetprops}, we also show the most probable value of $P_{n}$ ($P_{\rm marg}$) and the corresponding value of $n$ ($N_{\rm alias}$).

The transit events for all \Ncand\ duotransit candidates are plotted in Figures~\ref{fig:lightcurves0}-\ref{fig:lightcurves4}.  We plot the best fit transit and eclipse model for each event from the \texttt{monofind} algorithm (see Section~\ref{sec:monofind}). 
We also plot the sky distribution of the candidates in Figure~\ref{fig:SPOCall}.

\begin{figure*}
    \centering
    \includegraphics[width=\textwidth]{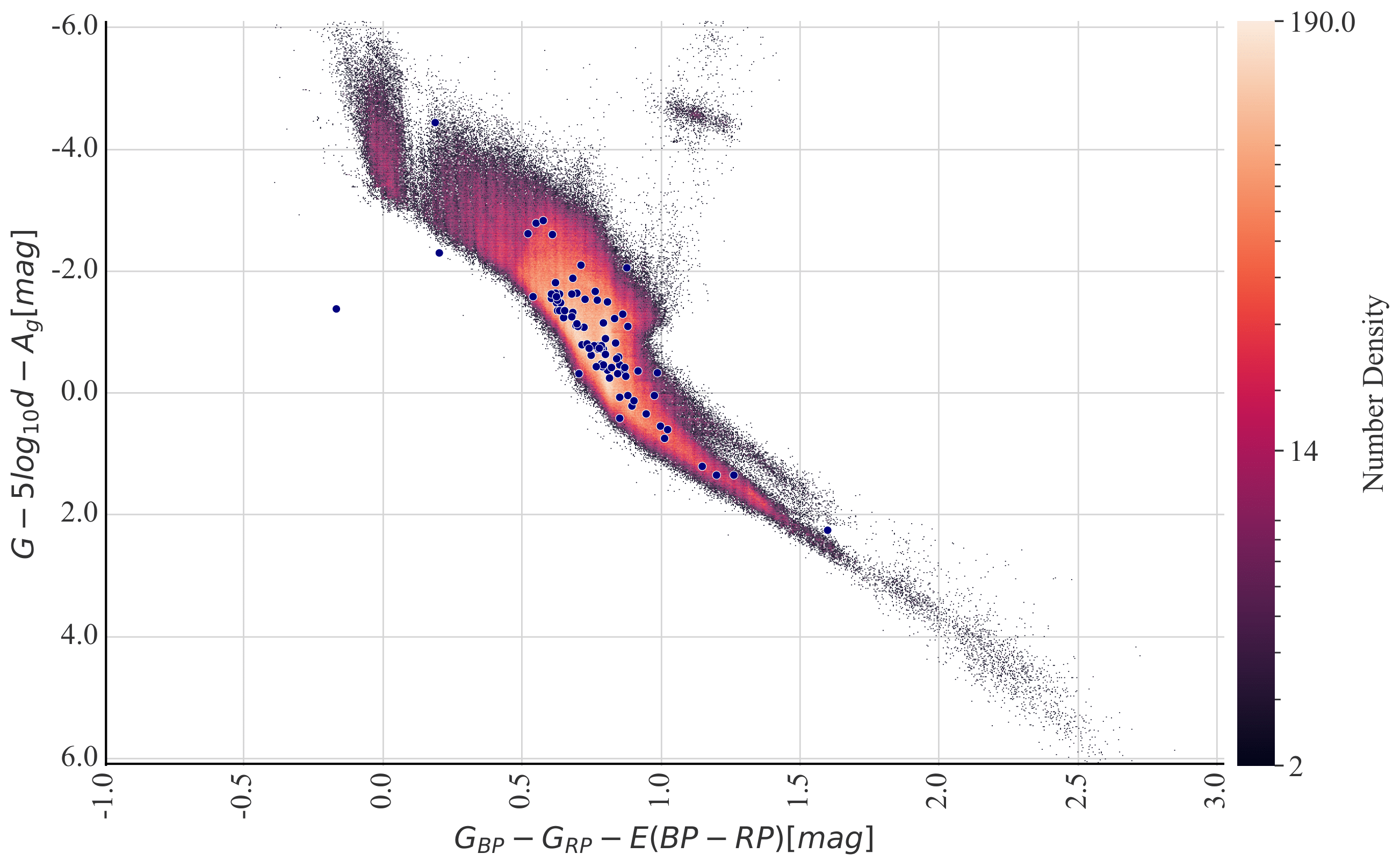}
    \caption{\gaia\ DR3 \citep{gaiadr3} colour-magnitude diagram (corrected for extinction and reddening) showing all stars with \tess\ full frame image SPOC lightcurves from Cycle 1 and Cycle 3 that have measured \gaia\ DR3 distances.  Blue circles indicate the positions of our \Ncand\ duotransit candidates. The colourbar shows the relative number density of stars at each location in the diagram.  For six of the duotransit candidates there was no \gaia\ DR3 distance measurement (see Table~\ref{tab:candidate_stellar}) so we estimate the distance from the parallax and do not correct for reddening or extinction.}
    \label{fig:HR}
\end{figure*}

Figure~\ref{fig:HR} shows where our duotransit candidate host stars lie on the HR diagram, with most on the main-sequence spanning spectral types A-K, some on the main-sequence turn off, and one in the hot sub-dwarf regime (see Section \ref{sec:individual}). Our candidates span a range of brightness between 8\,$<$\,\tmag\,$<$\,14, with most candidates lying between 10\,$<$\,\tmag\,$<$\,13; see Figure~\ref{fig:tmagdist}.  This magnitude range is largely governed by the magnitude distribution of the SPOC FFI lightcurves that cuts off at 
around 
\tmag\,=\,13.5 due to the number of lightcurves per sector being restricted to approximately 160,000 \citep{jenkinsSPOC2016}.

We calculated the total SNRs (Signal-to-Noise Ratios) for each of our candidates as a function of the SNRs of each individual transit event. The SNR of each transit event was estimated using the double box approximation model described by \cite{kipping2023}. Using this square well transit model, the SNR for each transit event with a depth $\delta$ in ppm, a duration $T_\text{dur}$ in hours, light curve noise $\sigma$ in ppm, and a contamination ratio $C$ was calculated using Equation~\ref{eq:singleSNR}:

\begin{equation}
    \text{SNR} = \frac{1}{C}\times\frac{\delta}{\sigma}\sqrt{T_\text{dur}} .
    \label{eq:singleSNR}
\end{equation}

The contamination ratio and noise value were both extracted directly from the headers of the SPOC lightcurves. In some cases these values were missing from the headers of the Cycle 3 lightcurves, in such instances we used the same noise as the Cycle 1 lightcurve. To combine the SNR across both transit events we add the individual SNR for each event in quadrature.

\begin{figure}
    \centering
    \includegraphics[width=0.48\textwidth]{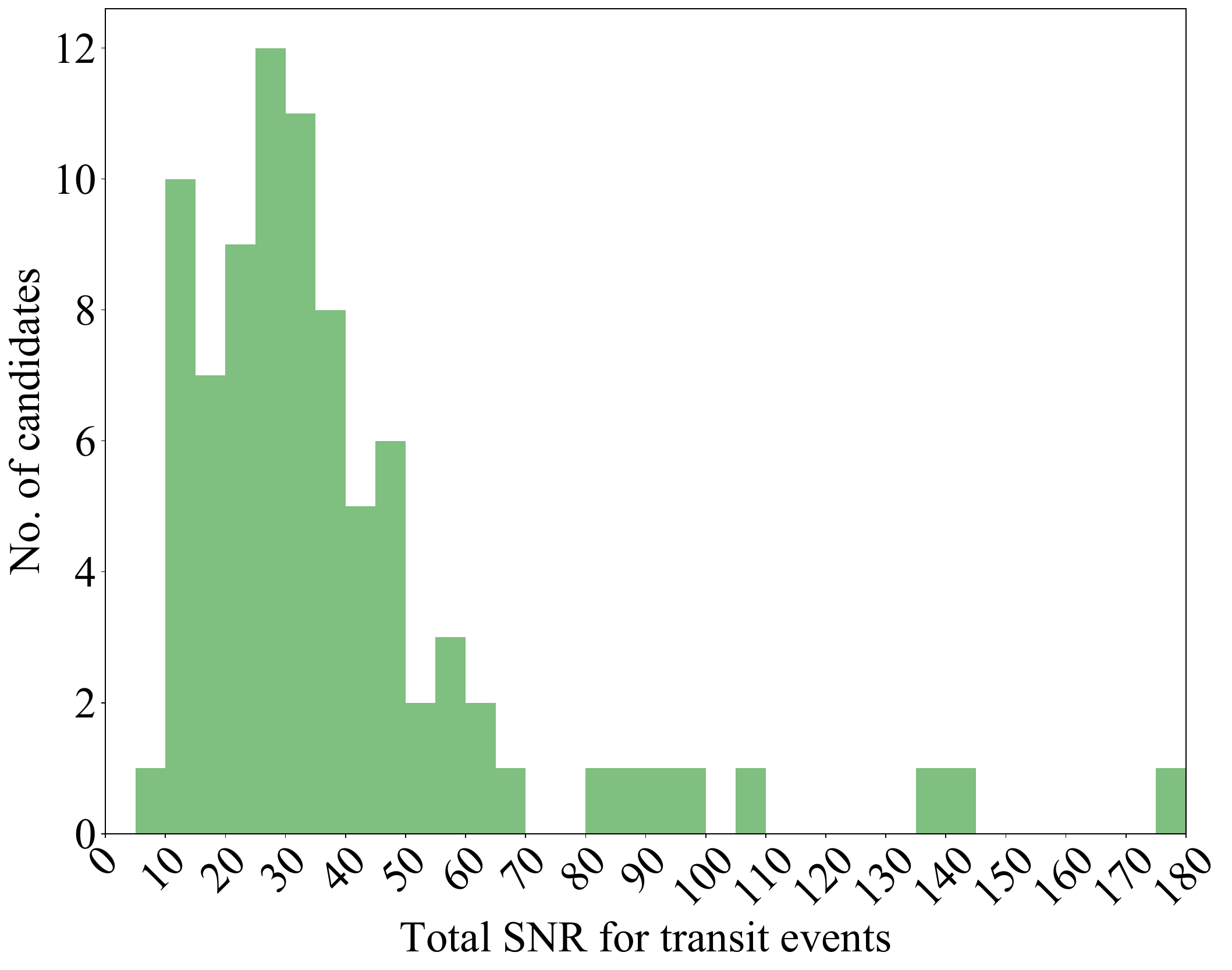}
    \caption{The distribution of total transit event SNR values (see Section~\ref{sec:candidates}) for our \Ncand\ duotransit candidates.}
    \label{fig:snrdist}
\end{figure}

We plot the distribution of the total SNR values for our \Ncand\ candidates in Figure~\ref{fig:snrdist}, and find that most candidates have a SNR between 10-70, with the distribution peaking at SNR values of 25-30. There is one candidate with a total SNR value below 10 and four candidates with a SNR value above 100.

\subsection{Candidates of Special Interest}
\label{sec:individual}

Based on our \gaia\ H-R diagram in Figure~\ref{fig:HR}, we identify three candidates that are distinct from the rest.

\subsubsection{A1V type host star, TIC-221915858}

TIC-\href{https://exofop.ipac.caltech.edu/tess/target.php?id=221915858}{221915858} (candidate 45) is the most luminous star in our sample and resides in the region of the colour-magnitude diagram where the main-sequence approaches the sub-giant branch (Figure~\ref{fig:HR}). This is complemented by \gaia\ DR3 stellar parameters suggesting that TIC-221915858 is a hot ($\teff$ = 9000\,K) A1 star with a RUWE of 0.814 indicative of a good astrometric solution. The fitted spectral energy distribution in TIC V8, which uses \gaia\ DR2 along with 2MASS colours, suggests TIC-221915858 is a main sequence A1 ($\teff = 9475 \pm 187$\,K) star with mass and radius of $2.41 \pm 0.33$\,\msun\ and $2.47 \pm 0.09$\,\rsun\ respectively. In addition to being the most luminous host star in our sample, TIC-221915858 is also the hottest on the main sequence and represents a under-sampled population of exoplanets around hot stars. If a planet is confirmed around TIC-221915858, it would be the hottest planet host star discovered by the \tess\ mission\footnote{exoplanetarchive.ipac.caltech.edu (as of 2023 Oct 10)}. The 5.74\,ppt transit depth is detectable with ground-based photometric facilities and so it is possible to recover the orbital period with ground-based instruments. However, recovering the spectroscopic orbit will be challenging, in-part due to the typical rotation and lack of absorption lines in A-type stars making it difficult to measure a precise CCF (Cross-Correlation Function) centre. Doppler tomography \citep[e.g.][]{2019MNRAS.490.1991W} has been used successfully to determine the radial-velocity semi-amplitude of fast-rotating stars with broad CCFs \citep{2018MNRAS.480.5307T,2019MNRAS.490.2467T} and could be applicable to TIC-221915858 once the orbital period is known. We provide the light curves and transit models for this candidate in Figure~\ref{fig:lightcurves1}.

\subsubsection{Late K-dwarf host, TOI-4310/TIC-303317324}
While the majority of our duotransit candidates orbit F, G and early K type dwarf stars, the candidate TOI-4310 (TIC-\href{https://exofop.ipac.caltech.edu/tess/target.php?id=303317324}{303317324}, candidate 65) is a late K-dwarf star with $\teff$\,=\,4159\,K.  This is an outlier amongst our candidates on the \gaia\ colour-magnitude diagram (Fig~\ref{fig:HR}), positioned much further down the main sequence than any of our other candidates.  The estimated radius of TOI-4130 is 0.72\,\rsun, and the transit model indicates the transiting planet would have a radius of 2.67\,\rearth, making it one of the smallest of our duotransit planet candidates.  We provide the light curves and transit models for this candidate in Appendix Figure~\ref{fig:lightcurves0}.

\subsubsection{Hot subdwarf host, TOI-709/TIC-396720998}
TOI-709 (TIC-\href{https://exofop.ipac.caltech.edu/tess/target.php?id=396720998}{396720998}, candidate 78) has two $\sim6$~ppt deep transits with durations of $\sim$4.3~hours. It is the only candidate in our list categorised as a hot subdwarf host star \citep[LB 1721;][]{culpan2022}, residing on the $G_{BP} - G_{RP} < 0$ region of the colour-magnitude diagram in Figure~\ref{fig:HR} outside of the main population of exoplanet hosts from our candidates. Hot subdwarfs are evolved compact stars, mainly resulting from enhanced mass-loss at the tip of the red giant branch, which likely occurred due to binary interaction \citep{han2002, han2003, maxted2001, pelisoli2020}. This makes them targets of interest for probing the survivability of planets to both stellar evolution and binary environments. Most hot subdwarf companions fall into two populations containing either close white dwarf or M-dwarf/brown dwarf companions or wide FGK-type companions \citep[e.g.][]{2022A&A...666A.182S}, with no planetary companions confirmed to date \citep[e.g.][]{vangrootel2021, thuillier2022}.

Parameters from TICv8 poorly constrain the mass and a radius of the host to $0.5\,\pm\,0.3$\,\msun\ and $0.15\,\pm\,0.11$\,\rsun\ ($\log~g = 5.8\pm0.9$), which are consistent with a hot subdwarf and make it the smallest candidate host star in our list. \citet{jeffery2021} obtained for this star a $\teff$ of $45,600\,\pm\,1000$\,K and a consistent \logg\ of $6.08\,\pm\,0.04$ and classified it as helium-rich sdO (He-sdO). They reported no radius or mass estimate. To derive a radius consistent with the precise $\teff$ and \logg\ of \citet{jeffery2021}, we performed a spectral energy distribution (SED) fit using {\tt speedyfit}\footnote{https://speedyfit.readthedocs.io/}. We employed spectral models from the T\"{u}bingen NLTE Model-Atmosphere Package \citep[TMAP,][]{werner1999, werner2003, rauch2003}. Initial attempts revealed the existence of excess flux towards the red/near-infrared compared to the hot subdwarf model. We therefore included a second component in the fit, modelled with ATLAS9 spectra \citep{atlas9}. The \logg\ was kept fixed at the spectroscopic value for the hot subdwarf (as it is poorly constrained by a SED fit), and a Gaussian prior was applied on the temperature. The temperature of the companion star was allowed to vary freely within 3500 and 6000~K \citep[restricted by the lack of contribution to the spectrum in][: higher temperatures would lead to visible lines in the spectrum; for lower temperatures, the hot subwarf would completely dominate and a companion would have no effect]{jeffery2021} and the \logg\ was left to vary in the range of 4 to 5. A prior on the {\it Gaia} parallax was also applied, which enables the determination of precise radii. The MCMC fit converged to radii of $0.18^{+0.06}_{-0.03}$\,\rsun\ and $0.88^{+0.30}_{-0.15}$\,\rsun\ for the hot subdwarf and the companion, respectively. The companion \teff\ was found to be $5100\pm120$~K, consistent with a K-type main sequence star. FGK-type companions are found with 30 per cent of hot subdwarfs \citep{stark2003}, and their periods are in the range of 100-1000~days \citep[e.g.][]{vos2019}, which could be consistent with the orbital period of the duotransit signal.   
We note that there is also another much deeper (5.8\%) transit/eclipse signal with a 32~day period that is reported in TFOP, and was the reason this candidate was flagged as a TOI. However that signal is distinct from the duotransit signal that we report in this paper.  We provide the light curves and transit models for this candidate in Appendix Figure~\ref{fig:lightcurves3}.
\subsection{Other notable candidates}

\subsubsection{Continuous Viewing Zone candidates}
Three of our candidates lie inside or close to the \tess\ Continuous Viewing Zone (CVZ), where targets are observed in every Sector. TIC-\href{https://exofop.ipac.caltech.edu/tess/target.php?id=25194908}{25194908} (candidate 8) was observed in Sectors 1-13 during Cycle 1, and Sectors 27-38 in Cycle 3, for a total of 24 Sectors, with only two transits detected. TIC-\href{https://exofop.ipac.caltech.edu/tess/target.php?id=294097549}{294097549} (candidate 62) was observed in multiple Sectors across Cycle 1 and Cycle 3 for a total of 13 Sectors, also with only two transits detected.
This additional TESS coverage rules out most alias periods, and only 2 and 3 periods are possible respectively for these candidates. These are likely to be among the longest periods from our sample, with the remaining allowed periods in the range 337--675\,d and 299--599\,d respectively (see Table\,\ref{tab:candidate_planetprops}). 

TIC-\href{https://exofop.ipac.caltech.edu/tess/target.php?id=349091983}{349091983} (candidate 74) was observed in all Sectors in Cycle 1 with the exception of Sector 5, and again in Cycle 3 in all Sectors with the exception of Sector 35, for a total of 24 Sectors. There is only one possible period for this candidate of approximately $\sim549$~days, and thus the true orbital period is solved from \tess\ alone.

\subsubsection{TOI and CTOI candidates}
\tess\ Objects of Interest (TOIs) are planet candidates that have been vetted by the \tess\ Science Office (TSO), the \tess\ Follow-up Observing Program (TFOP) and designated as promising candidates for follow-up and publication. 11 of our total \Ncand\ candidates have previously been flagged as \tess\ Objects of Interest (TOI), and these are labelled in Table~\ref{tab:candidate_stellar}.

Another 14 of our \Ncand\ candidates have previously been flagged as Community \tess\ Objects of Interest (CTOIs), and these are also labelled in Table~\ref{tab:candidate_stellar}. CTOIs are typically planet candidates submitted by additional community projects outside of the official TOI planet search pipelines, and CTOIs are reviewed by the \tess\ TOI team before being promoted to TOI status and assigned a number.

The majority of CTOIs have been identified by the  \textit{WINE} (Warm gIaNts with tEss) collaboration \citep[e.g.][]{schlecker2020,jordan2020,hobson2021}, the  \textit{PHT} (Planet Hunters \tess) project \citep{eisner2021}, and the \textit{STF} (Single-Transit Finder) project.

\subsubsection{Confirmed and published candidates}
\label{sec:published}
 
Two of our duotransit planet candidates have already been confirmed and published. 
TOI-5153 (candidate 25; TIC-\href{https://exofop.ipac.caltech.edu/tess/target.php?id=124029677}{124029677}) has been published by \citet{ulmermoll2022}. The planet is a large warm Jupiter (M\,=\,$3.26^{+0.18}_{-0.17}$\,\mjup, R\,=\,$1.06^{+0.04}_{-0.04}$\,\rjup) orbiting an F8-type star on a period of 20.33\,days. We find that our fitted planetary radius from Table~\ref{tab:candidate_planetprops} is in agreement with this published value. TIC-\href{https://exofop.ipac.caltech.edu/tess/target.php?id=466206508}{466206508} (candidate 85; TOI-5542) has also been published by \citet{grieves2022}. The planet is an old warm Jupiter orbiting a relatively metal-poor G-dwarf host star on a period of 75.12~days, with a mass of M\,=\,$1.32^{+0.10}_{-0.10}$\,\mjup\, and a radius of R\,=\,$1.01^{+0.04}_{-0.04}$\,\rjup. 
Other publications for these solved systems are forthcoming, e.g. TIC-\href{https://exofop.ipac.caltech.edu/tess/target.php?id=77437543}{77437543} (Henderson et al. 2023, in preparation), TIC-\href{https://exofop.ipac.caltech.edu/tess/target.php?id=77437543}{333736132} (Kendall et al. 2023, in preparation), and TIC-\href{https://exofop.ipac.caltech.edu/tess/target.php?id=224279805}{224279805}.

\subsection{Previous TOIs and CTOIs identified as false positives}
Some of the candidates that were detected in our transit search but rejected as false positives have previously been announced as TOIs or CTOIs. In all cases, we found these to be examples of passing asteroids that caused peaks in the SAP background time series and hence false dips in the target light curve (see Section~\ref{sec:vetting}).
These false positives are are TOI-4312 (TIC-\href{https://exofop.ipac.caltech.edu/tess/target.php?id=251086776}{251086776}) and the CTOIs TIC-\href{https://exofop.ipac.caltech.edu/tess/target.php?id=152070435}{152070435} and TIC-\href{https://exofop.ipac.caltech.edu/tess/target.php?id=275180352}{275180352}.

\section{Discussion} \label{sec:disc}
\begin{figure}
    \centering
    \includegraphics[width=0.48\textwidth]{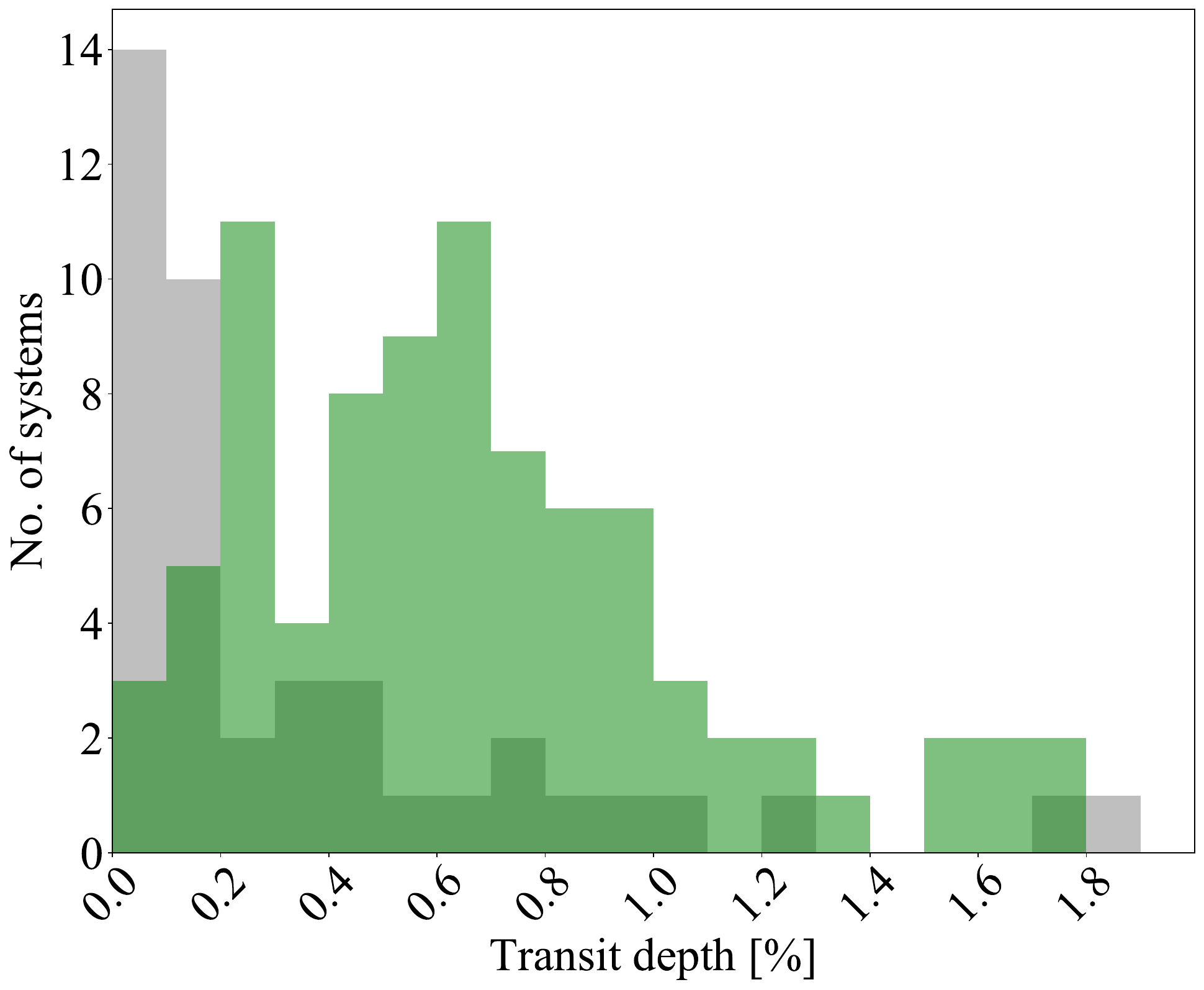}
    \caption{The distribution of transit depths for our duotransit candidates (green) and the confirmed \TESS\ discovered exoplanets from the Southern Ecliptic Hemisphere (grey).}
    \label{fig:depthdist}
\end{figure}

\begin{figure}
    \centering
    \includegraphics[width=0.48\textwidth]{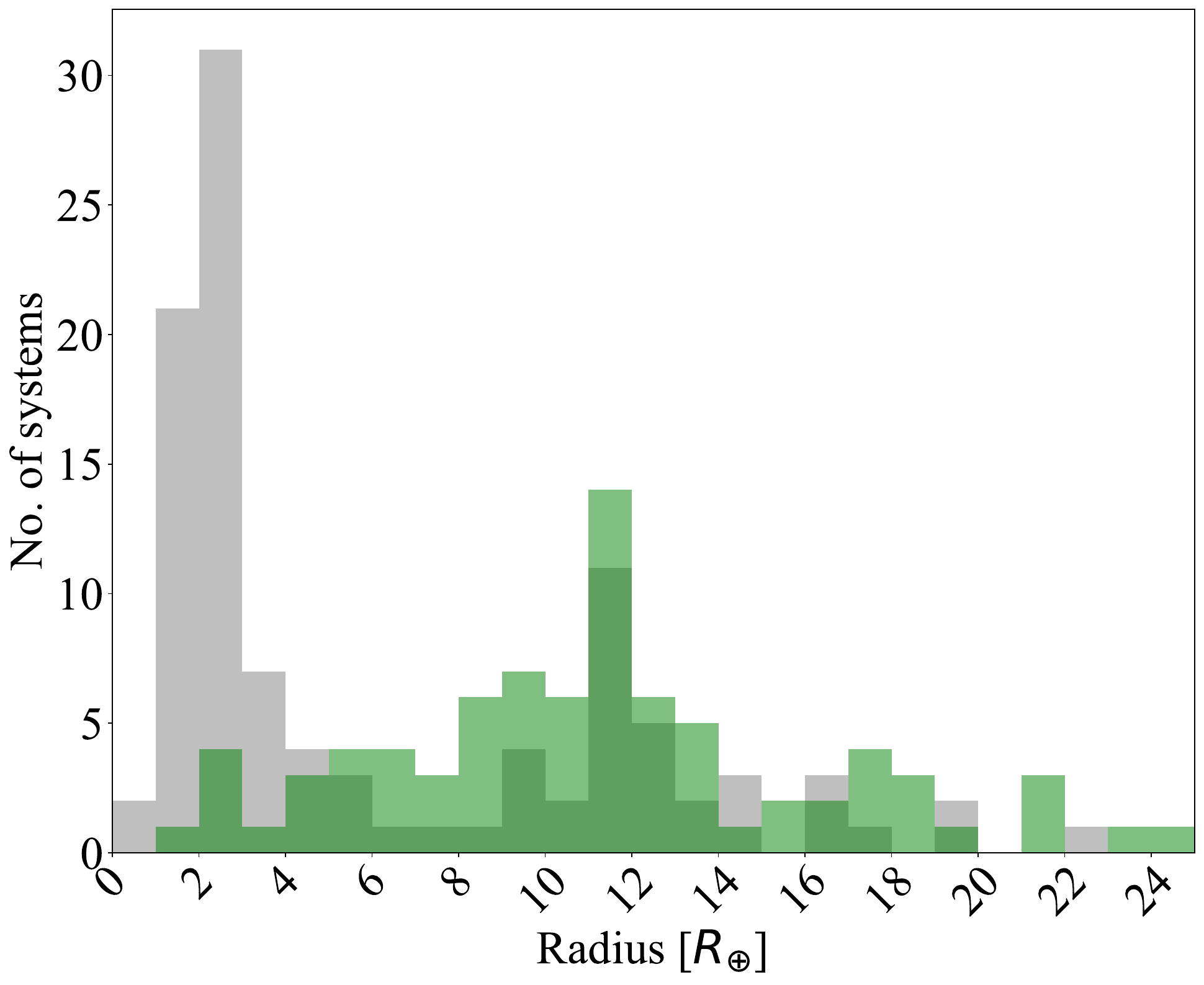}
    \caption{The distribution of planet radii for our duotransit candidates (green) and the confirmed \TESS\ discovered exoplanets from the Southern Ecliptic Hemisphere (grey).}
    \label{fig:radiusdist}
\end{figure}

\begin{figure}
    \centering
    \includegraphics[width=0.48\textwidth]{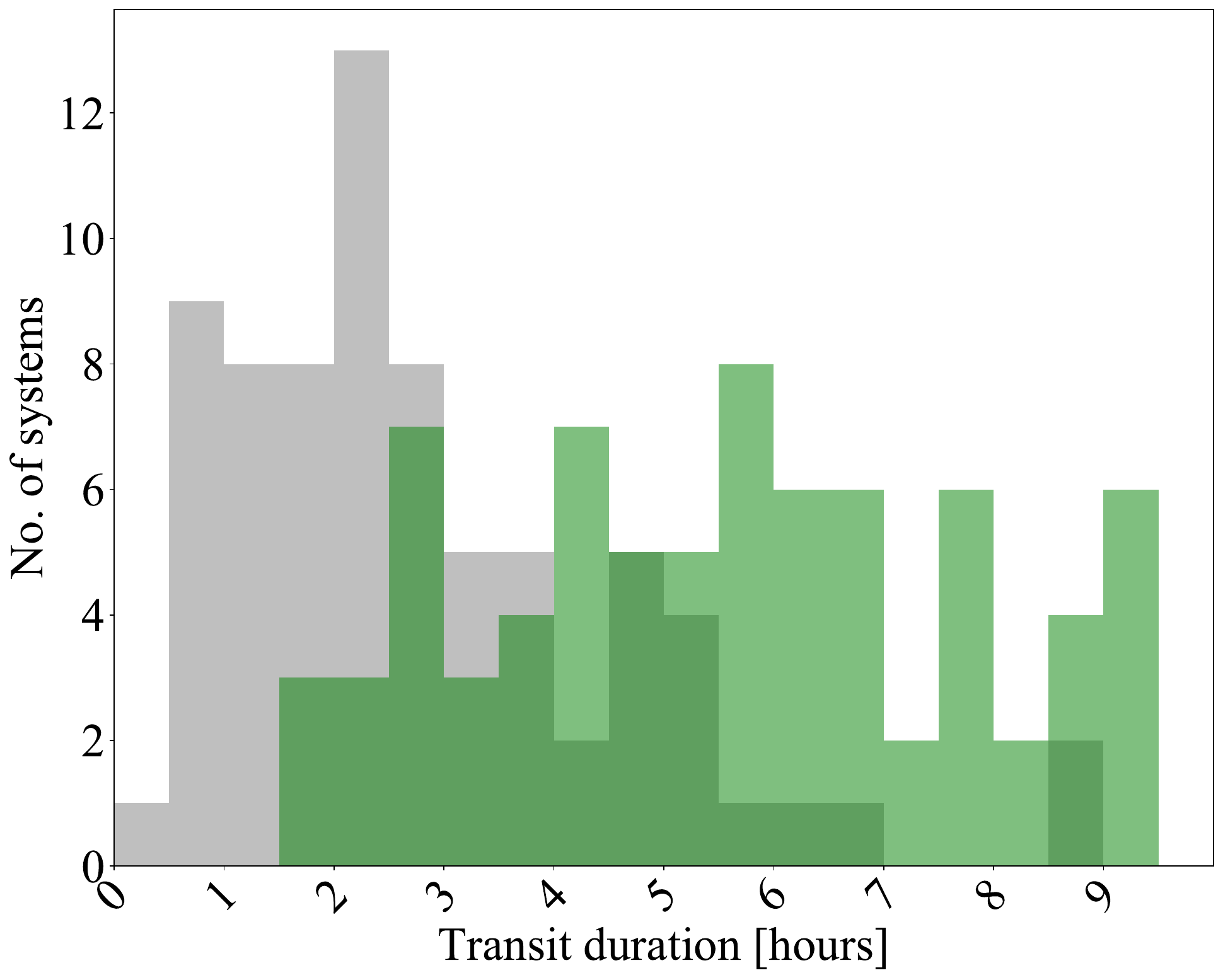}
    \caption{The distribution of transit durations for our duotransit candidates (green) and the confirmed \TESS\ discovered exoplanets from the Southern Ecliptic Hemisphere (grey).}
    \label{fig:durationdist}
\end{figure}

\begin{figure}
    \centering
    \includegraphics[width=0.48\textwidth]{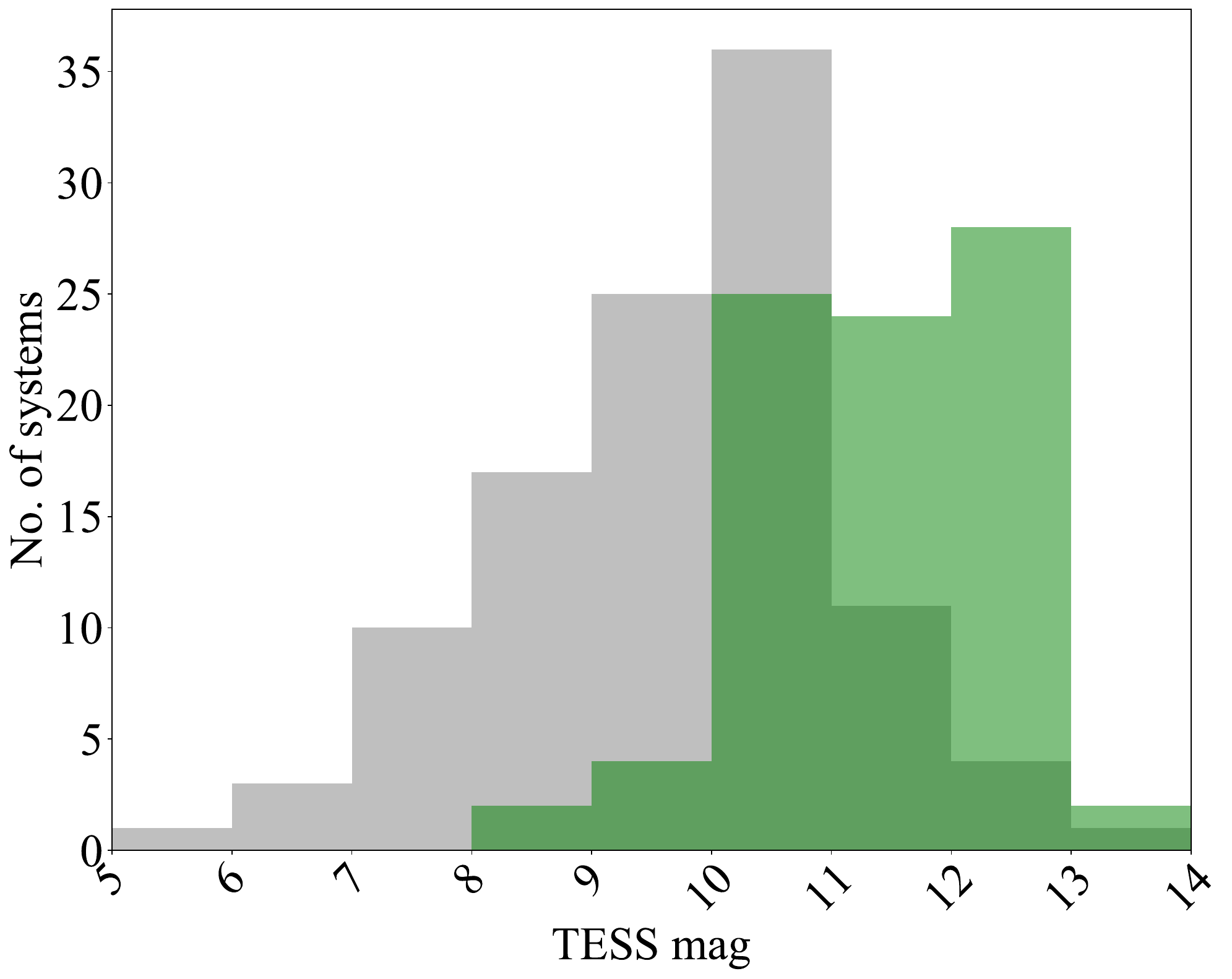}
    \caption{The distribution of \tess\ magnitudes for our duotransit candidate host stars (green) and the confirmed \TESS\ discovered exoplanet host stars from the Southern Ecliptic Hemisphere (grey).}
    \label{fig:tmagdist}
\end{figure}

\begin{figure}
    \centering
    \includegraphics[width=0.48\textwidth]{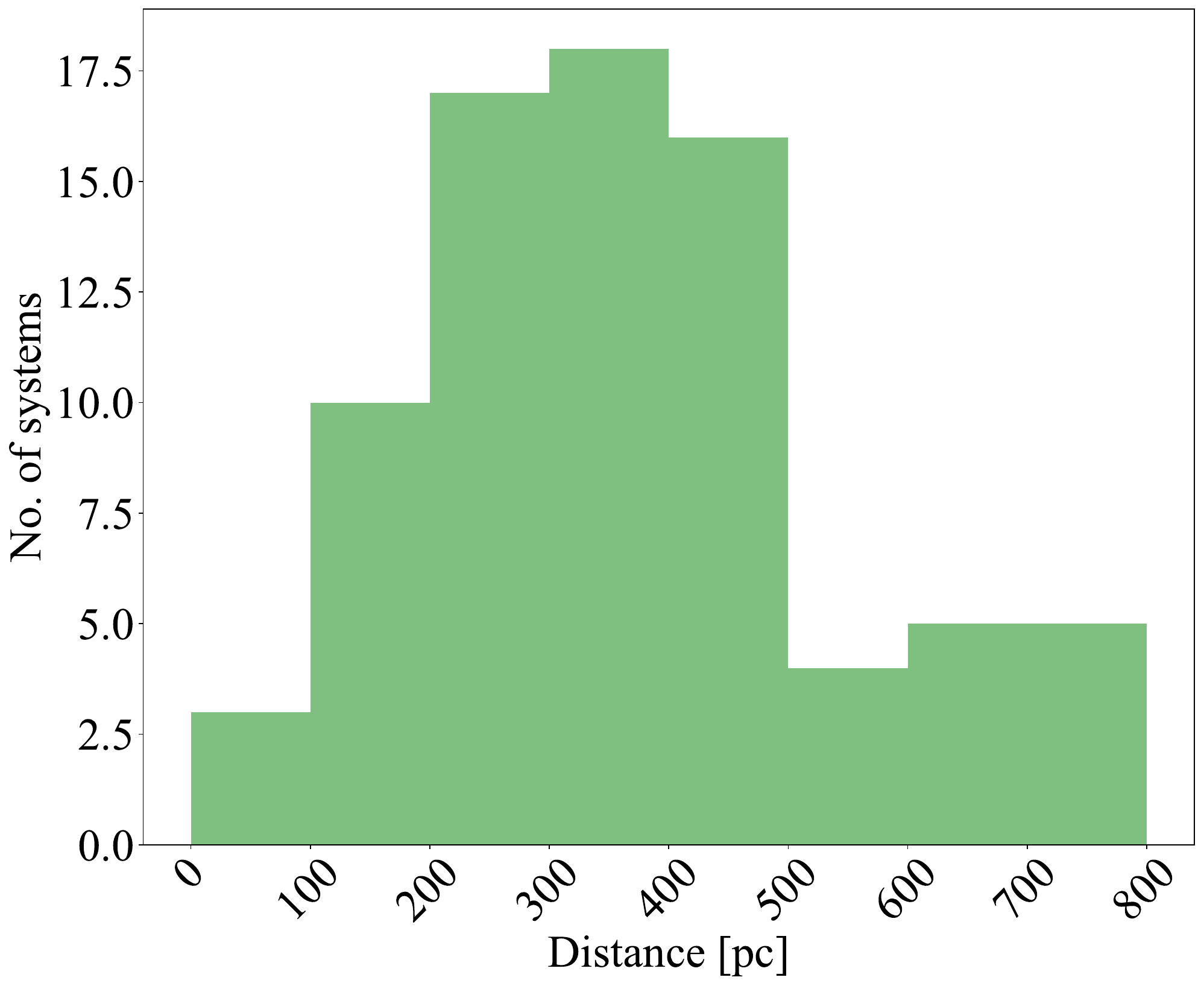}
    \caption{The distribution of \gaia\ DR3 distances (in parsecs) for our duotransit candidate host stars.}
    \label{fig:distdist}
\end{figure}

\begin{figure}
    \centering
    \includegraphics[width=0.48\textwidth]{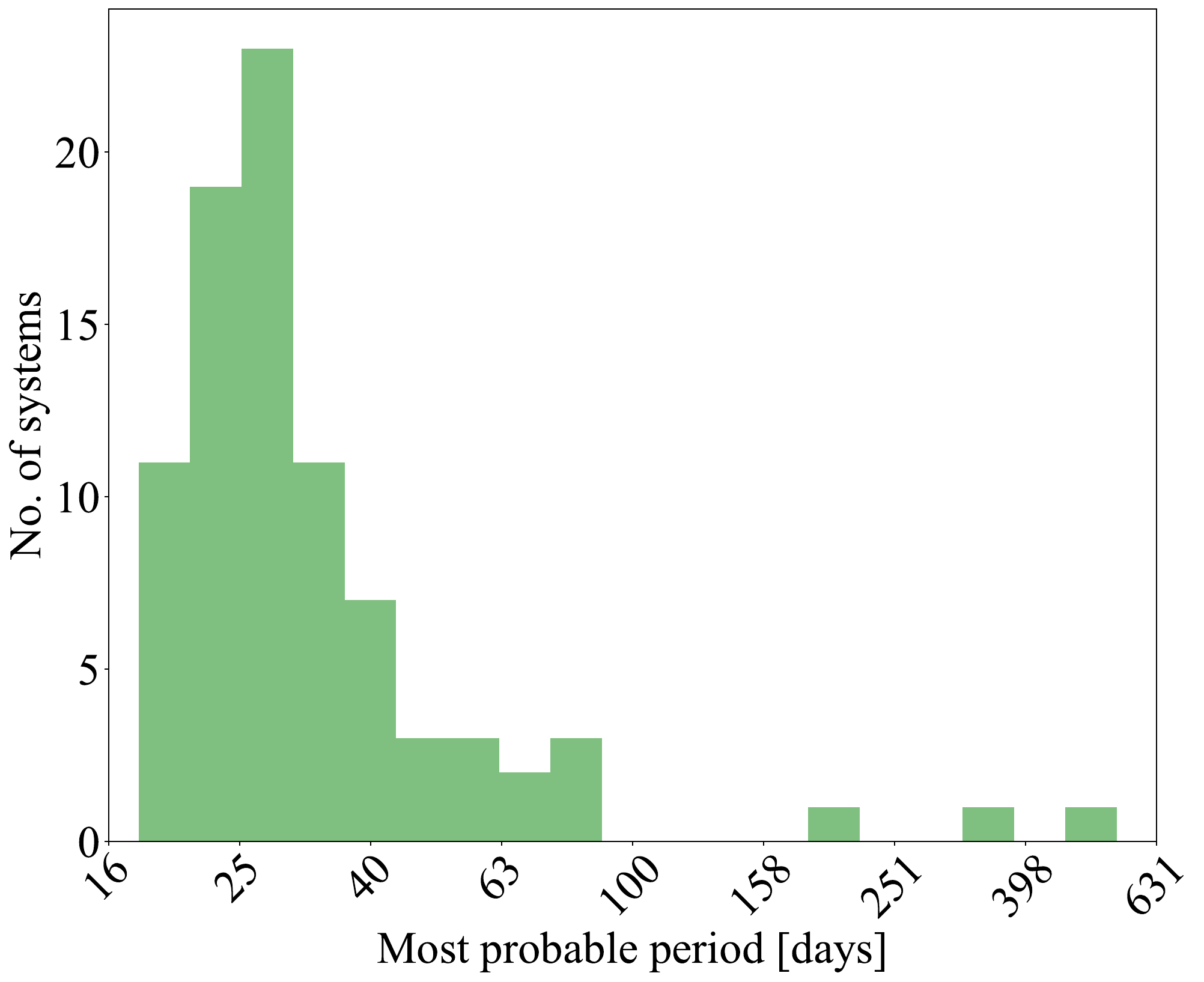}
    \caption{The distribution of most probable periods from modelling with \texttt{MonoTools} for our duotransit candidates.}
    \label{fig:perioddist}
\end{figure}

From our initial set of 1,422,473 unique TIC IDs, 9718 were identified as duotransit candidates from our \texttt{monofind} algorithm in the first instance, and after a quick visual inspection to reject obvious variable stars this was reduced to 736 candidates pre-vetting (see Section~\ref{sec:vetting}. From these candidates, 651 were determined to be caused by false-positive scenarios including blended sources, asteroid crossing events, eclipsing binaries, systematics and stellar variability. This leads to a final list of \Ncand\ duotransit candidates presented in this work (see Section~\ref{sec:results} and Tables~\ref{tab:candidate_stellar}-\ref{tab:candidate_planetprops}).

We have submitted all of our non-TOI/CTOI duotransit planet candidates to the TESS follow-up program (TFOP) (60 candidates).  We hope that this will allow the community to follow-up these interesting systems further with the aim of determining their true orbital periods and providing confirmation and full characterisations of new long-period planets. 

\subsection{Duotransit candidate properties}
\label{sec:properties}

We plot the transit depths of our \Ncand\ duotransit candidates in Figure~\ref{fig:depthdist}, and for comparison we overlay the \tess\ planets that have been confirmed in the Southern Ecliptic Hemisphere\footnote{NASA Exoplanet Archive; \url{https://exoplanetarchive.ipac.caltech.edu/}, accessed 2023 Jul 15}. 
The transit depths of our duotransit candidates range from approximately 0.1 percent to 1.8 percent, peaking around 0.7 percent.  This is significantly deeper than is typical for the confirmed \tess\ exoplanets, the majority of which are less than 0.2 percent. 
This depth difference is to be expected given that the \texttt{monofind} algorithm is designed to detect individual high-SNR transits, which will naturally be deeper than the general TOI population. 

In Figure~\ref{fig:radiusdist} we plot the best-fitting radii of our duotransit planet candidates, again in comparison with the \tess\ confirmed planet sample.
We find that our sample peaks at around the radius of Jupiter (11~\rearth), which is also to be expected for a sample on individual high-SNR transits. Our sample extends to radii exceeding 16~\rearth, and while the sample of confirmed \tess\ planets also extends to similar radii, these are hot Jupiters with radii inflated by irradiation, which we would not expect for long-period planets \citep{gu2003}. Nevertheless, rather than applying an arbitrary upper radius cut, we keep these candidates in our sample for completeness and will rely on follow-up observations to determine the upper radius limit of long-period planets. 





Figure~\ref{fig:durationdist} shows that most of our candidates have significantly longer transit durations than many of the confirmed \tess\ planets. This is expected as we are probing planets with orbital periods in excess of $\sim 17$ days, which is significantly longer than the median orbital period of \tess\ planets ($\sim 5.8$ days). 


In Figure~\ref{fig:tmagdist} we plot the \tess\ magnitudes of the host stars of our candidates, again comparing with the sample of confirmed \tess\ planets. Here it can be seen that our long-period planet candidates tend to orbit stars that are significantly fainter than the confirmed planet sample (typically \tmag\ of 10--13). This is for three reasons. First, the confirmed \tess\ sample is biased towards bright stars, partly because brighter stars make it easier to detect shallow transits, and partly because follow-up efforts tend to be focused on bright host stars. Second, the much smaller transit probabilities of wide-separation planets means that long-period transiting candidates are rare, and so the brightest detected examples will be around fainter stars, which are more numerous (the absence of brighter examples is not a selection effect because such systems would readily detected). Third, gas giant planets are inherently rarer than the sub-Neptune sized planets that make up the bulk of the confirmed \tess\ sample. The steep cut-off in our detected candidates at \tmag=13 simply reflects the selection criteria for the SPOC FFI lightcurves \citep{caldwell2020}.

In Figure~\ref{fig:distdist} we plot the distances of our duotransit candidate host stars from the \gaia\ DR3 catalog. We find that most candidate host stars are at distances from 200-500\,pc, with the furthest object being TIC-\href{https://exofop.ipac.caltech.edu/tess/target.php?id=218977148}{218977148} at a distance of $\sim826$\,pc, and the closest being TIC-\href{https://exofop.ipac.caltech.edu/tess/target.php?id=303317324}{303317324}/TOI-4310 at a distance of $\sim40$\,pc.

As described in Section~\ref{sec:modelling}, we use the shape and width of our detected transit events to determine the most probable orbital period for each candidate \citep[e.g.][]{2022A&A...664A.156O}.
The probability of each allowed alias is indicated on Figure\,\ref{fig:perioddist_monotools} and the most probable period for each candidate is listed in Table\,\ref{tab:candidate_planetprops}. We plot a histogram of these most probable period in Figure~\ref{fig:perioddist}. 
We find that our distribution peaks at periods around 20-30~days.
While these values are the most probable periods according to the information and data available, they are not necessarily the correct values, and the true orbital periods must be determined through follow-up observations. We also note the three candidates in our sample with the longest periods: TIC-\href{https://exofop.ipac.caltech.edu/tess/target.php?id=294097549}{294097549} with a value of $P_\text{marg}$ of 200~days, TIC-\href{https://exofop.ipac.caltech.edu/tess/target.php?id=25194908}{25194908} with a value of $P_\text{marg}$ of 338~days, and TIC-\href{https://exofop.ipac.caltech.edu/tess/target.php?id=349091983}{349091983} with a value of $P_\text{marg}$ of 549~days (see Table~\ref{tab:candidate_planetprops}).

\subsection{Follow-up programme}
\subsubsection{Photometric period determination}
\label{sec:NGTS}

The limited number of discrete period alias for duotransit candidates means that precise orbital periods can be determined efficiently with photometric follow-up targeted at predicted transit times for each alias. 

Our own team is actively following up the duotransit candidates presented here using the Next Generation Transit Survey \citep[\ngts ;][]{Wheatley2018}. \ngts\ is a ground-based facility located at the ESO Paranal Observatory in Chile that deploys an array of twelve telescopes optimised for photometric detection of exoplanet transits \citep[e.g.][]{Bryant2020}. Our monotransit programme with \ngts\ confirmed the first planet that was initially identified as a single-transit event with \tess\ \citep{gill2020a}, and has since confirmed several other long-period planets \citep[e.g.][Gill et al., submitted]{ulmermoll2022,grieves2022}.
We have also confirmed three long-period, low-mass eclipsing binary systems \citep{gill2020b,2020MNRAS.492.1761L,2020MNRAS.495.2713G}.

We are also searching \tess\ Cycle 5 data as it is released for third transits of our duotransit candidates. In some cases this will significantly reduce the number of period aliases to search, although with a similar two-year gap between Cycles 1, 3 and 5 a third transit usually still leaves multiple aliases to be searched.  

\subsubsection{Radial velocity detections}
\label{sec:follow-up}

Our team has also begun a spectroscopic campaign using the \coralie\ spectrograph \citep{coralie} on the Swiss 1.2~m Leonhard Euler Telescope at La Silla Observatory (program ID 500) to help confirm our candidates. With a first \CORALIE\ spectrum, we are able to identify and remove double-line spectroscopic binary (SB2) systems, and detect rapidly rotating stars from the broadening of the cross-correlation function. With a second \CORALIE\ spectrum, taken about one week later, we are able to exclude systems for which the change of measured radial velocity is too large to be consistent with a planetary companion ($>$1~\kms). Bright targets hosting relatively massive planetary companions exhibit radial velocity variations detectable with \CORALIE\ \citep[$\sim6$\ms\ for bright stars; e.g.][]{2013A&A...551A..90M} and for other targets we use \HARPS. 
Previous efforts to confirm these mono- and duotransiting planets using spectroscopic observations have been successful \citep[e.g.][]{ulmermoll2022,grieves2022,ulmermoll2023}.

\section{Conclusion} \label{sec:conc}
We have presented the discovery of \Ncand\ duotransit candidates from the \tess-SPOC FFI light curves from the Southern Ecliptic Hemisphere. The candidates were found using our custom-built \texttt{monofind} algorithm through which all SPOC pipeline data products were passed. We also employed a vetting procedure to rule out false-positive scenarios such as eclipsing binaries and asteroids in the sample. A total of 25 of our candidates have previously been flagged as either TOIs or CTOIs, and we found during our vetting process 3 TOI/CTOI objects outside of our duotransits falsely flagged as such that are asteroids. A dedicated program on the \ngts\ facility is now being used to determine the true orbital periods of these systems, and we welcome other community efforts in this difficult task.  We also have programs on instruments such as \CORALIE\ and \harps\ to acquire radial velocity measurements to confirm the planetary nature of these systems and to determine the mass of the exoplanets. Once confirmed, these planets will form the basis for characterisation studies into their atmospheric compositions and orbital obliquities, and hence into the formation, evolution and migration of warm and cool gas giant exoplanets.


\section*{Acknowledgements}

This paper includes data collected by the \TESS\ mission. Funding for the \TESS\ mission is provided by the NASA Explorer Program. Resources supporting this work were provided by the NASA High-End Computing (HEC) Program through the NASA Advanced Supercomputing (NAS) Division at Ames Research Center for the production of the SPOC data products. The TESS team shall assure that the masses of fifty (50) planets with radii less than 4 REarth are determined.

This work makes use of data from the European Space Agency (ESA) mission {\it Gaia} (\url{https://www.cosmos.esa.int/gaia}), processed by the {\it Gaia} Data Processing and Analysis Consortium (DPAC, \url{https://www.cosmos.esa.int/web/gaia/dpac/consortium}). Funding for the DPAC has been provided by national institutions, in particular the institutions participating in the {\it Gaia} Multilateral Agreement.

We acknowledge the use of public \TESS\ Alert data from pipelines at the \TESS\ Science Office and at the \TESS\ Science Processing Operations Center.

This research makes use of the Exoplanet Follow-up Observation Program website, which is operated by the California Institute of Technology, under contract with the National Aeronautics and Space Administration under the Exoplanet Exploration Program.

This paper includes data collected by the TESS mission that are publicly available from the Mikulski Archive for Space Telescopes (MAST).

FH, TR and AO are supported by STFC studentships. ML acknowledges support of the Swiss National Science Foundation under grant number PCEFP2\_194576. The contribution of MB, ML, HPO, \& SUM  have been carried out within the framework of the NCCR PlanetS supported by the Swiss National Science Foundation under grants 51NF40\_182901 and 51NF40\_205606. JSJ acknowledges support by Fondecyt grant 1201371 and by the ANID BASAL project FB210003. IP acknowledges a Warwick Astrophysics prize post-doctoral fellowship made possible thanks to a generous philanthropic donation. The contributions at the University of Warwick by SG, DB, PJW, RGW and DA have been supported by STFC through consolidated grants ST/P000495/1, ST/T000406/1 and ST/X001121/1.


\section*{Data Availability}
The \tess\ data is accessible via the MAST (Mikulski Archive for Space Telescopes) portal at \url{https://mast.stsci.edu/portal/Mashup/Clients/Mast/Portal.html}. The associated \texttt{python} scripts for analysis and plotting are available upon reasonable request to the authors.


\bibliographystyle{mnras}
\bibliography{bib}

\begin{thebibliography}{}
\makeatletter
\relax
\def\mn@urlcharsother{\let\do\@makeother \do\$\do\&\do\#\do\^\do\_\do\%\do\~}
\def\mn@doi{\begingroup\mn@urlcharsother \@ifnextchar [ {\mn@doi@}
  {\mn@doi@[]}}
\def\mn@doi@[#1]#2{\def\@tempa{#1}\ifx\@tempa\@empty \href
  {http://dx.doi.org/#2} {doi:#2}\else \href {http://dx.doi.org/#2} {#1}\fi
  \endgroup}
\def\mn@eprint#1#2{\mn@eprint@#1:#2::\@nil}
\def\mn@eprint@arXiv#1{\href {http://arxiv.org/abs/#1} {{\tt arXiv:#1}}}
\def\mn@eprint@dblp#1{\href {http://dblp.uni-trier.de/rec/bibtex/#1.xml}
  {dblp:#1}}
\def\mn@eprint@#1:#2:#3:#4\@nil{\def\@tempa {#1}\def\@tempb {#2}\def\@tempc
  {#3}\ifx \@tempc \@empty \let \@tempc \@tempb \let \@tempb \@tempa \fi \ifx
  \@tempb \@empty \def\@tempb {arXiv}\fi \@ifundefined
  {mn@eprint@\@tempb}{\@tempb:\@tempc}{\expandafter \expandafter \csname
  mn@eprint@\@tempb\endcsname \expandafter{\@tempc}}}

\bibitem[\protect\citeauthoryear{{Akeson} et~al.,}{{Akeson}
  et~al.}{2013}]{Akeson2013}
{Akeson} R.~L.,  et~al., 2013, \mn@doi [\pasp] {10.1086/672273}, \href
  {https://ui.adsabs.harvard.edu/abs/2013PASP..125..989A} {125, 989}

\bibitem[\protect\citeauthoryear{{Albrecht} et~al.,}{{Albrecht}
  et~al.}{2012}]{2012ApJ...757...18A}
{Albrecht} S.,  et~al., 2012, \mn@doi [\apj] {10.1088/0004-637X/757/1/18},
  \href {https://ui.adsabs.harvard.edu/abs/2012ApJ...757...18A} {757, 18}

\bibitem[\protect\citeauthoryear{{Borucki} et~al.,}{{Borucki}
  et~al.}{2010}]{kepler2010}
{Borucki} W.~J.,  et~al., 2010, \mn@doi [Science] {10.1126/science.1185402},
  \href {https://ui.adsabs.harvard.edu/abs/2010Sci...327..977B} {327, 977}

\bibitem[\protect\citeauthoryear{{Bryant} et~al.,}{{Bryant}
  et~al.}{2020}]{Bryant2020}
{Bryant} E.~M.,  et~al., 2020, \mn@doi [\mnras] {10.1093/mnras/staa2976}, \href
  {https://ui.adsabs.harvard.edu/abs/2020MNRAS.499.3139B} {499, 3139}

\bibitem[\protect\citeauthoryear{{Bryant}, {Bayliss}  \& {Van Eylen}}{{Bryant}
  et~al.}{2023}]{bryant2023}
{Bryant} E.~M.,  {Bayliss} D.,   {Van Eylen} V.,  2023, \mn@doi [\mnras]
  {10.1093/mnras/stad626}, \href
  {https://ui.adsabs.harvard.edu/abs/2023MNRAS.521.3663B} {521, 3663}

\bibitem[\protect\citeauthoryear{{Caldwell} et~al.,}{{Caldwell}
  et~al.}{2020}]{caldwell2020}
{Caldwell} D.~A.,  et~al., 2020, \mn@doi [Research Notes of the American
  Astronomical Society] {10.3847/2515-5172/abc9b3}, \href
  {https://ui.adsabs.harvard.edu/abs/2020RNAAS...4..201C} {4, 201}

\bibitem[\protect\citeauthoryear{{Castelli} \& {Kurucz}}{{Castelli} \&
  {Kurucz}}{2003}]{atlas9}
{Castelli} F.,  {Kurucz} R.~L.,  2003, in {Piskunov} N.,  {Weiss} W.~W.,
  {Gray} D.~F.,  eds,  International Astronomical Union 210th Symposium Posters
  Vol. 210, Modelling of Stellar Atmospheres. p.~A20 (\mn@eprint {arXiv}
  {astro-ph/0405087}), \mn@doi{10.48550/arXiv.astro-ph/0405087}

\bibitem[\protect\citeauthoryear{{Claret}}{{Claret}}{2017}]{Claret2017}
{Claret} A.,  2017, \mn@doi [\aap] {10.1051/0004-6361/201629705}, \href
  {https://ui.adsabs.harvard.edu/abs/2017A&A...600A..30C} {600, A30}

\bibitem[\protect\citeauthoryear{{Cooke}, {Pollacco}, {West}, {McCormac}  \&
  {Wheatley}}{{Cooke} et~al.}{2018}]{cooke2018}
{Cooke} B.~F.,  {Pollacco} D.,  {West} R.,  {McCormac} J.,   {Wheatley} P.~J.,
  2018, \mn@doi [\aap] {10.1051/0004-6361/201834014}, \href
  {https://ui.adsabs.harvard.edu/abs/2018A&A...619A.175C} {619, A175}

\bibitem[\protect\citeauthoryear{{Cooke} et~al.,}{{Cooke}
  et~al.}{2021}]{cooke2021}
{Cooke} B.~F.,  et~al., 2021, \mn@doi [\mnras] {10.1093/mnras/staa3569}, \href
  {https://ui.adsabs.harvard.edu/abs/2021MNRAS.500.5088C} {500, 5088}

\bibitem[\protect\citeauthoryear{{Culpan}, {Geier}, {Reindl}, {Pelisoli},
  {Gentile Fusillo}  \& {Vorontseva}}{{Culpan} et~al.}{2022}]{culpan2022}
{Culpan} R.,  {Geier} S.,  {Reindl} N.,  {Pelisoli} I.,  {Gentile Fusillo} N.,
   {Vorontseva} A.,  2022, \mn@doi [\aap] {10.1051/0004-6361/202243337}, \href
  {https://ui.adsabs.harvard.edu/abs/2022A&A...662A..40C} {662, A40}

\bibitem[\protect\citeauthoryear{{Dalba} et~al.,}{{Dalba}
  et~al.}{2020}]{dalba2020}
{Dalba} P.~A.,  et~al., 2020, \mn@doi [\aj] {10.3847/1538-3881/ab84e3}, \href
  {https://ui.adsabs.harvard.edu/abs/2020AJ....159..241D} {159, 241}

\bibitem[\protect\citeauthoryear{{Eisner} et~al.,}{{Eisner}
  et~al.}{2021}]{eisner2021}
{Eisner} N.~L.,  et~al., 2021, \mn@doi [\mnras] {10.1093/mnras/staa3739}, \href
  {https://ui.adsabs.harvard.edu/abs/2021MNRAS.501.4669E} {501, 4669}

\bibitem[\protect\citeauthoryear{{Espinoza} \& {Jord{\'a}n}}{{Espinoza} \&
  {Jord{\'a}n}}{2016}]{Espinoza2016}
{Espinoza} N.,  {Jord{\'a}n} A.,  2016, \mn@doi [\mnras]
  {10.1093/mnras/stw224}, \href
  {https://ui.adsabs.harvard.edu/abs/2016MNRAS.457.3573E} {457, 3573}

\bibitem[\protect\citeauthoryear{Foreman-Mackey, Agol, Ambikasaran  \&
  Angus}{Foreman-Mackey et~al.}{2017}]{foreman2017celerite}
Foreman-Mackey D.,  Agol E.,  Ambikasaran S.,   Angus R.,  2017, Astrophysics
  Source Code Library, pp ascl--1709

\bibitem[\protect\citeauthoryear{Foreman-Mackey et~al.,}{Foreman-Mackey
  et~al.}{2021}]{exoplanet:exoplanet}
Foreman-Mackey D.,  et~al., 2021, exoplanet-dev/exoplanet v0.4.5,
  \mn@doi{10.5281/zenodo.1998447}, \url
  {https://doi.org/10.5281/zenodo.1998447}

\bibitem[\protect\citeauthoryear{{Gaia Collaboration}}{{Gaia
  Collaboration}}{2022}]{gaiadr3}
{Gaia Collaboration} 2022, VizieR Online Data Catalog, \href
  {https://ui.adsabs.harvard.edu/abs/2022yCat.1355....0G} {p. I/355}

\bibitem[\protect\citeauthoryear{{Gill} et~al.,}{{Gill}
  et~al.}{2020a}]{gill2020b}
{Gill} S.,  et~al., 2020a, \mn@doi [\mnras] {10.1093/mnras/stz3212}, \href
  {https://ui.adsabs.harvard.edu/abs/2020MNRAS.491.1548G} {491, 1548}

\bibitem[\protect\citeauthoryear{{Gill} et~al.,}{{Gill}
  et~al.}{2020b}]{2020MNRAS.495.2713G}
{Gill} S.,  et~al., 2020b, \mn@doi [\mnras] {10.1093/mnras/staa1248}, \href
  {https://ui.adsabs.harvard.edu/abs/2020MNRAS.495.2713G} {495, 2713}

\bibitem[\protect\citeauthoryear{{Gill} et~al.,}{{Gill}
  et~al.}{2020c}]{gill2020a}
{Gill} S.,  et~al., 2020c, \mn@doi [\apjl] {10.3847/2041-8213/ab9eb9}, \href
  {https://ui.adsabs.harvard.edu/abs/2020ApJ...898L..11G} {898, L11}

\bibitem[\protect\citeauthoryear{{Grieves} et~al.,}{{Grieves}
  et~al.}{2022}]{grieves2022}
{Grieves} N.,  et~al., 2022, arXiv e-prints, \href
  {https://ui.adsabs.harvard.edu/abs/2022arXiv220914830G} {p. arXiv:2209.14830}

\bibitem[\protect\citeauthoryear{{Gu}, {Lin}  \& {Bodenheimer}}{{Gu}
  et~al.}{2003}]{gu2003}
{Gu} P.-G.,  {Lin} D. N.~C.,   {Bodenheimer} P.~H.,  2003, \mn@doi [\apj]
  {10.1086/373920}, \href
  {https://ui.adsabs.harvard.edu/abs/2003ApJ...588..509G} {588, 509}

\bibitem[\protect\citeauthoryear{{Han}, {Podsiadlowski}, {Maxted}, {Marsh}  \&
  {Ivanova}}{{Han} et~al.}{2002}]{han2002}
{Han} Z.,  {Podsiadlowski} P.,  {Maxted} P.~F.~L.,  {Marsh} T.~R.,   {Ivanova}
  N.,  2002, \mn@doi [\mnras] {10.1046/j.1365-8711.2002.05752.x}, \href
  {https://ui.adsabs.harvard.edu/abs/2002MNRAS.336..449H} {336, 449}

\bibitem[\protect\citeauthoryear{{Han}, {Podsiadlowski}, {Maxted}  \&
  {Marsh}}{{Han} et~al.}{2003}]{han2003}
{Han} Z.,  {Podsiadlowski} P.,  {Maxted} P.~F.~L.,   {Marsh} T.~R.,  2003,
  \mn@doi [\mnras] {10.1046/j.1365-8711.2003.06451.x}, \href
  {https://ui.adsabs.harvard.edu/abs/2003MNRAS.341..669H} {341, 669}

\bibitem[\protect\citeauthoryear{{Hattori}, {Foreman-Mackey}, {Hogg}, {Montet},
  {Angus}, {Pritchard}, {Curtis}  \& {Sch{\"o}lkopf}}{{Hattori}
  et~al.}{2022}]{2022AJ....163..284H}
{Hattori} S.,  {Foreman-Mackey} D.,  {Hogg} D.~W.,  {Montet} B.~T.,  {Angus}
  R.,  {Pritchard} T.~A.,  {Curtis} J.~L.,   {Sch{\"o}lkopf} B.,  2022, \mn@doi
  [\aj] {10.3847/1538-3881/ac625a}, \href
  {https://ui.adsabs.harvard.edu/abs/2022AJ....163..284H} {163, 284}

\bibitem[\protect\citeauthoryear{{Hobson} et~al.,}{{Hobson}
  et~al.}{2021}]{hobson2021}
{Hobson} M.~J.,  et~al., 2021, \mn@doi [\aj] {10.3847/1538-3881/abeaa1}, \href
  {https://ui.adsabs.harvard.edu/abs/2021AJ....161..235H} {161, 235}

\bibitem[\protect\citeauthoryear{{Hsu}, {Ford}, {Ragozzine}  \& {Ashby}}{{Hsu}
  et~al.}{2019}]{hsu2019}
{Hsu} D.~C.,  {Ford} E.~B.,  {Ragozzine} D.,   {Ashby} K.,  2019, \mn@doi [\aj]
  {10.3847/1538-3881/ab31ab}, \href
  {https://ui.adsabs.harvard.edu/abs/2019AJ....158..109H} {158, 109}

\bibitem[\protect\citeauthoryear{{Jeffery}, {Miszalski}  \&
  {Snowdon}}{{Jeffery} et~al.}{2021}]{jeffery2021}
{Jeffery} C.~S.,  {Miszalski} B.,   {Snowdon} E.,  2021, \mn@doi [\mnras]
  {10.1093/mnras/staa3648}, \href
  {https://ui.adsabs.harvard.edu/abs/2021MNRAS.501..623J} {501, 623}

\bibitem[\protect\citeauthoryear{{Jenkins} et~al.,}{{Jenkins}
  et~al.}{2016}]{jenkinsSPOC2016}
{Jenkins} J.~M.,  et~al., 2016, in Software and Cyberinfrastructure for
  Astronomy IV. p. 99133E, \mn@doi{10.1117/12.2233418}

\bibitem[\protect\citeauthoryear{{Jord{\'a}n} et~al.,}{{Jord{\'a}n}
  et~al.}{2020}]{jordan2020}
{Jord{\'a}n} A.,  et~al., 2020, \mn@doi [\aj] {10.3847/1538-3881/ab6f67}, \href
  {https://ui.adsabs.harvard.edu/abs/2020AJ....159..145J} {159, 145}

\bibitem[\protect\citeauthoryear{{Kipping}}{{Kipping}}{2013}]{Kipping2013_ecc}
{Kipping} D.~M.,  2013, \mn@doi [\mnras] {10.1093/mnrasl/slt075}, \href
  {https://ui.adsabs.harvard.edu/abs/2013MNRAS.434L..51K} {434, L51}

\bibitem[\protect\citeauthoryear{{Kipping}}{{Kipping}}{2023}]{kipping2023}
{Kipping} D.,  2023, \mn@doi [\mnras] {10.1093/mnras/stad1492}, \href
  {https://ui.adsabs.harvard.edu/abs/2023MNRAS.523.1182K} {523, 1182}

\bibitem[\protect\citeauthoryear{{Lendl} et~al.,}{{Lendl}
  et~al.}{2020a}]{lendl2020a}
{Lendl} M.,  et~al., 2020a, \mn@doi [\mnras] {10.1093/mnras/stz3545}, \href
  {https://ui.adsabs.harvard.edu/abs/2020MNRAS.492.1761L} {492, 1761}

\bibitem[\protect\citeauthoryear{{Lendl} et~al.,}{{Lendl}
  et~al.}{2020b}]{2020MNRAS.492.1761L}
{Lendl} M.,  et~al., 2020b, \mn@doi [\mnras] {10.1093/mnras/stz3545}, \href
  {https://ui.adsabs.harvard.edu/abs/2020MNRAS.492.1761L} {492, 1761}

\bibitem[\protect\citeauthoryear{{Lopez} \& {Fortney}}{{Lopez} \&
  {Fortney}}{2016}]{lopez2016}
{Lopez} E.~D.,  {Fortney} J.~J.,  2016, \mn@doi [\apj]
  {10.3847/0004-637X/818/1/4}, \href
  {https://ui.adsabs.harvard.edu/abs/2016ApJ...818....4L} {818, 4}

\bibitem[\protect\citeauthoryear{{Madhusudhan}, {Bitsch}, {Johansen}  \&
  {Eriksson}}{{Madhusudhan} et~al.}{2017}]{2017MNRAS.469.4102M}
{Madhusudhan} N.,  {Bitsch} B.,  {Johansen} A.,   {Eriksson} L.,  2017, \mn@doi
  [\mnras] {10.1093/mnras/stx1139}, \href
  {https://ui.adsabs.harvard.edu/abs/2017MNRAS.469.4102M} {469, 4102}

\bibitem[\protect\citeauthoryear{{Marmier} et~al.,}{{Marmier}
  et~al.}{2013}]{2013A&A...551A..90M}
{Marmier} M.,  et~al., 2013, \mn@doi [\aap] {10.1051/0004-6361/201219639},
  \href {https://ui.adsabs.harvard.edu/abs/2013A&A...551A..90M} {551, A90}

\bibitem[\protect\citeauthoryear{{Maxted}, {Heber}, {Marsh}  \&
  {North}}{{Maxted} et~al.}{2001}]{maxted2001}
{Maxted} P.~F.~L.,  {Heber} U.,  {Marsh} T.~R.,   {North} R.~C.,  2001, \mn@doi
  [\mnras] {10.1111/j.1365-2966.2001.04714.x}, \href
  {https://ui.adsabs.harvard.edu/abs/2001MNRAS.326.1391M} {326, 1391}

\bibitem[\protect\citeauthoryear{{Osborn}}{{Osborn}}{2022}]{Osborn2022code}
{Osborn} H.~P.,  2022, {MonoTools: Planets of uncertain periods detector and
  modeler}, Astrophysics Source Code Library, record ascl:2204.020 (\mn@eprint
  {ascl} {2204.020})

\bibitem[\protect\citeauthoryear{{Osborn} et~al.,}{{Osborn}
  et~al.}{2021}]{osborn2021}
{Osborn} H.~P.,  et~al., 2021, \mn@doi [\mnras] {10.1093/mnras/stab182}, \href
  {https://ui.adsabs.harvard.edu/abs/2021MNRAS.502.4842O} {502, 4842}

\bibitem[\protect\citeauthoryear{{Osborn} et~al.,}{{Osborn}
  et~al.}{2022a}]{osborn2022}
{Osborn} H.~P.,  et~al., 2022a, \mn@doi [\aap] {10.1051/0004-6361/202243065},
  \href {https://ui.adsabs.harvard.edu/abs/2022A&A...664A.156O} {664, A156}

\bibitem[\protect\citeauthoryear{{Osborn} et~al.,}{{Osborn}
  et~al.}{2022b}]{2022A&A...664A.156O}
{Osborn} H.~P.,  et~al., 2022b, \mn@doi [\aap] {10.1051/0004-6361/202243065},
  \href {https://ui.adsabs.harvard.edu/abs/2022A&A...664A.156O} {664, A156}

\bibitem[\protect\citeauthoryear{{Pelisoli}, {Vos}, {Geier}, {Schaffenroth}  \&
  {Baran}}{{Pelisoli} et~al.}{2020}]{pelisoli2020}
{Pelisoli} I.,  {Vos} J.,  {Geier} S.,  {Schaffenroth} V.,   {Baran} A.~S.,
  2020, \mn@doi [\aap] {10.1051/0004-6361/202038473}, \href
  {https://ui.adsabs.harvard.edu/abs/2020A&A...642A.180P} {642, A180}

\bibitem[\protect\citeauthoryear{{Queloz} et~al.,}{{Queloz}
  et~al.}{1999}]{coralie}
{Queloz} D.,  et~al., 1999, \mn@doi [arXiv e-prints]
  {10.48550/arXiv.astro-ph/9910223}, \href
  {https://ui.adsabs.harvard.edu/abs/1999astro.ph.10223Q} {pp
  astro--ph/9910223}

\bibitem[\protect\citeauthoryear{{Rauch} \& {Deetjen}}{{Rauch} \&
  {Deetjen}}{2003}]{rauch2003}
{Rauch} T.,  {Deetjen} J.~L.,  2003, in {Hubeny} I.,  {Mihalas} D.,   {Werner}
  K.,  eds,  Astronomical Society of the Pacific Conference Series Vol. 288,
  Stellar Atmosphere Modeling. p.~103 (\mn@eprint {arXiv} {astro-ph/0403239}),
  \mn@doi{10.48550/arXiv.astro-ph/0403239}

\bibitem[\protect\citeauthoryear{{Rice} et~al.,}{{Rice}
  et~al.}{2022}]{rice2022}
{Rice} M.,  et~al., 2022, \mn@doi [\aj] {10.3847/1538-3881/ac8153}, \href
  {https://ui.adsabs.harvard.edu/abs/2022AJ....164..104R} {164, 104}

\bibitem[\protect\citeauthoryear{{Ricker} et~al.,}{{Ricker}
  et~al.}{2015}]{Ricker:2015}
{Ricker} G.~R.,  et~al., 2015, \mn@doi [Journal of Astronomical Telescopes,
  Instruments, and Systems] {10.1117/1.JATIS.1.1.014003}, \href
  {http://adsabs.harvard.edu/abs/2015JATIS...1a4003R} {1, 014003}

\bibitem[\protect\citeauthoryear{Salvatier, Wiecki  \& Fonnesbeck}{Salvatier
  et~al.}{2016}]{exoplanet:pymc3}
Salvatier J.,  Wiecki T.~V.,   Fonnesbeck C.,  2016, PeerJ Computer Science, 2,
  e55

\bibitem[\protect\citeauthoryear{{Schaffenroth}, {Pelisoli}, {Barlow}, {Geier}
  \& {Kupfer}}{{Schaffenroth} et~al.}{2022}]{2022A&A...666A.182S}
{Schaffenroth} V.,  {Pelisoli} I.,  {Barlow} B.~N.,  {Geier} S.,   {Kupfer} T.,
   2022, \mn@doi [\aap] {10.1051/0004-6361/202244214}, \href
  {https://ui.adsabs.harvard.edu/abs/2022A&A...666A.182S} {666, A182}

\bibitem[\protect\citeauthoryear{{Schlecker} et~al.,}{{Schlecker}
  et~al.}{2020}]{schlecker2020}
{Schlecker} M.,  et~al., 2020, \mn@doi [\aj] {10.3847/1538-3881/abbe03}, \href
  {https://ui.adsabs.harvard.edu/abs/2020AJ....160..275S} {160, 275}

\bibitem[\protect\citeauthoryear{{Seabroke} et~al.,}{{Seabroke}
  et~al.}{2021}]{seabroke2021}
{Seabroke} G.~M.,  et~al., 2021, \mn@doi [\aap] {10.1051/0004-6361/202141008},
  \href {https://ui.adsabs.harvard.edu/abs/2021A&A...653A.160S} {653, A160}

\bibitem[\protect\citeauthoryear{{Smith} et~al.,}{{Smith}
  et~al.}{2012}]{Smith2012}
{Smith} J.~C.,  et~al., 2012, \mn@doi [\pasp] {10.1086/667697}, \href
  {https://ui.adsabs.harvard.edu/abs/2012PASP..124.1000S} {124, 1000}

\bibitem[\protect\citeauthoryear{{Stark} \& {Wade}}{{Stark} \&
  {Wade}}{2003}]{stark2003}
{Stark} M.~A.,  {Wade} R.~A.,  2003, \mn@doi [\aj] {10.1086/377017}, \href
  {https://ui.adsabs.harvard.edu/abs/2003AJ....126.1455S} {126, 1455}

\bibitem[\protect\citeauthoryear{{Stassun} et~al.,}{{Stassun}
  et~al.}{2019}]{Stassun2019}
{Stassun} K.~G.,  et~al., 2019, \mn@doi [\aj] {10.3847/1538-3881/ab3467}, \href
  {https://ui.adsabs.harvard.edu/abs/2019AJ....158..138S} {158, 138}

\bibitem[\protect\citeauthoryear{{Stumpe} et~al.,}{{Stumpe}
  et~al.}{2012}]{Stumpe2012}
{Stumpe} M.~C.,  et~al., 2012, \mn@doi [\pasp] {10.1086/667698}, \href
  {https://ui.adsabs.harvard.edu/abs/2012PASP..124..985S} {124, 985}

\bibitem[\protect\citeauthoryear{{Stumpe}, {Smith}, {Catanzarite}, {Van Cleve},
  {Jenkins}, {Twicken}  \& {Girouard}}{{Stumpe} et~al.}{2014}]{Stumpe2014}
{Stumpe} M.~C.,  {Smith} J.~C.,  {Catanzarite} J.~H.,  {Van Cleve} J.~E.,
  {Jenkins} J.~M.,  {Twicken} J.~D.,   {Girouard} F.~R.,  2014, \mn@doi [\pasp]
  {10.1086/674989}, \href
  {https://ui.adsabs.harvard.edu/abs/2014PASP..126..100S} {126, 100}

\bibitem[\protect\citeauthoryear{{Temple} et~al.,}{{Temple}
  et~al.}{2018}]{2018MNRAS.480.5307T}
{Temple} L.~Y.,  et~al., 2018, \mn@doi [\mnras] {10.1093/mnras/sty2197}, \href
  {https://ui.adsabs.harvard.edu/abs/2018MNRAS.480.5307T} {480, 5307}

\bibitem[\protect\citeauthoryear{{Temple} et~al.,}{{Temple}
  et~al.}{2019}]{2019MNRAS.490.2467T}
{Temple} L.~Y.,  et~al., 2019, \mn@doi [\mnras] {10.1093/mnras/stz2632}, \href
  {https://ui.adsabs.harvard.edu/abs/2019MNRAS.490.2467T} {490, 2467}

\bibitem[\protect\citeauthoryear{{Thuillier}, {Van Grootel},
  {D{\'e}vora-Pajares}, {Pozuelos}, {Charpinet}  \& {Siess}}{{Thuillier}
  et~al.}{2022}]{thuillier2022}
{Thuillier} A.,  {Van Grootel} V.,  {D{\'e}vora-Pajares} M.,  {Pozuelos} F.~J.,
   {Charpinet} S.,   {Siess} L.,  2022, arXiv e-prints, \href
  {https://ui.adsabs.harvard.edu/abs/2022arXiv220902437T} {p. arXiv:2209.02437}

\bibitem[\protect\citeauthoryear{{Twicken} et~al.,}{{Twicken}
  et~al.}{2019}]{twicken2019}
{Twicken} J.,  et~al., 2019, in American Astronomical Society Meeting Abstracts
  \#233. p. 140.03

\bibitem[\protect\citeauthoryear{{Ulmer-Moll} et~al.,}{{Ulmer-Moll}
  et~al.}{2022}]{ulmermoll2022}
{Ulmer-Moll} S.,  et~al., 2022, \mn@doi [\aap] {10.1051/0004-6361/202243583},
  \href {https://ui.adsabs.harvard.edu/abs/2022A&A...666A..46U} {666, A46}

\bibitem[\protect\citeauthoryear{{Ulmer-Moll} et~al.,}{{Ulmer-Moll}
  et~al.}{2023}]{ulmermoll2023}
{Ulmer-Moll} S.,  et~al., 2023, \mn@doi [\aap] {10.1051/0004-6361/202245478},
  \href {https://ui.adsabs.harvard.edu/abs/2023A&A...674A..43U} {674, A43}

\bibitem[\protect\citeauthoryear{{Van Grootel} et~al.,}{{Van Grootel}
  et~al.}{2021}]{vangrootel2021}
{Van Grootel} V.,  et~al., 2021, \mn@doi [\aap] {10.1051/0004-6361/202140381},
  \href {https://ui.adsabs.harvard.edu/abs/2021A&A...650A.205V} {650, A205}

\bibitem[\protect\citeauthoryear{{Villanueva}, {Dragomir}  \&
  {Gaudi}}{{Villanueva} et~al.}{2019}]{2019AJ....157...84V}
{Villanueva} Steven J.,  {Dragomir} D.,   {Gaudi} B.~S.,  2019, \mn@doi [\aj]
  {10.3847/1538-3881/aaf85e}, \href
  {https://ui.adsabs.harvard.edu/abs/2019AJ....157...84V} {157, 84}

\bibitem[\protect\citeauthoryear{{Vos}, {Vu{\v{c}}kovi{\'c}}, {Chen}, {Han},
  {Boudreaux}, {Barlow}, {{\O}stensen}  \& {N{\'e}meth}}{{Vos}
  et~al.}{2019}]{vos2019}
{Vos} J.,  {Vu{\v{c}}kovi{\'c}} M.,  {Chen} X.,  {Han} Z.,  {Boudreaux} T.,
  {Barlow} B.~N.,  {{\O}stensen} R.,   {N{\'e}meth} P.,  2019, \mn@doi [\mnras]
  {10.1093/mnras/sty3017}, \href
  {https://ui.adsabs.harvard.edu/abs/2019MNRAS.482.4592V} {482, 4592}

\bibitem[\protect\citeauthoryear{{Wang} et~al.,}{{Wang}
  et~al.}{2021}]{wang2021}
{Wang} S.,  et~al., 2021, \mn@doi [\aj] {10.3847/1538-3881/ac0626}, \href
  {https://ui.adsabs.harvard.edu/abs/2021AJ....162...50W} {162, 50}

\bibitem[\protect\citeauthoryear{{Watson}, {de Mooij}, {Steeghs}, {Marsh},
  {Brogi}, {Gibson}  \& {Matthews}}{{Watson}
  et~al.}{2019}]{2019MNRAS.490.1991W}
{Watson} C.~A.,  {de Mooij} E.~J.~W.,  {Steeghs} D.,  {Marsh} T.~R.,  {Brogi}
  M.,  {Gibson} N.~P.,   {Matthews} S.,  2019, \mn@doi [\mnras]
  {10.1093/mnras/stz2679}, \href
  {https://ui.adsabs.harvard.edu/abs/2019MNRAS.490.1991W} {490, 1991}

\bibitem[\protect\citeauthoryear{{Werner} \& {Dreizler}}{{Werner} \&
  {Dreizler}}{1999}]{werner1999}
{Werner} K.,  {Dreizler} S.,  1999, \mn@doi [Journal of Computational and
  Applied Mathematics] {10.48550/arXiv.astro-ph/9906130}, \href
  {https://ui.adsabs.harvard.edu/abs/1999JCoAM.109...65W} {109, 65}

\bibitem[\protect\citeauthoryear{{Werner}, {Deetjen}, {Dreizler}, {Nagel},
  {Rauch}  \& {Schuh}}{{Werner} et~al.}{2003}]{werner2003}
{Werner} K.,  {Deetjen} J.~L.,  {Dreizler} S.,  {Nagel} T.,  {Rauch} T.,
  {Schuh} S.~L.,  2003, in {Hubeny} I.,  {Mihalas} D.,   {Werner} K.,  eds,
  Astronomical Society of the Pacific Conference Series Vol. 288, Stellar
  Atmosphere Modeling. p.~31 (\mn@eprint {arXiv} {astro-ph/0209535}),
  \mn@doi{10.48550/arXiv.astro-ph/0209535}

\bibitem[\protect\citeauthoryear{{Wheatley} et~al.,}{{Wheatley}
  et~al.}{2018}]{Wheatley2018}
{Wheatley} P.~J.,  et~al., 2018, \mn@doi [\mnras] {10.1093/mnras/stx2836},
  \href {https://ui.adsabs.harvard.edu/abs/2018MNRAS.475.4476W} {475, 4476}

\makeatother
\end{thebibliography}


\newpage
\appendix
\section{Candidates and Light Curves} \label{sec:lightcurves}
\setcounter{table}{0}
\renewcommand{\thetable}{A\arabic{table}}
\clearpage 

\onecolumn

\begin{longtable}{cccccccccc}
\caption{Properties of the host stars of our duotransit candidates.}\\
\toprule
\textbf{ID} & \textbf{TIC ID} & \textbf{TOI/CTOI} & \textbf{RA (deg)} & \textbf{Dec (deg)} & \textbf{\tmag} & \textbf{\rstar\ (\rsun)} & \textbf{\teff\ (K)} & \textbf{Distance (pc)} & \textbf{N\textsubscript{sec}} \endhead
\hline

1  & \href{https://exofop.ipac.caltech.edu/tess/target.php?id=1504460}{1504460}     & --       & 131.9658 & -23.9895 & 11.63 & 1.5  & 6218  & $ 451.71^{+20.73}_{-15.90} $ & 2  \\
2  & \href{https://exofop.ipac.caltech.edu/tess/target.php?id=2669681}{2669681}     & --       & 289.368  & -28.5803 & 11.47 & 1.88 & 5160  & $ 382.06^{+12.79}_{-9.96} $ & 2  \\
3  & \href{https://exofop.ipac.caltech.edu/tess/target.php?id=9844069}{9844069}     & --       & 48.03923 & -11.6808 & 12.96 & 1.61 & 5665  & $ 777.93^{+11.74}_{-11.59} $ & 2  \\
4  & \href{https://exofop.ipac.caltech.edu/tess/target.php?id=13072758}{13072758}   & CTOI     & 76.08837 & -29.0342 & 12.22 & 1.43 & 6313  & $ 601.42^{+2.95}_{-3.19} $ & 2  \\
5  & \href{https://exofop.ipac.caltech.edu/tess/target.php?id=13713700}{13713700}   & --       & 133.6396 & -7.94134 & 12.29 & 0.97 & 5864  & $ 391.66^{+2.67}_{-2.25} $ & 2  \\
6  & \href{https://exofop.ipac.caltech.edu/tess/target.php?id=14445414}{14445414}   & --       & 110.8209 & 12.66321 & 11.02 & 1.31 & 6425  & $ 346.17^{+12.45}_{-3.76} $ & 2  \\
7  & \href{https://exofop.ipac.caltech.edu/tess/target.php?id=20904104}{20904104}   & --       & 7.41368  & -19.3663 & 9.76  & 0.88 & 5628  & $ 105.79^{+0.24}_{-0.22} $ & 2  \\
8  & \href{https://exofop.ipac.caltech.edu/tess/target.php?id=25194908}{25194908}   & --       & 63.90318 & -66.3519 & 12.46 & 1.32 & 6309  & $ 666.61^{+32.34}_{-8.91} $ & 24 \\
9  & \href{https://exofop.ipac.caltech.edu/tess/target.php?id=32179255}{32179255}   & --       & 332.0396 & -29.6093 & 10.52 & 1.75 & 5989  & $ 330.44^{+14.95}_{-7.29} $ & 2  \\
10 & \href{https://exofop.ipac.caltech.edu/tess/target.php?id=38138512}{38138512}   & --       & 170.4183 & -4.19721 & 12.94 & 1.34 & 6300  & $ 769.66^{+10.80}_{-9.15} $ & 2  \\
11 & \href{https://exofop.ipac.caltech.edu/tess/target.php?id=39167176}{39167176}   & --       & 88.32738 & -3.79767 & 12.01 & 1.56 & 8416  & $ 486.84^{+3.95}_{-1.46} $ & 2  \\
12 & \href{https://exofop.ipac.caltech.edu/tess/target.php?id=42428568}{42428568}   & --       & 96.29347 & -3.76657 & 10.83 & 2.16 & 6686  & $ 567.87^{+8.12}_{-10.63} $ & 2  \\
13 & \href{https://exofop.ipac.caltech.edu/tess/target.php?id=52195587}{52195587}   & --       & 19.24834 & -66.1636 & 12.25 & 1.2  & 5736  & $ 450.99^{+2.50}_{-2.59} $ & 4  \\
14 & \href{https://exofop.ipac.caltech.edu/tess/target.php?id=66439839}{66439839}   & --       & 14.47851 & -30.3231 & 11.97 & 1.03 & 5882  & $ 356.25^{+2.47}_{-3.35} $ & 3  \\
15 & \href{https://exofop.ipac.caltech.edu/tess/target.php?id=67599025}{67599025}   & CTOI     & 17.69268 & -31.5055 & 11.96 & 0.78 & 4816  & $ 177.82^{+1.19}_{-1.34} $ & 2  \\
16 & \href{https://exofop.ipac.caltech.edu/tess/target.php?id=70561926}{70561926}   & --       & 139.2559 & -26.7007 & 11    & 1.35 & 6386  & $ 353.55^{+5.09}_{-3.76} $ & 2  \\
17 & \href{https://exofop.ipac.caltech.edu/tess/target.php?id=71028120}{71028120}   & --       & 66.12741 & -11.9835 & 10.44 & 1.37 & 5596  & $ 215.31^{+0.82}_{-0.84} $ & 2  \\
18 & \href{https://exofop.ipac.caltech.edu/tess/target.php?id=77437543}{77437543}   & TOI-2490 & 73.12441 & -36.2572 & 11.28 & 1.17 & 5459  & $ 264.92^{+1.04}_{-0.96} $ & 3  \\
19 & \href{https://exofop.ipac.caltech.edu/tess/target.php?id=81089255}{81089255}   & CTOI     & 122.6942 & -43.8535 & 11.84 & 1.43 & 7459  & $ 547.70^{+2.71}_{-2.72} $ & 4  \\
20 & \href{https://exofop.ipac.caltech.edu/tess/target.php?id=100776118}{100776118} & CTOI     & 88.31784 & -34.5185 & 11.51 & 1.25 & 6439  &  & 4  \\
21 & \href{https://exofop.ipac.caltech.edu/tess/target.php?id=101824521}{101824521} & --       & 206.004  & -26.4978 & 12.61 & 1.23 & 5553  & $ 481.79^{+4.21}_{-3.97} $ & 2  \\
22 & \href{https://exofop.ipac.caltech.edu/tess/target.php?id=107113345}{107113345} & --       & 183.0341 & -32.6628 & 10.34 & 1.31 & 6696  & $ 267.74^{+1.41}_{-1.63} $ & 2  \\
23 & \href{https://exofop.ipac.caltech.edu/tess/target.php?id=118339710}{118339710} & CTOI     & 129.572  & -23.5502 & 10.35 & 1.56 & 6119  & $ 285.48^{+15.23}_{-4.22} $ & 3  \\
24 & \href{https://exofop.ipac.caltech.edu/tess/target.php?id=123763494}{123763494} & --       & 90.30269 & -21.2359 & 11.73 & 1.22 & 5568  & $ 353.10^{+29.67}_{-9.36} $ & 2  \\
25 & \href{https://exofop.ipac.caltech.edu/tess/target.php?id=124029677}{124029677} & TOI-5153 & 91.54028 & -19.9533 & 11.21 & 1.38 & 6408  & $ 366.92^{+2.94}_{-2.71} $ & 2  \\
26 & \href{https://exofop.ipac.caltech.edu/tess/target.php?id=140215502}{140215502} & CTOI     & 85.15071 & -34.7664 & 11.23 & 0.93 & 5528  & $ 210.03^{+0.62}_{-0.56} $ & 4  \\
27 & \href{https://exofop.ipac.caltech.edu/tess/target.php?id=140750416}{140750416} & --       & 116.4382 & -23.143  & 11.81 & 0.87 & 5331  & $ 238.74^{+5.45}_{-2.41} $ & 2  \\
28 & \href{https://exofop.ipac.caltech.edu/tess/target.php?id=142278054}{142278054} & --       & 180.7626 & -33.343  & 10.28 & 2.39 & 6400  & $ 418.42^{+1.64}_{-2.10} $ & 2  \\
29 & \href{https://exofop.ipac.caltech.edu/tess/target.php?id=145006304}{145006304} & --       & 193.3345 & -39.7347 & 11.3  & 1.12 & 6340  & $ 309.68^{+1.88}_{-1.72} $ & 2  \\
30 & \href{https://exofop.ipac.caltech.edu/tess/target.php?id=145913596}{145913596} & --       & 123.1298 & -30.5555 & 10.17 & 1.45 & 6467  & $ 261.13^{+1.63}_{-0.99} $ & 4  \\
31 & \href{https://exofop.ipac.caltech.edu/tess/target.php?id=153838604}{153838604} & --       & 68.07396 & -47.6088 & 12.48 & 0.91 & 5649  & $ 387.07^{+22.34}_{-7.33} $ & 4  \\
32 & \href{https://exofop.ipac.caltech.edu/tess/target.php?id=156716001}{156716001} & --       & 103.1063 & -39.6876 & 10.9  & 1.08 & 6129  & $ 240.55^{+2.60}_{-11.28} $ & 4  \\
33 & \href{https://exofop.ipac.caltech.edu/tess/target.php?id=157119927}{157119927} & --       & 175.2155 & -31.1374 & 11.64 & 1.04 & 5765  & $ 299.69^{+8.62}_{-2.98} $ & 2  \\
34 & \href{https://exofop.ipac.caltech.edu/tess/target.php?id=157698565}{157698565} & TOI-2589 & 107.4883 & -37.231  & 10.72 & 1.07 & 5938  & $ 198.90^{+23.29}_{-1.84} $ & 3  \\
35 & \href{https://exofop.ipac.caltech.edu/tess/target.php?id=159490807}{159490807} & --       & 220.2349 & -39.3757 & 10.57 & 1.69 & 6556  & $ 341.73^{+3.68}_{-3.50} $ & 2  \\
36 & \href{https://exofop.ipac.caltech.edu/tess/target.php?id=161169240}{161169240} & CTOI     & 339.3573 & -53.3188 & 12.12 & 1.19 & 5952  & $ 445.30^{+6.33}_{-36.66} $ & 2  \\
37 & \href{https://exofop.ipac.caltech.edu/tess/target.php?id=176518126}{176518126} & --       & 87.1832  & -0.49503 & 10.48 & 1.36 & 7927  & $ 281.19^{+1.57}_{-1.56} $ & 2  \\
38 & \href{https://exofop.ipac.caltech.edu/tess/target.php?id=188620407}{188620407} & CTOI     & 350.0511 & -13.0494 & 11.57 & 1.31 & 6169  & $ 404.71^{+3.12}_{-2.81} $ & 3  \\
39 & \href{https://exofop.ipac.caltech.edu/tess/target.php?id=193318850}{193318850} & --       & 177.7302 & -16.7417 & 12.76 & 0.79 & 5560  & $ 340.21^{+1.78}_{-1.68} $ & 2  \\
40 & \href{https://exofop.ipac.caltech.edu/tess/target.php?id=207783865}{207783865} & --       & 207.0609 & -54.4206 & 10.54 & 1.04 & 5936  & $ 180.65^{+2.63}_{-0.98} $ & 2  \\
41 & \href{https://exofop.ipac.caltech.edu/tess/target.php?id=211409161}{211409161} & --       & 321.7768 & -30.5273 & 10.44 & 0.78 & 4660  &  & 2  \\
42 & \href{https://exofop.ipac.caltech.edu/tess/target.php?id=215402824}{215402824} & --       & 43.0624  & -36.2887 & 12.97 & 1.33 & 6180  & $ 761.55^{+11.13}_{-8.01} $ & 4  \\
43 & \href{https://exofop.ipac.caltech.edu/tess/target.php?id=218977148}{218977148} & --       & 349.5331 & -33.1369 & 12.58 & 1.61 & 6214  & $ 826.08^{+12.07}_{-12.51} $ & 2  \\
44 & \href{https://exofop.ipac.caltech.edu/tess/target.php?id=220622886}{220622886} & --       & 84.89306 & -42.6811 & 12    & 2.56 & 5671  & $ 796.01^{+3.48}_{-16.0} $ & 3  \\
45 & \href{https://exofop.ipac.caltech.edu/tess/target.php?id=221915858}{221915858} & --       & 249.513  & -54.3579 & 10.07 & 2.47 & 9475  &  & 2  \\
46 & \href{https://exofop.ipac.caltech.edu/tess/target.php?id=224279805}{224279805} & CTOI     & 355.7386 & -40.7722 & 12.15 & 1.2  & 5591  & $ 422.59^{+3.22}_{-3.1} $ & 2  \\
47 & \href{https://exofop.ipac.caltech.edu/tess/target.php?id=235058563}{235058563} & --       & 83.51699 & -50.1216 & 12.43 & 1.47 & 6438  & $ 706.77^{+5.66}_{-5.64} $ & 6  \\
48 & \href{https://exofop.ipac.caltech.edu/tess/target.php?id=237605045}{237605045} & --       & 103.2896 & 2.05465  & 10.29 & 2.92 & 6577  & $ 471.34^{+7.62}_{-7.43} $ & 2  \\
49 & \href{https://exofop.ipac.caltech.edu/tess/target.php?id=242241304}{242241304} & CTOI     & 211.5851 & -43.0447 & 11.81 & 1.14 & 5781  & $ 346.35^{+1.79}_{-1.77} $ & 2  \\
50 & \href{https://exofop.ipac.caltech.edu/tess/target.php?id=251057075}{251057075} & --       & 36.17507 & -5.85401 & 11.05 & 0.84 & 5225  & $ 157.35^{+0.52}_{-0.54} $ & 2  \\
51 & \href{https://exofop.ipac.caltech.edu/tess/target.php?id=256912435}{256912435} & --       & 200.5462 & -55.2184 & 11.8  & 1.19 & 5650  & $ 333.69^{+3.17}_{-3.45} $ & 2  \\
52 & \href{https://exofop.ipac.caltech.edu/tess/target.php?id=265465927}{265465927} & CTOI     & 320.9166 & -62.9238 & 12.08 & 1.16 & 6099  & $ 418.77^{+65.41}_{-56.26} $ & 2  \\
53 & \href{https://exofop.ipac.caltech.edu/tess/target.php?id=265466589}{265466589} & --       & 321.0794 & -64.4743 & 12.61 & 1.03 & 5834  &  & 2  \\
54 & \href{https://exofop.ipac.caltech.edu/tess/target.php?id=268534931}{268534931} & CTOI     & 15.12618 & -23.1795 & 12.22 & 0.78 & 5179  & $ 246.72^{+7.21}_{-2.25} $ & 2  \\
55 & \href{https://exofop.ipac.caltech.edu/tess/target.php?id=269333648}{269333648} & TOI-2529 & 118.9944 & -52.3549 & 10.67 & 1.75 & 5822  & $ 295.40^{+2.04}_{-1.85} $ & 6  \\
56 & \href{https://exofop.ipac.caltech.edu/tess/target.php?id=275878706}{275878706} & --       & 207.9018 & -44.3144 & 12.17 & 0.93 & 5038  & $ 247.70^{+2.42}_{-2.18} $ & 2  \\
57 & \href{https://exofop.ipac.caltech.edu/tess/target.php?id=279727635}{279727635} & --       & 50.59176 & -64.4444 & 12.5  & 1.05 & 5851  & $ 447.75^{+3.66}_{-3.51} $ & 6  \\
58 & \href{https://exofop.ipac.caltech.edu/tess/target.php?id=287204963}{287204963} & --       & 30.28722 & -11.8835 & 13.25 & 1.04 & 5344  & $ 551.78^{+10.06}_{-5.87} $ & 2  \\
59 & \href{https://exofop.ipac.caltech.edu/tess/target.php?id=289840544}{289840544} & --       & 317.5455 & -26.3999 & 10.64 & 1.75 & 5989  & $ 329.46^{+3.72}_{-3.65} $ & 2  \\
60 & \href{https://exofop.ipac.caltech.edu/tess/target.php?id=290165539}{290165539} & --       & 319.8521 & -25.5971 & 11.99 & 1.33 & 5895  & $ 453.86^{+6.32}_{-15.61} $ & 2  \\
61 & \href{https://exofop.ipac.caltech.edu/tess/target.php?id=292719109}{292719109} & --       & 107.416  & 3.56756  & 10.1  & 2.08 & 6216  & $ 326.24^{+9.07}_{-9.63} $ & 2  \\
62 & \href{https://exofop.ipac.caltech.edu/tess/target.php?id=294097549}{294097549} & --       & 107.67   & -55.7967 & 10.47 & 1.15 & 5872  & $ 197.78^{+0.46}_{-0.5} $ & 13 \\
63 & \href{https://exofop.ipac.caltech.edu/tess/target.php?id=296737508}{296737508} & CTOI     & 142.1793 & -14.6864 & 9.2   & 1.27 & 5700  & $ 117.37^{+0.23}_{-0.22} $ & 2  \\
64 & \href{https://exofop.ipac.caltech.edu/tess/target.php?id=300394149}{300394149} & --       & 170.747  & -52.6986 & 10.13 & 1.42 & 5997  & $ 218.59^{+0.60}_{-0.59} $ & 2  \\
65 & \href{https://exofop.ipac.caltech.edu/tess/target.php?id=303317324}{303317324} & TOI-4310 & 351.9037 & -25.5081 & 9.5   & 0.72 & 4159  & $ 40.13^{+0.43}_{-0.10} $ & 2  \\
66 & \href{https://exofop.ipac.caltech.edu/tess/target.php?id=304339227}{304339227} & TOI-4629 & 275.9439 & -68.3449 & 8.66  & 1.13 & 6012  & $ 91.13^{+0.23}_{-0.15} $ & 2  \\
67 & \href{https://exofop.ipac.caltech.edu/tess/target.php?id=306249066}{306249066} & --       & 166.7285 & -54.7388 & 10.54 & 1.14 & 6139  & $ 221.63^{+1.01}_{-2.35} $ & 2  \\
68 & \href{https://exofop.ipac.caltech.edu/tess/target.php?id=313671132}{313671132} & --       & 202.1582 & -62.1699 & 10.08 & 1.22 & 5878  & $ 175.65^{+0.41}_{-0.43} $ & 2  \\
69 & \href{https://exofop.ipac.caltech.edu/tess/target.php?id=317923092}{317923092} & --       & 89.85183 & -14.9061 & 12.26 & 1.34 & 7604  & $ 668.65^{+21.27}_{-10.4} $ & 2  \\
70 & \href{https://exofop.ipac.caltech.edu/tess/target.php?id=323295479}{323295479} & TOI-1861 & 130.7987 & -83.061  & 9.95  & 1.06 & 5684  & $ 134.32^{+0.24}_{-0.23} $ & 5  \\
71 & \href{https://exofop.ipac.caltech.edu/tess/target.php?id=332697924}{332697924} & --       & 63.26121 & -12.4593 & 11    & 1.32 & 6266  & $ 312.19^{+3.28}_{-9.2} $ & 2  \\
72 & \href{https://exofop.ipac.caltech.edu/tess/target.php?id=333736132}{333736132} & CTOI     & 170.2765 & -26.0666 & 12.07 & 0.76 & 4722  & $ 178.23^{+0.58}_{-1.38} $ & 2  \\
73 & \href{https://exofop.ipac.caltech.edu/tess/target.php?id=339399841}{339399841} & --       & 296.0705 & -62.8136 & 12.35 & 1.55 & 5686  & $ 596.49^{+4.8}_{-5.11} $ & 2  \\
74 & \href{https://exofop.ipac.caltech.edu/tess/target.php?id=349091983}{349091983} & --       & 107.7051 & -61.3524 & 12.29 & 1.5  & 6083  &  & 24 \\
75 & \href{https://exofop.ipac.caltech.edu/tess/target.php?id=381553868}{381553868} & --       & 269.7743 & -50.9212 & 11.85 & 1.59 & 6355  &  & 2  \\
76 & \href{https://exofop.ipac.caltech.edu/tess/target.php?id=393229954}{393229954} & --       & 86.74486 & -46.7119 & 11.77 & 0.9  & 5571  & $ 262.19^{+5.47}_{-1.15} $ & 5  \\
77 & \href{https://exofop.ipac.caltech.edu/tess/target.php?id=394340183}{394340183} & --       & 39.32208 & -79.4583 & 10.95 & 1.04 & 6325  & $ 253.84^{+0.64}_{-0.65} $ & 2  \\
78 & \href{https://exofop.ipac.caltech.edu/tess/target.php?id=396720998}{396720998} & TOI-709  & 65.65532 & -54.1472 & 13.99 & 0.15 & 50000 &  & 6  \\
79 & \href{https://exofop.ipac.caltech.edu/tess/target.php?id=404518509}{404518509} & TOI-4320 & 51.64036 & -43.6136 & 8.6   & 1.04 & 5871  & $ 79.30^{+0.07}_{-0.07} $ & 4  \\
80 & \href{https://exofop.ipac.caltech.edu/tess/target.php?id=412386707}{412386707} & --       & 110.4871 & -23.414  & 10.92 & 1.96 & 6304  & $ 425.23^{+5.01}_{-4.39} $ & 2  \\
81 & \href{https://exofop.ipac.caltech.edu/tess/target.php?id=437293313}{437293313} & --       & 191.2234 & -22.6177 & 12.97 & 0.94 & 5536  & $ 416.28^{+7.26}_{-6.35} $ & 2  \\
82 & \href{https://exofop.ipac.caltech.edu/tess/target.php?id=439491923}{439491923} & --       & 21.38365 & -18.1471 & 12.87 & 1.26 & 5944  & $ 631.09^{+11.04}_{-9.74} $ & 2  \\
83 & \href{https://exofop.ipac.caltech.edu/tess/target.php?id=442893494}{442893494} & --       & 80.00938 & -16.2193 & 12.41 & 1.39 & 6322  & $ 622.15^{+5.22}_{-5.04} $ & 3  \\
84 & \href{https://exofop.ipac.caltech.edu/tess/target.php?id=457649900}{457649900} & TOI-4958 & 217.9991 & -51.0744 & 10.75 & 2.11 & 6713  & $ 479.53^{+4.71}_{-3.72} $ & 2  \\
85 & \href{https://exofop.ipac.caltech.edu/tess/target.php?id=466206508}{466206508} & TOI-5542 & 302.7985 & -61.1355 & 11.85 & 1.27 & 5393  & $ 342.66^{+1.69}_{-1.51} $ & 2 \\
\bottomrule
\label{tab:candidate_stellar}
\end{longtable}

\pagebreak
\begin{longtable}{cccccc}
\caption{The transit event properties of our duotransit candidates.} \\
\toprule
\textbf{ID} &
  \textbf{TIC ID} &
  \textbf{T\textsubscript{c1}} &
  \textbf{T\textsubscript{c2}} &
  \textbf{Depth} &
  \textbf{Duration} \\
  --- &
  --- &
  \textbf{(TBJD, [Sector])} &
  \textbf{(TBJD, [Sector])} &
  \textbf{(ppt)} &
  \textbf{(hours)}
  \endhead
\hline

1  & \href{https://exofop.ipac.caltech.edu/tess/target.php?id=1504460}{1504460}     & 1528.39739 [08] & 2261.11990 [35] & 3.367 [08], 3.415 [35]     & 3.78  \\
2  & \href{https://exofop.ipac.caltech.edu/tess/target.php?id=2669681}{2669681}     & 1676.73302 [13] & 2044.45589 [27] & 7.879 [13], 7.037 [27]     & 9.69  \\
3  & \href{https://exofop.ipac.caltech.edu/tess/target.php?id=9844069}{9844069}     & 1431.16511 [04] & 2148.92146 [31] & 13.452 [04], 12.224 [31]   & 7.20  \\
4  & \href{https://exofop.ipac.caltech.edu/tess/target.php?id=13072758}{13072758}   & 1455.59404 [05] & 2193.21295 [32] & 7.023 [05], 6.755 [32]     & 9.25  \\
5  & \href{https://exofop.ipac.caltech.edu/tess/target.php?id=13713700}{13713700}   & 1536.98226 [08] & 2245.82903 [34] & 7.234 [08], 4.700   [34]   & 3.26  \\
6  & \href{https://exofop.ipac.caltech.edu/tess/target.php?id=14445414}{14445414}   & 1497.55590 [07] & 2210.97620 [33] & 2.317 [07], 2.179   [33]   & 2.86  \\
7  & \href{https://exofop.ipac.caltech.edu/tess/target.php?id=20904104}{20904104}   & 1390.11189 [03] & 2105.36177 [29] & 0.738 [03], 1.065   [29]   & 4.61  \\
8  & \href{https://exofop.ipac.caltech.edu/tess/target.php?id=25194908}{25194908}   & 1402.29491 [03] & 2077.28343 [28] & 6.590 [03], 6.266 [28]     & 9.16  \\
9  & \href{https://exofop.ipac.caltech.edu/tess/target.php?id=32179255}{32179255}   & 1345.87922 [01] & 2068.45418 [28] & 3.619 [01], 4.682   [28]   & 3.71  \\
10 & \href{https://exofop.ipac.caltech.edu/tess/target.php?id=38138512}{38138512}   & 1559.95080 [09] & 2300.02560 [36] & 17.907 [09], 18.148   [36] & 6.73  \\
11 & \href{https://exofop.ipac.caltech.edu/tess/target.php?id=39167176}{39167176}   & 1470.62312 [06] & 2211.42630 [33] & 7.282 [06], 8.112   [33]   & 6.25  \\
12 & \href{https://exofop.ipac.caltech.edu/tess/target.php?id=42428568}{42428568}   & 1475.69109 [06] & 2224.32460 [33] & 5.380 [06], 4.939   [33]   & 10.20 \\
13 & \href{https://exofop.ipac.caltech.edu/tess/target.php?id=52195587}{52195587}   & 1350.27811 [01] & 2107.61083 [29] & 5.775 [01], 6.138   [29]   & 9.07  \\
14 & \href{https://exofop.ipac.caltech.edu/tess/target.php?id=66439839}{66439839}   & 1386.34406 [03] & 2137.03203 [30] & 6.227 [03], 5.420   [30]   & 2.16  \\
15 & \href{https://exofop.ipac.caltech.edu/tess/target.php?id=67599025}{67599025}   & 1404.02101 [03] & 2138.82557 [30] & 6.684 [03], 7.345   [30]   & 3.27  \\
16 & \href{https://exofop.ipac.caltech.edu/tess/target.php?id=70561926}{70561926}   & 1523.06353 [08] & 2275.36040 [35] & 16.290 [08], 16.053   [35] & 5.91  \\
17 & \href{https://exofop.ipac.caltech.edu/tess/target.php?id=71028120}{71028120}   & 1445.81123 [05] & 2189.95412 [32] & 1.238 [05], 1.025   [32]   & 6.11  \\
18 & \href{https://exofop.ipac.caltech.edu/tess/target.php?id=77437543}{77437543}   & 1456.70241 [05] & 2180.69226 [32] & 9.825 [05], 9.326   [32]   & 7.93  \\
19 & \href{https://exofop.ipac.caltech.edu/tess/target.php?id=81089255}{81089255}   & 1509.94498 [07] & 2260.20690 [35] & 7.318 [07], 6.372   [35]   & 5.14  \\
20 & \href{https://exofop.ipac.caltech.edu/tess/target.php?id=100776118}{100776118} & 1472.81029 [06] & 2219.67770 [33] & 13.264 [06], 13.070   [33] & 22.34 \\
21 & \href{https://exofop.ipac.caltech.edu/tess/target.php?id=101824521}{101824521} & 1620.35423 [11] & 2331.52820 [37] & 9.396 [11], 9.103   [37]   & 8.52  \\
22 & \href{https://exofop.ipac.caltech.edu/tess/target.php?id=107113345}{107113345} & 1576.32832 [10] & 2317.29675 [37] & 1.771 [10], 2.035   [37]   & 1.77  \\
23 & \href{https://exofop.ipac.caltech.edu/tess/target.php?id=118339710}{118339710} & 1525.64667 [08] & 2249.54641 [34] & 4.329 [08], 3.977   [34]   & 8.68  \\
24 & \href{https://exofop.ipac.caltech.edu/tess/target.php?id=123763494}{123763494} & 1486.92891 [06] & 2209.76758 [33] & 10.645 [06], 11.728   [33] & 2.47  \\
25 & \href{https://exofop.ipac.caltech.edu/tess/target.php?id=124029677}{124029677} & 1486.12090 [06] & 2218.00409 [33] & 6.368 [06], 6.105   [33]   & 4.58  \\
26 & \href{https://exofop.ipac.caltech.edu/tess/target.php?id=140215502}{140215502} & 1460.54584 [05] & 2194.44033 [32] & 2.959 [05], 4.468   [32]   & 7.14  \\
27 & \href{https://exofop.ipac.caltech.edu/tess/target.php?id=140750416}{140750416} & 1504.80151 [07] & 2245.39501 [34] & 8.912 [07], 9.197   [34]   & 3.85  \\
28 & \href{https://exofop.ipac.caltech.edu/tess/target.php?id=142278054}{142278054} & 1595.02504 [10] & 2310.35369 [37] & 2.177 [10], 2.022   [37]   & 4.82  \\
29 & \href{https://exofop.ipac.caltech.edu/tess/target.php?id=145006304}{145006304} & 1579.50384 [10] & 2318.24443 [37] & 2.048 [10], 2.772   [37]   & 5.19  \\
30 & \href{https://exofop.ipac.caltech.edu/tess/target.php?id=145913596}{145913596} & 1539.51696 [08] & 2242.69675 [34] & 1.728 [08], 2.163   [34]   & 8.50  \\
31 & \href{https://exofop.ipac.caltech.edu/tess/target.php?id=153838604}{153838604} & 1447.10227 [05] & 2176.13853 [32] & 15.209 [05], 16.237   [32] & 2.61  \\
32 & \href{https://exofop.ipac.caltech.edu/tess/target.php?id=156716001}{156716001} & 1485.40631 [06] & 2224.51833 [33] & 2.652 [06], 3.037   [33]   & 9.11  \\
33 & \href{https://exofop.ipac.caltech.edu/tess/target.php?id=157119927}{157119927} & 1591.62175 [10] & 2291.00613 [36] & 5.445 [10], 6.897   [36]   & 2.57  \\
34 & \href{https://exofop.ipac.caltech.edu/tess/target.php?id=157698565}{157698565} & 1494.57973 [07] & 2234.11360 [34] & 9.825 [07], 9.338   [34]   & 6.38  \\
35 & \href{https://exofop.ipac.caltech.edu/tess/target.php?id=159490807}{159490807} & 1606.22551 [11] & 2355.93796 [38] & 5.155 [11], 5.184   [38]   & 5.38  \\
36 & \href{https://exofop.ipac.caltech.edu/tess/target.php?id=161169240}{161169240} & 1328.25945 [01] & 2083.31754 [28] & 9.707 [01], 10.134   [28]  & 7.82  \\
37 & \href{https://exofop.ipac.caltech.edu/tess/target.php?id=176518126}{176518126} & 1489.05533 [06] & 2219.53761 [33] & 3.565 [06], 3.244   [33]   & 5.61  \\
38 & \href{https://exofop.ipac.caltech.edu/tess/target.php?id=188620407}{188620407} & 1375.48660 [02] & 2461.70715 [42] & 5.467 [02], 4.512   [42]   & 6.08  \\
39 & \href{https://exofop.ipac.caltech.edu/tess/target.php?id=193318850}{193318850} & 1553.77031 [09] & 2292.51011 [36] & 6.916 [09], 7.143   [36]   & 3.89  \\
40 & \href{https://exofop.ipac.caltech.edu/tess/target.php?id=207783865}{207783865} & 1608.93692 [11] & 2354.11345 [38] & 17.211 [11], 17.729   [38] & 1.99  \\
41 & \href{https://exofop.ipac.caltech.edu/tess/target.php?id=211409161}{211409161} & 1335.10057 [01] & 2070.36583 [28] & 13.629 [01], 12.232   [28] & 2.14  \\
42 & \href{https://exofop.ipac.caltech.edu/tess/target.php?id=215402824}{215402824} & 1413.62655 [04] & 2169.04205 [31] & 5.589 [04], 9.955   [31]   & 1.87  \\
43 & \href{https://exofop.ipac.caltech.edu/tess/target.php?id=218977148}{218977148} & 1355.47069 [02] & 2090.09772 [29] & 5.597 [02], 4.944   [29]   & 8.13  \\
44 & \href{https://exofop.ipac.caltech.edu/tess/target.php?id=220622886}{220622886} & 1479.52544 [06] & 2204.58502 [33] & 4.324 [06], 3.501   [33]   & 4.43  \\
45 & \href{https://exofop.ipac.caltech.edu/tess/target.php?id=221915858}{221915858} & 1643.35799 [12] & 2372.55118 [39] & 5.737 [12], 5.526   [39]   & 3.05  \\
46 & \href{https://exofop.ipac.caltech.edu/tess/target.php?id=224279805}{224279805} & 1369.72941 [02] & 2105.21571 [29] & 6.392 [02], 6.618   [29]   & 5.93  \\
47 & \href{https://exofop.ipac.caltech.edu/tess/target.php?id=235058563}{235058563} & 1416.61066 [04] & 2216.44259 [33] & 6.164 [04], 5.630   [33]   & 10.17 \\
48 & \href{https://exofop.ipac.caltech.edu/tess/target.php?id=237605045}{237605045} & 1489.04268 [06] & 2210.29724 [33] & 6.212 [06], 7.014   [33]   & 4.42  \\
49 & \href{https://exofop.ipac.caltech.edu/tess/target.php?id=242241304}{242241304} & 1623.07095 [11] & 2345.37403 [38] & 6.158 [11], 6.186   [38]   & 4.72  \\
50 & \href{https://exofop.ipac.caltech.edu/tess/target.php?id=251057075}{251057075} & 1426.31846 [04] & 2154.02335 [31] & 1.986 [04], 2.348   [31]   & 4.42  \\
51 & \href{https://exofop.ipac.caltech.edu/tess/target.php?id=256912435}{256912435} & 1623.43775 [11] & 2338.70979 [38] & 20.396 [11], 21.562   [38] & 2.72  \\
52 & \href{https://exofop.ipac.caltech.edu/tess/target.php?id=265465927}{265465927} & 1342.64749 [01] & 2058.60913 [27] & 8.678 [01], 8.720   [27]   & 6.05  \\
53 & \href{https://exofop.ipac.caltech.edu/tess/target.php?id=265466589}{265466589} & 1352.52439 [01] & 2037.86660 [27] & 6.818 [01], 6.664   [27]   & 4.14  \\
54 & \href{https://exofop.ipac.caltech.edu/tess/target.php?id=268534931}{268534931} & 1398.29475 [03] & 2139.09941 [30] & 5.829 [03], 6.133   [30]   & 7.82  \\
55 & \href{https://exofop.ipac.caltech.edu/tess/target.php?id=269333648}{269333648} & 1538.52419 [08] & 2249.05416 [34] & 4.162 [08], 4.032   [34]   & 9.46  \\
56 & \href{https://exofop.ipac.caltech.edu/tess/target.php?id=275878706}{275878706} & 1621.08010 [11] & 2341.78257 [38] & 6.544 [11], 5.320   [38]   & 4.38  \\
57 & \href{https://exofop.ipac.caltech.edu/tess/target.php?id=279727635}{279727635} & 1343.22676 [01] & 2124.62930 [30] & 7.181 [01], 6.789   [30]   & 5.71  \\
58 & \href{https://exofop.ipac.caltech.edu/tess/target.php?id=287204963}{287204963} & 1402.93640 [03] & 2132.33438 [30] & 11.498 [03], 10.182   [30] & 4.73  \\
59 & \href{https://exofop.ipac.caltech.edu/tess/target.php?id=289840544}{289840544} & 1331.28977 [01] & 2083.28790 [28] & 2.702 [01], 1.997   [28]   & 2.99  \\
60 & \href{https://exofop.ipac.caltech.edu/tess/target.php?id=290165539}{290165539} & 1332.75615 [01] & 2067.97094 [28] & 13.627 [01], 15.056   [28] & 11.50 \\
61 & \href{https://exofop.ipac.caltech.edu/tess/target.php?id=292719109}{292719109} & 1499.70159 [07] & 2222.31933 [33] & 2.859 [07], 2.677   [33]   & 9.08  \\
62 & \href{https://exofop.ipac.caltech.edu/tess/target.php?id=294097549}{294097549} & 1657.74149 [13] & 2256.34250 [35] & 1.977 [13], 1.426   [35]   & 7.96  \\
63 & \href{https://exofop.ipac.caltech.edu/tess/target.php?id=296737508}{296737508} & 1538.00622 [08] & 2271.99796 [35] & 2.021 [08], 1.854   [35]   & 5.65  \\
64 & \href{https://exofop.ipac.caltech.edu/tess/target.php?id=300394149}{300394149} & 1576.88642 [10] & 2325.42588 [37] & 2.051 [10], 2.195   [37]   & 13.40 \\
65 & \href{https://exofop.ipac.caltech.edu/tess/target.php?id=303317324}{303317324} & 1365.17582 [02] & 2104.34337 [29] & 1.360 [02], 1.323   [29]   & 5.74  \\
66 & \href{https://exofop.ipac.caltech.edu/tess/target.php?id=304339227}{304339227} & 1673.33918 [13] & 2388.01911 [39] & 0.522 [13], 0.464   [39]   & 5.79  \\
67 & \href{https://exofop.ipac.caltech.edu/tess/target.php?id=306249066}{306249066} & 1593.97611 [10] & 2326.51397 [37] & 7.873 [10], 7.061   [37]   & 4.03  \\
68 & \href{https://exofop.ipac.caltech.edu/tess/target.php?id=313671132}{313671132} & 1604.35452 [11] & 2337.72693 [38] & 3.408 [11], 3.506   [38]   & 5.70  \\
69 & \href{https://exofop.ipac.caltech.edu/tess/target.php?id=317923092}{317923092} & 1479.81772 [06] & 2209.42617 [33] & 4.358 [06], 6.715   [33]   & 12.96 \\
70 & \href{https://exofop.ipac.caltech.edu/tess/target.php?id=323295479}{323295479} & 1622.92483 [11] & 2365.44036 [39] & 8.727 [11], 8.609   [39]   & 6.51  \\
71 & \href{https://exofop.ipac.caltech.edu/tess/target.php?id=332697924}{332697924} & 1445.74977 [05] & 2178.22194 [32] & 1.042 [05], 1.423   [32]   & 6.63  \\
72 & \href{https://exofop.ipac.caltech.edu/tess/target.php?id=333736132}{333736132} & 1551.55292 [09] & 2283.54681 [36] & 15.856 [09], 16.440   [36] & 2.96  \\
73 & \href{https://exofop.ipac.caltech.edu/tess/target.php?id=339399841}{339399841} & 1662.10984 [13] & 2043.59486 [27] & 4.773 [13], 6.471   [27]   & 6.66  \\
74 & \href{https://exofop.ipac.caltech.edu/tess/target.php?id=349091983}{349091983} & 1603.10253 [11] & 2152.16754 [31] & 8.301 [11], 8.133   [31]   & 9.52  \\
75 & \href{https://exofop.ipac.caltech.edu/tess/target.php?id=381553868}{381553868} & 1677.32594 [13] & 2367.57869 [39] & 5.653 [13], 6.030   [39]   & 6.49  \\
76 & \href{https://exofop.ipac.caltech.edu/tess/target.php?id=393229954}{393229954} & 1446.79953 [05] & 2204.55168 [33] & 8.172 [05], 6.652   [33]   & 2.67  \\
77 & \href{https://exofop.ipac.caltech.edu/tess/target.php?id=394340183}{394340183} & 1664.84261 [13] & 2389.11243 [39] & 11.255 [13], 10.820   [39] & 8.65  \\
78 & \href{https://exofop.ipac.caltech.edu/tess/target.php?id=396720998}{396720998} & 1399.21966 [03] & 2176.27178 [30] & 6.360 [03], 6.082   [30]   & 4.33  \\
79 & \href{https://exofop.ipac.caltech.edu/tess/target.php?id=404518509}{404518509} & 1431.26615 [04] & 2134.88294 [30] & 0.842 [04], 0.598   [30]   & 5.02  \\
80 & \href{https://exofop.ipac.caltech.edu/tess/target.php?id=412386707}{412386707} & 1493.95365 [07] & 2240.57066 [34] & 3.442 [07], 3.587   [34]   & 5.23  \\
81 & \href{https://exofop.ipac.caltech.edu/tess/target.php?id=437293313}{437293313} & 1578.44737 [10] & 2328.08914 [37] & 9.891 [10], 8.462   [37]   & 6.58  \\
82 & \href{https://exofop.ipac.caltech.edu/tess/target.php?id=439491923}{439491923} & 1405.41115 [03] & 2116.79038 [30] & 11.608 [03], 11.766   [30] & 6.70  \\
83 & \href{https://exofop.ipac.caltech.edu/tess/target.php?id=442893494}{442893494} & 1472.65885 [06] & 2194.38454 [32] & 5.776 [06], 6.458   [32]   & 7.70  \\
84 & \href{https://exofop.ipac.caltech.edu/tess/target.php?id=457649900}{457649900} & 1606.96984 [11] & 2355.79522 [38] & 2.857 [11], 2.945   [38]   & 7.61  \\
85 & \href{https://exofop.ipac.caltech.edu/tess/target.php?id=466206508}{466206508} & 1679.34576 [13] & 2054.97133 [27] & 11.214 [13], 9.274   [27]  & 8.03

\\ \bottomrule
\label{tab:candidate_events}
\end{longtable}
\begin{landscape}
\begin{longtable}{cccccccccc}
\caption{Properties of our duotransit candidates. $N_{\rm alias}$ is the alias number corresponding to the most probable orbital period $P_{\rm marg}$.} \\
\toprule
\textbf{ID} &
  \textbf{TIC ID} &
  \textbf{$R_p$} &
  \textbf{P\textsubscript{min}, P\textsubscript{max}} &
  \textbf{$N_{\rm alias}$} &
  \textbf{$P_{\rm marg}$} &
  \textbf{$b$} &
  \textbf{\gaia} &
  \textbf{\gaia\ RV} &
  \textbf{Total SNR} \\
--- &
  --- &
\textbf{($R_\oplus$)} &
  \textbf{(days)} &
  \textbf{---} &
  \textbf{(days)} &
  \textbf{---} &
  \textbf{NSS} &
  \textbf{amp (\kms)} &
  \textbf{---}
  \endhead
\\ \hline

1  & \href{https://exofop.ipac.caltech.edu/tess/target.php?id=1504460}{1504460}     & $ 9.6 \pm 1.0 $             & 19.80333, 732.72251  & 24 & $   20.0^{+19.0}_{-1.0} $  & $ 0.78^{+0.12}_{-0.37} $      & 0 & 13.84 & 14.82  \\
2  & \href{https://exofop.ipac.caltech.edu/tess/target.php?id=2669681}{2669681}     & $ 17.0 \pm 1.2 $            & 21.63067, 367.72287  & 18 & $ 24.5^{+8.9}_{-2.9} $     & $ 0.66^{+0.09}_{-0.24}   $    & 0 & 36.15 & 41.8   \\
3  & \href{https://exofop.ipac.caltech.edu/tess/target.php?id=9844069}{9844069}     & $ 18.9 \pm 1.4 $            & 22.42991, 717.75635  & 33 & $ 25.6^{+7.0}_{-3.2} $     & $ 0.62^{+0.12}_{-0.3}   $     & 0 & --    & 46.12  \\
4  & \href{https://exofop.ipac.caltech.edu/tess/target.php?id=13072758}{13072758}   & $   11.78^{+0.67}_{-0.65} $ & 19.93567, 737.61891  & 38 & $ 26.0 \pm 10.0 $          & $ 0.31 \pm 0.22 $             & 0 & 20.36 & 35.92  \\
5  & \href{https://exofop.ipac.caltech.edu/tess/target.php?id=13713700}{13713700}   & $ 8.3 \pm 1.0 $             & 19.69018, 708.84677  & 37 & $ 22.9^{+8.0}_{-3.2} $     & $ 0.79^{+0.11}_{-0.28}   $    & 0 & --    & 19.73  \\
6  & \href{https://exofop.ipac.caltech.edu/tess/target.php?id=14445414}{14445414}   & $ 7.5^{+3.9}_{-1.2}   $     & 19.28163, 713.42031  & 38 & $ 22.3^{+7.4}_{-3.0} $     & $ 0.93^{+0.06}_{-0.18}   $    & 0 & 4.78  & 12.08  \\
7  & \href{https://exofop.ipac.caltech.edu/tess/target.php?id=20904104}{20904104}   & $ 2.6 \pm 0.19 $            & 17.88103, 715.24989  & 41 & $ 23.0^{+11.0}_{-4.0} $    & $ 0.27^{+0.25}_{-0.19}   $    & 0 & 2.9   & 11.84  \\
8  & \href{https://exofop.ipac.caltech.edu/tess/target.php?id=25194908}{25194908}   & $ 11.18 \pm 0.68 $          & 337.49490, 674.98980 & 2  &  $337.49490$                    & $ 0.7^{+0.1}_{-0.28} $        & 0 & --    & 31.43  \\
9  & \href{https://exofop.ipac.caltech.edu/tess/target.php?id=32179255}{32179255}   & $ 21.0 \pm 8.0 $            & 20.64500, 722.57496  & 36 & $ 22.6^{+6.3}_{-1.9} $     & $   0.999^{+0.07}_{-0.045} $  & 2 & 49.01 & 36.77  \\
10 & \href{https://exofop.ipac.caltech.edu/tess/target.php?id=38138512}{38138512}   & $ 17.8 \pm 1.0 $            & 18.50185, 740.07480  & 41 & $ 21.1^{+6.3}_{-2.2} $     & $ 0.32 \pm 0.2 $              & 0 & --    & 49.35  \\
11 & \href{https://exofop.ipac.caltech.edu/tess/target.php?id=39167176}{39167176}   & $ 18.8^{+8.9}_{-2.9}   $    & 20.57787, 740.80317  & 37 & $ 26.0^{+20.0}_{-5.0} $    & $   0.947^{+0.081}_{-0.044} $ & 0 & 81.59 & 40.79  \\
12 & \href{https://exofop.ipac.caltech.edu/tess/target.php?id=42428568}{42428568}   & $ 24.0^{+11.0}_{-5.0}   $   & 24.14954, 748.63351  & 32 & $ 31.0^{+22.0}_{-6.0}   $  & $   0.981^{+0.066}_{-0.038} $ & 0 & 16.42 & 56.65  \\
13 & \href{https://exofop.ipac.caltech.edu/tess/target.php?id=52195587}{52195587}   & $ 9.28 \pm 0.52 $           & 34.42428, 757.33272  & 18 & $ 50.0 \pm 17.0 $          & $ 0.29 \pm 0.22 $             & 0 & --    & 35.19  \\
14 & \href{https://exofop.ipac.caltech.edu/tess/target.php?id=66439839}{66439839}   & $ 9.1^{+2.8}_{-1.0}   $     & 37.53440, 750.68798  & 17 & $ 38.0^{+20.0}_{-2.0} $    & $ 0.87^{+0.09}_{-0.29}   $    & 0 & 30.94 & 19.55  \\
15 & \href{https://exofop.ipac.caltech.edu/tess/target.php?id=67599025}{67599025}   & $ 6.86 \pm 0.57 $           & 24.49349, 734.80456  & 31 & $ 27.2^{+7.8}_{-2.7} $     & $ 0.62^{+0.13}_{-0.35}   $    & 0 & 9.86  & 29.54  \\
16 & \href{https://exofop.ipac.caltech.edu/tess/target.php?id=70561926}{70561926}   & $ 17.51 \pm 0.82   $        & 20.33234, 752.29687  & 38 & $ 22.1^{+5.7}_{-1.8} $     & $ 0.15 \pm 0.12 $             & 2 & 18.35 & 144.8  \\
17 & \href{https://exofop.ipac.caltech.edu/tess/target.php?id=71028120}{71028120}   & $ 4.74 \pm 0.31 $           & 18.60345, 744.14289  & 41 & $ 21.3^{+6.3}_{-2.7} $     & $ 0.38^{+0.22}_{-0.25}   $    & 0 & 2.53  & 14.28  \\
18 & \href{https://exofop.ipac.caltech.edu/tess/target.php?id=77437543}{77437543}   & $   10.98^{+0.66}_{-0.63} $ & 22.62465, 723.98986  & 22 & $ 38.0^{+14.0}_{-15.0} $   & $ 0.31 \pm 0.22 $             & 2 & 14.02 & 83.13  \\
19 & \href{https://exofop.ipac.caltech.edu/tess/target.php?id=81089255}{81089255}   & $ 12.19 \pm 0.69   $        & 34.10282, 750.26191  & 23 & $ 37.5^{+9.4}_{-4.9} $     & $ 0.76^{+0.07}_{-0.26}   $    & 0 & 24.24 & 25.14  \\
20 & \href{https://exofop.ipac.caltech.edu/tess/target.php?id=100776118}{100776118} & $ 16.19 \pm 0.74   $        & 49.79069, 746.86040  & 16 & $ 62.0^{+31.0}_{-12.0}   $ & $ 0.842 \pm 0.01   $          & 3 & 30.51 & 177.47 \\
21 & \href{https://exofop.ipac.caltech.edu/tess/target.php?id=101824521}{101824521} & $ 12.17 \pm 0.72   $        & 24.52324, 711.17397  & 30 & $ 31.0 \pm 10.0 $          & $ 0.18^{+0.18}_{-0.13}   $    & 0 & --    & 34.34  \\
22 & \href{https://exofop.ipac.caltech.edu/tess/target.php?id=107113345}{107113345} & $ 9.2^{+5.1}_{-2.3}   $     & 20.02618, 740.96843  & 38 & $ 22.5^{+6.0}_{-2.4} $     & $ 0.976 \pm 0.048   $         & 0 & 19.1  & 12.5   \\
23 & \href{https://exofop.ipac.caltech.edu/tess/target.php?id=118339710}{118339710} & $   10.64^{+0.65}_{-0.61} $ & 21.29118, 723.89974  & 26 & $ 33.0 \pm 12.0 $          & $ 0.48 \pm 0.29 $             & 0 & 4.13  & 62.67  \\
24 & \href{https://exofop.ipac.caltech.edu/tess/target.php?id=123763494}{123763494} & $ 19.8^{+7.1}_{-4.5}   $    & 19.53618, 722.83868  & 38 & $ 21.9^{+5.9}_{-2.4} $     & $   0.97^{+0.076}_{-0.062} $  & 0 & 15.94 & 37.36  \\
25 & \href{https://exofop.ipac.caltech.edu/tess/target.php?id=124029677}{124029677} & $ 11.35 \pm 0.59   $        & 18.76624, 731.88319  & 40 & $ 20.9^{+5.2}_{-2.1} $     & $ 0.43^{+0.22}_{-0.28}   $    & 0 & 11.35 & 46.16  \\
26 & \href{https://exofop.ipac.caltech.edu/tess/target.php?id=140215502}{140215502} & $ 5.78^{+0.38}_{-0.34}   $  & 34.94727, 733.89270  & 22 & $ 43.0^{+18.0}_{-8.0}   $  & $ 0.31^{+0.3}_{-0.21}   $     & 0 & 9.83  & 32.98  \\
27 & \href{https://exofop.ipac.caltech.edu/tess/target.php?id=140750416}{140750416} & $ 8.55 \pm 0.54 $           & 16.83165, 740.59260  & 45 & $ 19.0^{+4.9}_{-2.2}   $   & $ 0.36 \pm 0.22 $             & 0 & 11.19 & 30.13  \\
28 & \href{https://exofop.ipac.caltech.edu/tess/target.php?id=142278054}{142278054} & $   11.49^{+0.96}_{-0.91} $ & 24.66649, 715.32865  & 30 & $ 27.5^{+6.6}_{-2.8} $     & $ 0.79^{+0.11}_{-0.38}   $    & 0 & 7.72  & 23.25  \\
29 & \href{https://exofop.ipac.caltech.edu/tess/target.php?id=145006304}{145006304} & $ 5.33^{+0.47}_{-0.43} $    & 16.78954, 738.74000  & 45 & $ 18.9^{+5.7}_{-2.2} $     & $ 0.62^{+0.17}_{-0.35} $      & 0 & 6.03  & 13.84  \\
30 & \href{https://exofop.ipac.caltech.edu/tess/target.php?id=145913596}{145913596} & $ 6.9^{+2.7}_{-1.1}   $     & 54.09067, 703.17978  & 14 & $ 59.0^{+29.0}_{-8.0} $    & $ 0.91^{+0.07}_{-0.32}   $    & 0 & 5.83  & 25.32  \\
31 & \href{https://exofop.ipac.caltech.edu/tess/target.php?id=153838604}{153838604} & $ 17.1^{+4.7}_{-3.3}   $    & 38.37032, 729.03626  & 20 & $ 41.0^{+12.0}_{-4.0} $    & $   0.961^{+0.068}_{-0.062} $ & 0 & --    & 42.87  \\
32 & \href{https://exofop.ipac.caltech.edu/tess/target.php?id=156716001}{156716001} & $ 5.97 \pm 0.3 $            & 33.59600, 739.11201  & 23 & $ 43.0^{+30.0}_{-10.0} $   & $ 0.28 \pm 0.22 $             & 0 & 4.73  & 33.42  \\
33 & \href{https://exofop.ipac.caltech.edu/tess/target.php?id=157119927}{157119927} & $ 13.4^{+5.9}_{-3.1}   $    & 21.19346, 699.38439  & 34 & $ 24.1^{+6.3}_{-2.9} $     & $   0.983^{+0.071}_{-0.048} $ & 2 & 64.82 & 25.01  \\
34 & \href{https://exofop.ipac.caltech.edu/tess/target.php?id=157698565}{157698565} & $ 11.66 \pm 0.59   $        & 35.21588, 739.53387  & 22 & $ 46.0^{+28.0}_{-11.0} $   & $   0.827^{+0.025}_{-0.041} $ & 0 & 3.49  & 105.68 \\
35 & \href{https://exofop.ipac.caltech.edu/tess/target.php?id=159490807}{159490807} & $ 15.4^{+4.7}_{-1.6}   $    & 22.71853, 749.71245  & 34 & $ 29.0 \pm 8.0 $           & $   0.938^{+0.048}_{-0.025} $ & 0 & 6.08  & 45.05  \\
36 & \href{https://exofop.ipac.caltech.edu/tess/target.php?id=161169240}{161169240} & $   12.19^{+0.77}_{-0.73} $ & 26.03638, 755.05809  & 30 & $ 31.0^{+13.0}_{-4.0} $    & $ 0.47^{+0.21}_{-0.29}   $    & 0 & 32.37 & 52.27  \\
37 & \href{https://exofop.ipac.caltech.edu/tess/target.php?id=176518126}{176518126} & $ 8.3 \pm 1.1 $             & 21.48477, 730.48210  & 35 & $ 24.3^{+7.4}_{-2.9}   $   & $ 0.3 \pm 0.22 $              & 0 & 9.38  & 32.89  \\
38 & \href{https://exofop.ipac.caltech.edu/tess/target.php?id=188620407}{188620407} & $ 9.58^{+0.57}_{-0.54}   $  & 24.13819, 1086.21870 & 35 & $ 28.0 \pm 10.0 $          & $ 0.39 \pm 0.24 $             & 0 & 11.56 & 33.51  \\
39 & \href{https://exofop.ipac.caltech.edu/tess/target.php?id=193318850}{193318850} & $ 7.08 \pm 0.47 $           & 15.07635, 738.73979  & 50 & $ 17.6^{+5.5}_{-2.2} $     & $ 0.31 \pm 0.22 $             & 0 & --    & 17.05  \\
40 & \href{https://exofop.ipac.caltech.edu/tess/target.php?id=207783865}{207783865} & $ 15.31 \pm 0.76   $        & 21.29077, 745.17689  & 36 & $ 23.3^{+6.5}_{-2.0}   $   & $   0.802^{+0.023}_{-0.035} $ & 2 & 13.91 & 95.83  \\
41 & \href{https://exofop.ipac.caltech.edu/tess/target.php?id=211409161}{211409161} & $ 13.1^{+4.5}_{-2.6}   $    & 18.85295, 735.26527  & 40 & $ 21.0^{+5.3}_{-2.2} $     & $   0.943^{+0.077}_{-0.066} $ & 0 & 38.64 & 38.49  \\
42 & \href{https://exofop.ipac.caltech.edu/tess/target.php?id=215402824}{215402824} & $ 16.0^{+10.0}_{-3.0}   $   & 41.96752, 755.41550  & 16 & $ 42.0^{+21.0}_{-2.0} $    & $ 0.93^{+0.11}_{-0.22}   $    & 0 & --    & 12.73  \\
43 & \href{https://exofop.ipac.caltech.edu/tess/target.php?id=218977148}{218977148} & $ 11.9 \pm 0.82 $           & 27.20845, 734.62703  & 28 & $ 30.6^{+8.1}_{-3.4} $     & $ 0.4^{+0.22}_{-0.26}   $     & 0 & --    & 22.83  \\
44 & \href{https://exofop.ipac.caltech.edu/tess/target.php?id=220622886}{220622886} & $ 28.0^{+16.0}_{-8.0}   $   & 42.65054, 725.05958  & 18 & $ 29.0^{+23.0}_{-0.0}   $  & $   0.994^{+0.071}_{-0.048} $ & 0 & 65.91 & 18.89  \\
45 & \href{https://exofop.ipac.caltech.edu/tess/target.php?id=221915858}{221915858} & $ 21.3^{+3.7}_{-1.4}   $    & 17.78520, 729.19319  & 42 & $ 19.7^{+5.4}_{-1.9} $     & $   0.911^{+0.044}_{-0.025} $ & 0 & 60.96 & 33.79  \\
46 & \href{https://exofop.ipac.caltech.edu/tess/target.php?id=224279805}{224279805} & $ 10.55 \pm 0.64   $        & 17.51161, 735.48631  & 43 & $ 19.9^{+5.5}_{-1.9} $     & $ 0.24 \pm 0.18 $             & 0 & 17.77 & 30.43  \\
47 & \href{https://exofop.ipac.caltech.edu/tess/target.php?id=235058563}{235058563} & $ 11.47 \pm 0.61   $        & 88.87019, 799.83193  & 10 & $ 89.0^{+25.0}_{-9.0} $    & $ 0.38 \pm 0.23 $             & 0 & --    & 27.17  \\
48 & \href{https://exofop.ipac.caltech.edu/tess/target.php?id=237605045}{237605045} & $ 44.0^{+17.0}_{-11.0}   $  & 21.85618, 721.25400  & 34 & $ 24.0^{+6.0}_{-2.2}   $   & $   1.01^{+0.067}_{-0.051} $  & 0 & 36.88 & 68.84  \\
49 & \href{https://exofop.ipac.caltech.edu/tess/target.php?id=242241304}{242241304} & $ 9.7 \pm 1.0 $             & 23.30012, 722.30308  & 32 & $ 28.0^{+10.0}_{-4.0} $    & $   0.855^{+0.061}_{-0.071} $ & 0 & 7.03  & 28.72  \\
50 & \href{https://exofop.ipac.caltech.edu/tess/target.php?id=251057075}{251057075} & $ 3.88 \pm 0.3 $            & 16.53870, 727.70489  & 45 & $ 20.2^{+7.8}_{-3.3} $     & $ 0.27 \pm 0.22 $             & 0 & 7.82  & 15.54  \\
51 & \href{https://exofop.ipac.caltech.edu/tess/target.php?id=256912435}{256912435} & $ 25.5^{+4.6}_{-3.8}   $    & 24.66455, 715.27204  & 30 & $ 27.5^{+6.6}_{-2.8}   $   & $ 0.95 \pm 0.051   $          & 0 & 46.85 & 57.7   \\
52 & \href{https://exofop.ipac.caltech.edu/tess/target.php?id=265465927}{265465927} & $ 11.73 \pm 0.63   $        & 23.86539, 715.96164  & 31 & $ 27.5^{+8.3}_{-3.7} $     & $ 0.27 \pm 0.19 $             & 0 & 19.67 & 38.93  \\
53 & \href{https://exofop.ipac.caltech.edu/tess/target.php?id=265466589}{265466589} & $ 8.57 \pm 0.59 $           & 28.55588, 685.34120  & 25 & $ 31.2^{+9.2}_{-3.7} $     & $ 0.49^{+0.21}_{-0.33} $      & 0 & --    & 13.71  \\
54 & \href{https://exofop.ipac.caltech.edu/tess/target.php?id=268534931}{268534931} & $ 6.12 \pm 0.39 $           & 24.69345, 740.80465  & 31 & $ 32.0^{+25.0}_{-7.0} $    & $ 0.29 \pm 0.23 $             & 0 & 25.23 & 29.74  \\
55 & \href{https://exofop.ipac.caltech.edu/tess/target.php?id=269333648}{269333648} & $ 11.72 \pm 0.67   $        & 64.59377, 710.52997  & 12 & $ 65.0^{+24.0}_{-5.0} $    & $ 0.64^{+0.1}_{-0.25}   $     & 0 & 3.2   & 54.46  \\
56 & \href{https://exofop.ipac.caltech.edu/tess/target.php?id=275878706}{275878706} & $ 8.1^{+3.1}_{-1.3}   $     & 22.52196, 720.70247  & 33 & $ 27.0^{+11.0}_{-4.0} $    & $ 0.88^{+0.09}_{-0.21}   $    & 1 & 21.81 & 20.68  \\
57 & \href{https://exofop.ipac.caltech.edu/tess/target.php?id=279727635}{279727635} & $ 10.5^{+4.0}_{-1.2}   $    & 65.11699, 781.40254  & 13 & $ 71.0^{+41.0}_{-19.0} $   & $ 0.905 \pm 0.078   $         & 0 & --    & 28.36  \\
58 & \href{https://exofop.ipac.caltech.edu/tess/target.php?id=287204963}{287204963} & $ 11.3 \pm 2.2 $            & 17.79021, 729.39798  & 42 & $ 20.3^{+5.8}_{-2.5} $     & $ 0.41^{+0.25}_{-0.28}   $    & 0 & --    & 17.86  \\
59 & \href{https://exofop.ipac.caltech.edu/tess/target.php?id=289840544}{289840544} & $ 14.0^{+9.1}_{-3.6}   $    & 22.78782, 751.99813  & 34 & $ 25.1^{+6.3}_{-2.3} $     & $   0.984^{+0.063}_{-0.045} $ & 0 & 6.43  & 17.85  \\
60 & \href{https://exofop.ipac.caltech.edu/tess/target.php?id=290165539}{290165539} & $ 23.4^{+6.3}_{-3.4}   $    & 21.62396, 735.21479  & 35 & $ 27.0^{+19.0}_{-5.0}   $  & $   0.95^{+0.066}_{-0.046} $  & 0 & 45.38 & 92.47  \\
61 & \href{https://exofop.ipac.caltech.edu/tess/target.php?id=292719109}{292719109} & $ 11.07 \pm 0.59   $        & 21.89747, 722.61775  & 34 & $ 24.1^{+7.3}_{-2.2} $     & $ 0.34 \pm 0.23 $             & 0 & 2.5   & 45.22  \\
62 & \href{https://exofop.ipac.caltech.edu/tess/target.php?id=294097549}{294097549} & $ 4.79 \pm 0.33 $           & 199.535169, 598.60600 & 3  & $199.5^{+99.8}_{-0}$  & $ 0.6^{+0.17}_{-0.37} $       & 0 & 4.14  & 22.63  \\
63 & \href{https://exofop.ipac.caltech.edu/tess/target.php?id=296737508}{296737508} & $ 5.88 \pm 0.36 $           & 20.97145, 733.99174  & 36 & $ 23.7^{+6.9}_{-2.7} $     & $ 0.6^{+0.12}_{-0.26}   $     & 0 & 1.12  & 41.08  \\
64 & \href{https://exofop.ipac.caltech.edu/tess/target.php?id=300394149}{300394149} & $ 6.46 \pm 0.34 $           & 19.69843, 748.53947  & 39 & $ 26.0^{+16.0}_{-6.0} $    & $ 0.27 \pm 0.22 $             & 0 & 4.92  & 36.29  \\
65 & \href{https://exofop.ipac.caltech.edu/tess/target.php?id=303317324}{303317324} & $ 2.67 \pm 0.26 $           & 16.79925, 739.16710  & 45 & $ 18.5^{+4.6}_{-1.7}   $   & $ 0.12 \pm 0.11 $             & 0 & 1.16  & 22.71  \\
66 & \href{https://exofop.ipac.caltech.edu/tess/target.php?id=304339227}{304339227} & $ 2.65 \pm 0.17 $           & 27.48783, 714.68370  & 27 & $ 31.1^{+8.6}_{-3.6}   $   & $ 0.34 \pm 0.24 $             & 0 & 2.19  & 13.81  \\
67 & \href{https://exofop.ipac.caltech.edu/tess/target.php?id=306249066}{306249066} & $ 11.09 \pm 0.61   $        & 23.63027, 732.53786  & 32 & $ 27.1^{+7.8}_{-3.5} $     & $ 0.867 \pm 0.021   $         & 0 & 7.34  & 49.32  \\
68 & \href{https://exofop.ipac.caltech.edu/tess/target.php?id=313671132}{313671132} & $ 7.78^{+0.67}_{-0.62}   $  & 23.65724, 733.37241  & 32 & $ 28.0^{+13.0}_{-5.0} $    & $ 0.81^{+0.08}_{-0.22}   $    & 0 & 7.02  & 27.41  \\
69 & \href{https://exofop.ipac.caltech.edu/tess/target.php?id=317923092}{317923092} & $ 12.5^{+4.8}_{-1.4}   $    & 19.19983, 729.59400  & 39 & $ 27.0^{+19.0}_{-5.0}   $  & $   0.928^{+0.062}_{-0.048} $ & 0 & 35.1  & 30.15  \\
70 & \href{https://exofop.ipac.caltech.edu/tess/target.php?id=323295479}{323295479} & $ 11.28 \pm 0.6 $           & 67.50129, 742.51554  & 12 & $ 83.0^{+41.0}_{-15.0} $   & $ 0.85 \pm 0.013   $          & 0 & 2.43  & 138.65 \\
71 & \href{https://exofop.ipac.caltech.edu/tess/target.php?id=332697924}{332697924} & $ 4.69 \pm 0.33 $           & 22.88951, 732.47218  & 33 & $ 26.2^{+7.1}_{-3.3} $     & $ 0.34 \pm 0.23 $             & 0 & 5.6   & 11.96  \\
72 & \href{https://exofop.ipac.caltech.edu/tess/target.php?id=333736132}{333736132} & $   10.25^{+0.87}_{-0.82} $ & 23.61270, 731.99390  & 32 & $ 26.1^{+7.1}_{-2.5} $     & $   0.751^{+0.045}_{-0.091} $ & 0 & 13.92 & 55.96  \\
73 & \href{https://exofop.ipac.caltech.edu/tess/target.php?id=339399841}{339399841} & $ 13.0^{+1.7}_{-1.1}   $    & 22.44020, 381.48503  & 18 & $ 25.4^{+9.2}_{-4.2} $     & $ 0.73^{+0.15}_{-0.34}   $    & 0 & --    & 21.13  \\
74 & \href{https://exofop.ipac.caltech.edu/tess/target.php?id=349091983}{349091983} & $ 21.3^{+8.8}_{-4.8}   $    & 549.0649, 549.06501  & 1  & 549.0649                        & $   0.975^{+0.075}_{-0.058} $ & 0 & --    & 44.11  \\
75 & \href{https://exofop.ipac.caltech.edu/tess/target.php?id=381553868}{381553868} & $ 12.41 \pm 0.84   $        & 20.91681, 690.25274  & 33 & $ 23.8^{+6.2}_{-2.9} $     & $ 0.39^{+0.22}_{-0.26}   $    & 0 & 9.94  & 22     \\
76 & \href{https://exofop.ipac.caltech.edu/tess/target.php?id=393229954}{393229954} & $ 9.3^{+4.1}_{-1.2}   $     & 75.77521, 757.75214  & 11 & $ 45.0^{+31.0}_{-0.0} $    & $   0.907^{+0.082}_{-0.098} $ & 0 & 12.97 & 33.63  \\
77 & \href{https://exofop.ipac.caltech.edu/tess/target.php?id=394340183}{394340183} & $ 10.87 \pm 0.61   $        & 60.35608, 724.27290  & 10 & $ 60.0^{+30.0}_{-18.0}   $ & $ 0.29 \pm 0.21 $             & 0 & 6.77  & 85.43  \\
78 & \href{https://exofop.ipac.caltech.edu/tess/target.php?id=396720998}{396720998} & $ 1.45^{+0.96}_{-0.85}   $  & 70.64108, 777.05180  & 12 & $ 78.0^{+19.0}_{-13.0}   $ & $ 0.46 \pm 0.32 $             & 0 & --    & 6.44   \\
79 & \href{https://exofop.ipac.caltech.edu/tess/target.php?id=404518509}{404518509} & $ 2.72 \pm 0.17 $           & 23.45386, 703.61679  & 18 & $ 23.0^{+12.0}_{-1.0}   $  & $ 0.35^{+0.28}_{-0.24}   $    & 0 & 0.93  & 22.52  \\
80 & \href{https://exofop.ipac.caltech.edu/tess/target.php?id=412386707}{412386707} & $ 12.37 \pm 0.81   $        & 23.33181, 746.61701  & 33 & $ 25.7^{+6.7}_{-2.4} $     & $ 0.65^{+0.15}_{-0.38}   $    & 0 & 6.18  & 28.97  \\
81 & \href{https://exofop.ipac.caltech.edu/tess/target.php?id=437293313}{437293313} & $ 8.57^{+0.56}_{-0.52}   $  & 20.26060, 749.64177  & 38 & $ 27.0 \pm 9.0 $           & $ 0.25 \pm 0.2 $              & 0 & --    & 26.34  \\
82 & \href{https://exofop.ipac.caltech.edu/tess/target.php?id=439491923}{439491923} & $ 13.9 \pm 1.3 $            & 26.34748, 711.37923  & 28 & $ 31.0^{+14.0}_{-5.0} $    & $ 0.7^{+0.14}_{-0.3}   $      & 0 & --    & 36.2   \\
83 & \href{https://exofop.ipac.caltech.edu/tess/target.php?id=442893494}{442893494} & $   11.55^{+0.67}_{-0.63} $ & 21.87047, 721.72570  & 22 & $ 36.0 \pm 12.0 $          & $ 0.34^{+0.25}_{-0.23}   $    & 0 & --    & 24.76  \\
84 & \href{https://exofop.ipac.caltech.edu/tess/target.php?id=457649900}{457649900} & $ 18.0^{+9.8}_{-3.9}   $    & 23.40078, 748.82538  & 33 & $ 30.0^{+20.0}_{-6.0} $    & $   0.986^{+0.058}_{-0.033} $ & 0 & 13.11 & 29.26  \\
85 & \href{https://exofop.ipac.caltech.edu/tess/target.php?id=466206508}{466206508} & $ 13.55 \pm 0.82   $        & 25.04186, 375.62557  & 16 & $ 31.0^{+16.0}_{-6.0} $    & $ 0.37 \pm 0.2 $              & 0 & 12.81 & 64.68

\\ \bottomrule
\label{tab:candidate_planetprops}
\end{longtable}
\end{landscape}
\begin{center}
\begin{table}
    \centering
    \caption{A sample of events from \texttt{monofind} that were not planetary in nature. This table is available in its entirety online.}
    \label{tab:extracand}
    \begin{tabularx}{0.465\columnwidth}{c | c | c c}
    \toprule
    \textbf{TIC ID} & \textbf{Designation} & \textbf{T\textsubscript{c1}}         & \textbf{T\textsubscript{c2}} \\
                    &                      & \textbf{(TBJD, [Sector])}            & \textbf{(TBJD, [Sector])} \\
    \hline
    \href{https://exofop.ipac.caltech.edu/tess/target.php?id=82286}{82286}            & Asteroid    & 1606.56693 [11]   & 2337.34989 [38] \\
    \href{https://exofop.ipac.caltech.edu/tess/target.php?id=109979}{109979}          & Asteroid    & 1601.53746 [11]   & 2355.66990 [38] \\
    \href{https://exofop.ipac.caltech.edu/tess/target.php?id=2490309}{2490309}        & Asteroid    & 1577.58862 [10]   & 2309.44366 [37] \\
    \href{https://exofop.ipac.caltech.edu/tess/target.php?id=3835932}{3835932}        & Asteroid    & 1392.54454 [03]   & 2133.24553 [30] \\
    \href{https://exofop.ipac.caltech.edu/tess/target.php?id=5740937}{5740937}        & Asteroid    & 1378.87982 [02]   & 2097.43075 [29] \\
    ...             & ...                  & ...    & ...             \\
    \bottomrule
    \end{tabularx}
\end{table}
\end{center}

\begin{figure}
    \centering
    \includegraphics[width=0.865\textwidth]{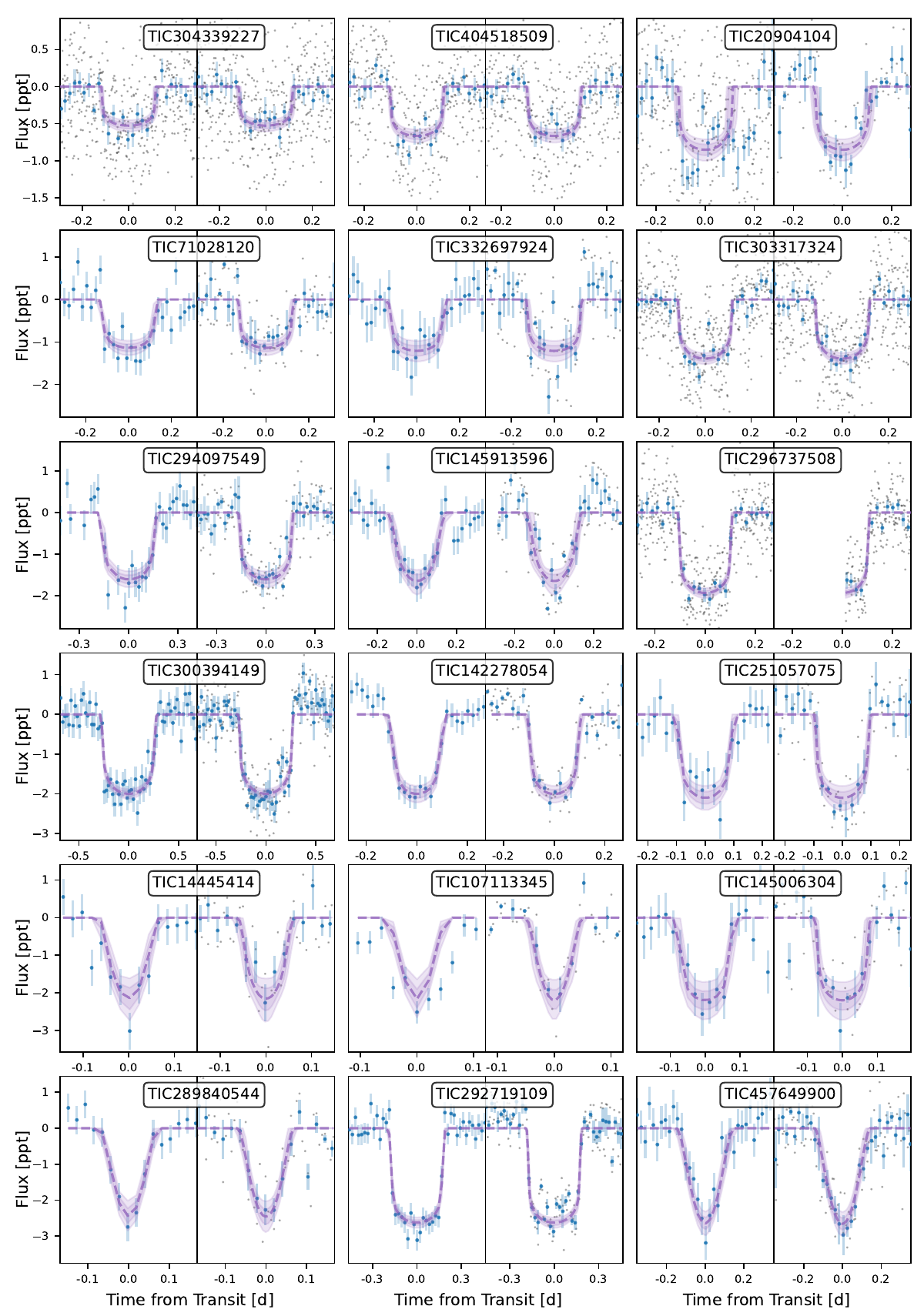}
    \caption{Zoomed-in transit events for duotransit candidates TIC-304339227 to TIC-156716001 ordered by transit depth. The left panel shows the Year 1 transit, and the right panel shows the Year 3 transit. The data in blue is the \tess\ PDCSAP photometry (see Section~\ref{sec:obs}). In cases with $<15$\,minute cadence blue points represent $30$\,minute bins with raw photometry shown in grey. The dashed purple lines show the best fit transit model with 1 and 2 standard deviations as shaded purple regions.}
    \label{fig:lightcurves0}
\end{figure}

\begin{figure}
    \centering
    \includegraphics[width=0.865\textwidth]{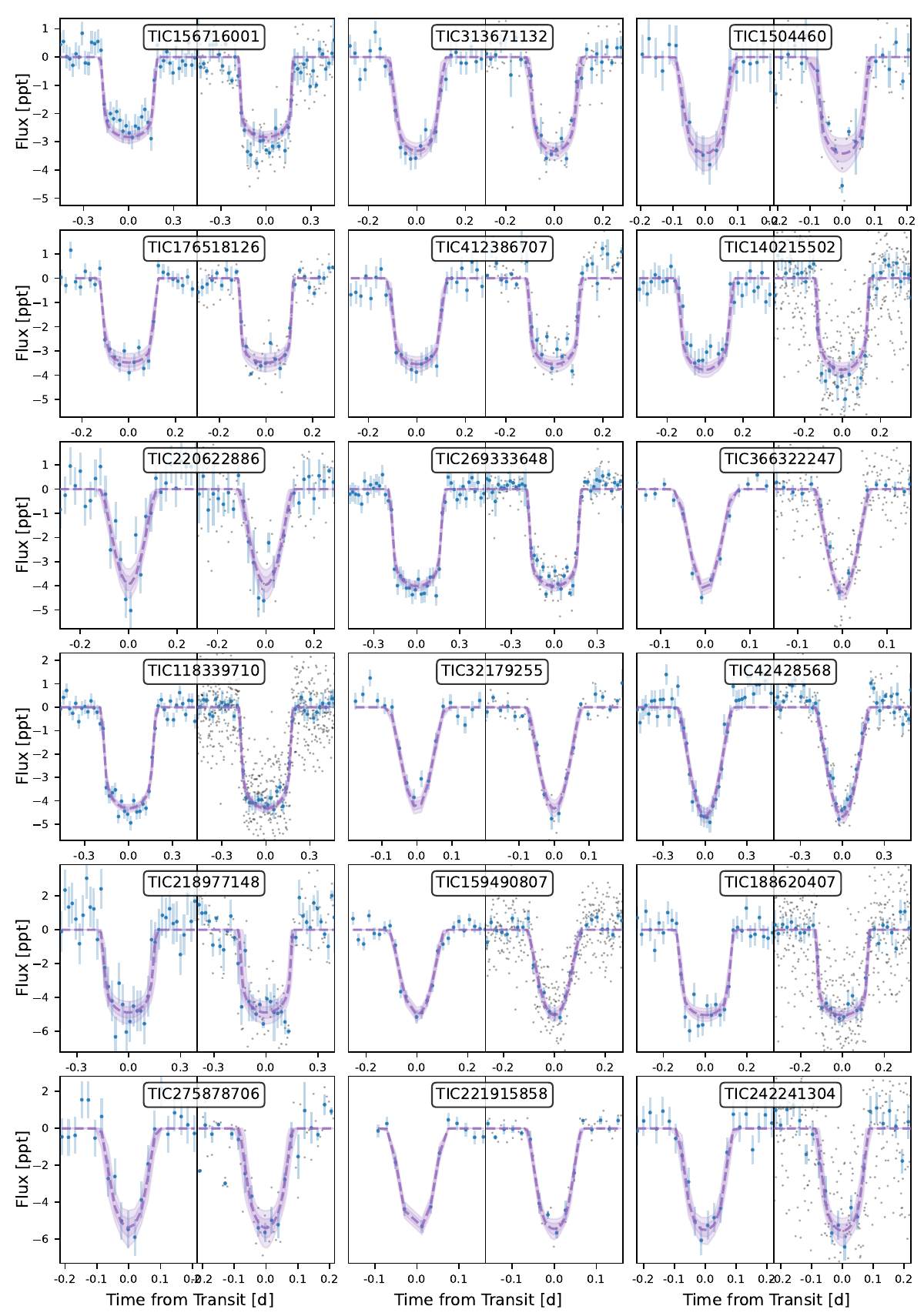}
    \caption{Zoomed-in transit events for duotransit candidates TIC-313671132 to TIC-381553868 ordered by transit depth. The left panel shows the Year 1 transit, and the right panel shows the Year 3 transit. The data in blue is the \tess\ PDCSAP photometry (see Section~\ref{sec:obs}). In cases with $<15$\,minute cadence blue points represent $30$\,minute bins with raw photometry shown in grey. The dashed purple lines show the best fit transit model with 1 and 2 standard deviations as shaded purple regions.}
    \label{fig:lightcurves1}
\end{figure}

\begin{figure}
    \centering
    \includegraphics[width=0.88\textwidth]{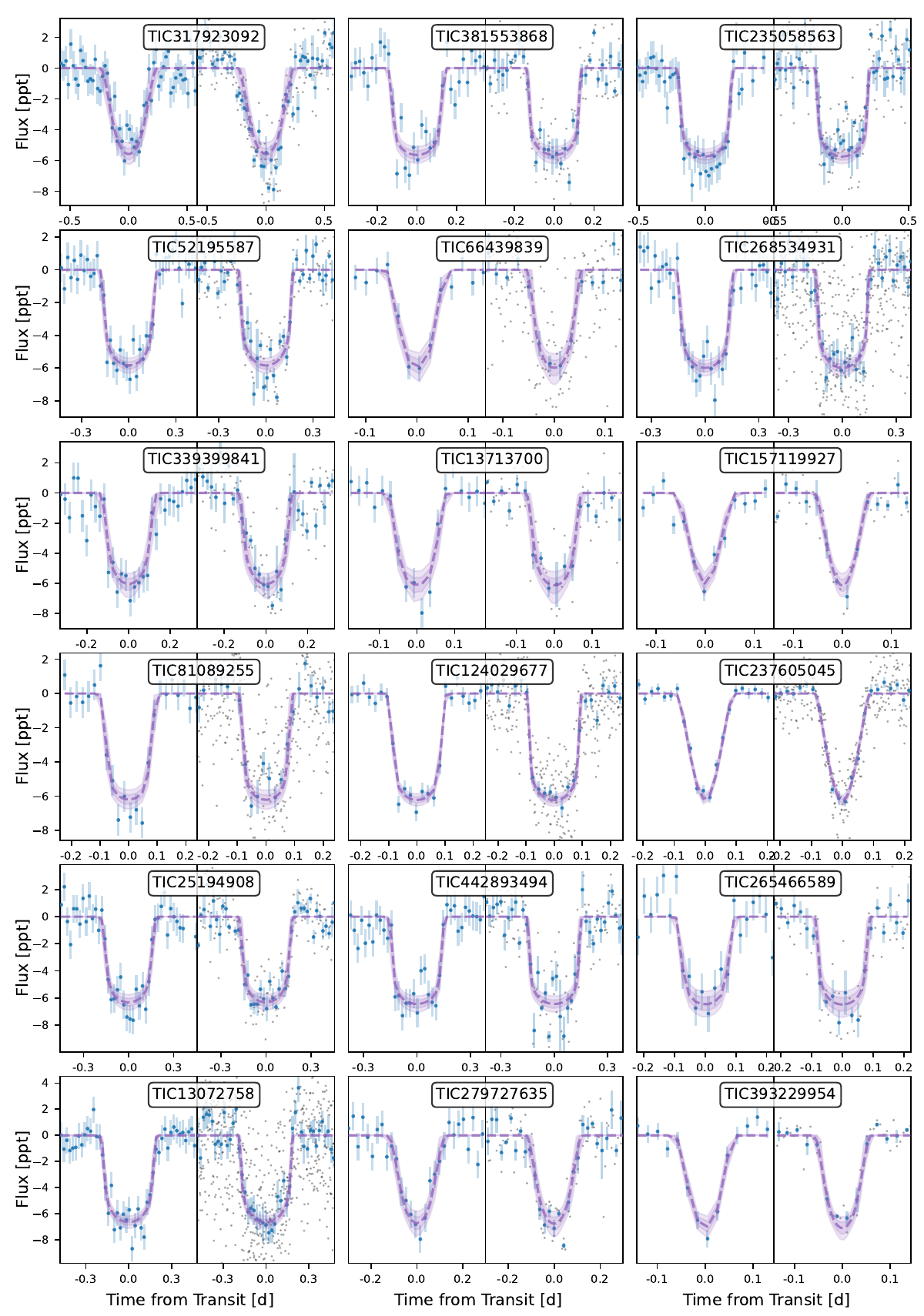}
    \caption{Zoomed-in transit events for duotransit candidates TIC-235058563 to TIC-306249066. The left panel shows the Year 1 transit, and the right panel shows the Year 3 transit. The data in blue is the \tess\ PDCSAP photometry (see Section~\ref{sec:obs}). The dashed purple line is the best fit transit model with 1 and 2 standard deviations either side.}
    \label{fig:lightcurves2}
\end{figure}

\begin{figure}
    \centering
    \includegraphics[width=0.88\textwidth]{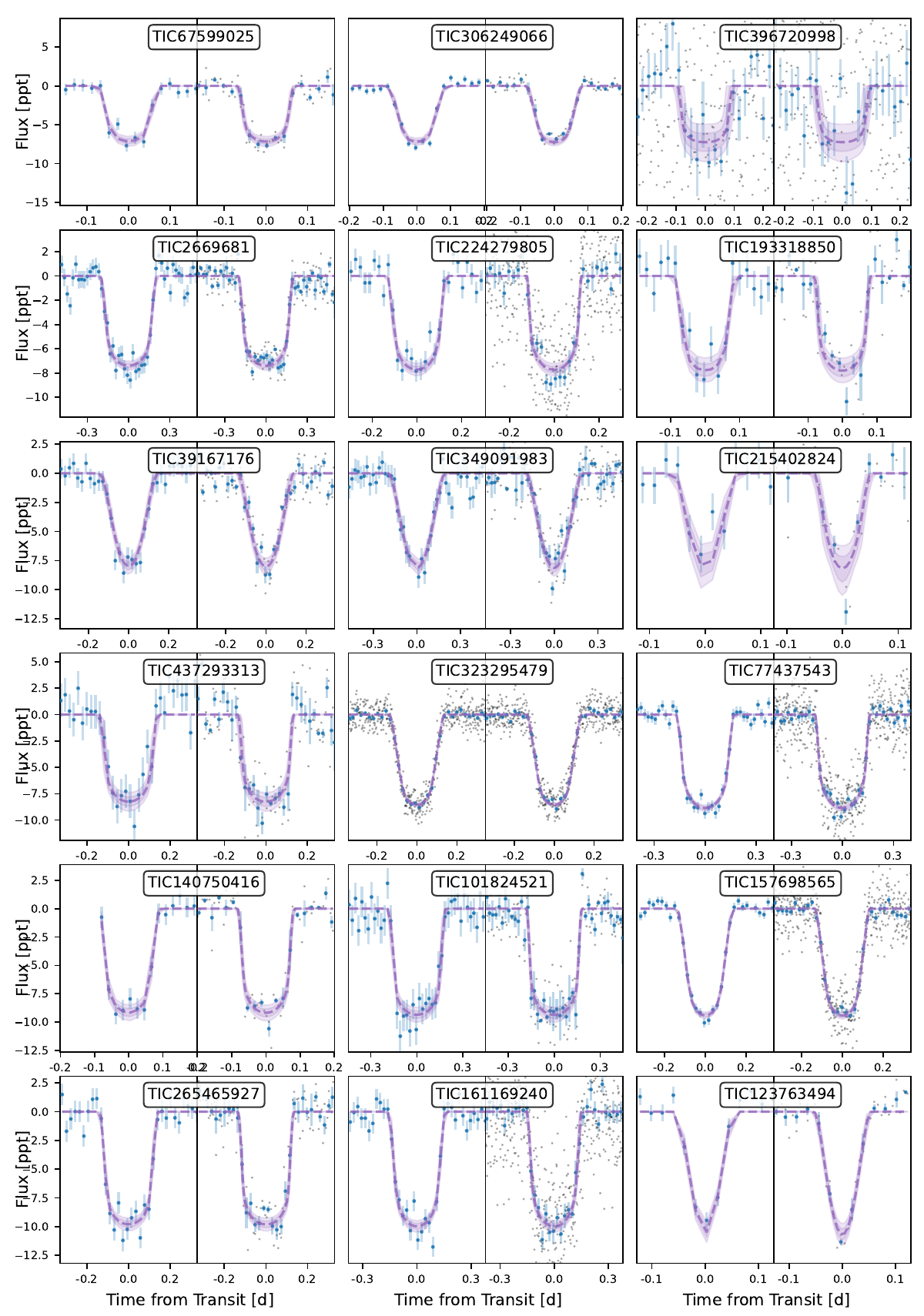}
    \caption{Zoomed-in transit events for duotransit candidates TIC-396720998 to TIC-439491923. The left panel shows the Year 1 transit, and the right panel shows the Year 3 transit. The data in blue is the \tess\ PDCSAP photometry (see Section~\ref{sec:obs}). The dashed purple line is the best fit transit model with 1 and 2 standard deviations either side.}
    \label{fig:lightcurves3}
\end{figure}

\begin{figure}
    \centering
    \includegraphics[width=0.88\textwidth]{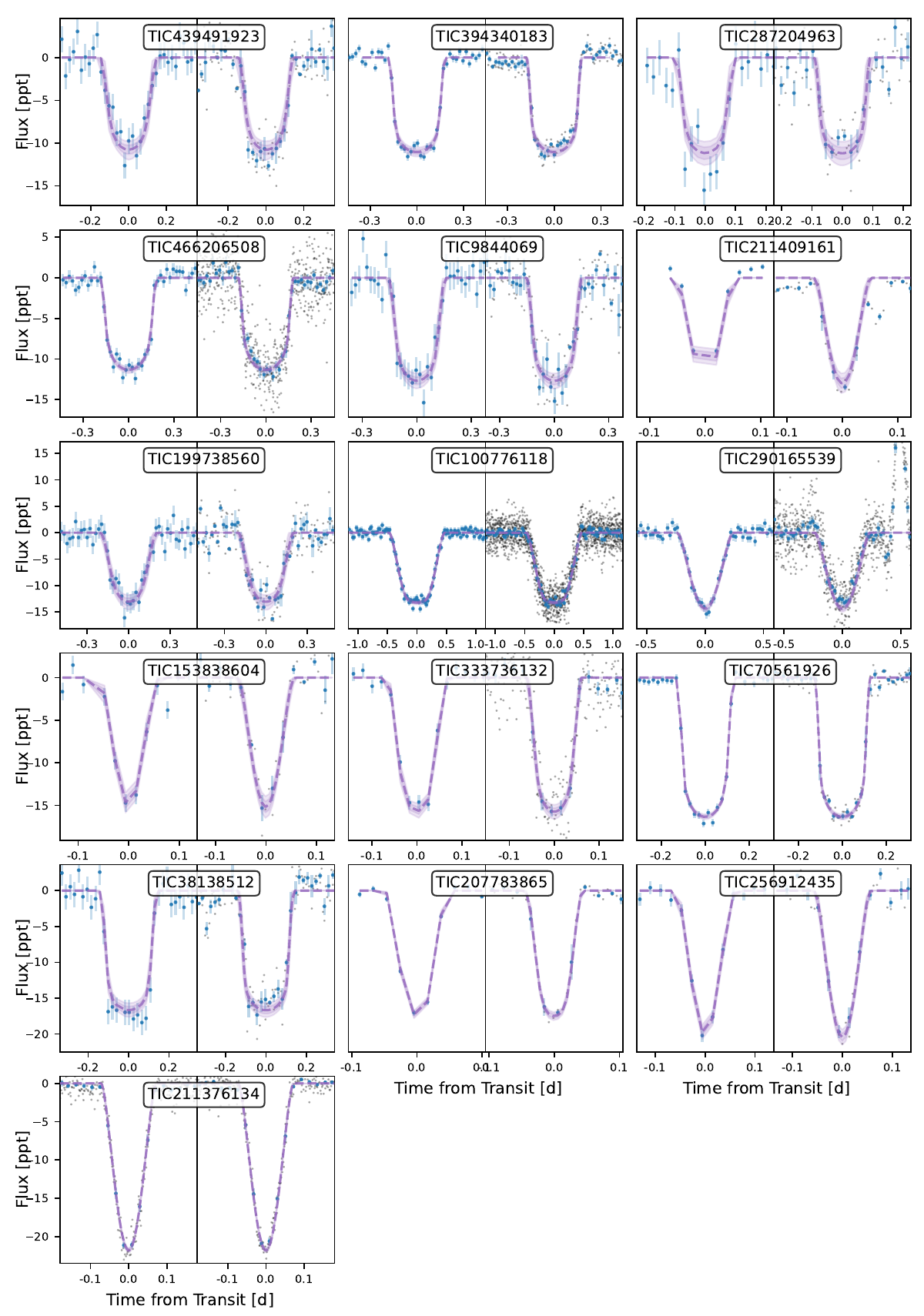}
    \caption{Zoomed-in transit events for duotransit candidates TIC-287204963 to TIC-256912435. The left panel shows the Year 1 transit, and the right panel shows the Year 3 transit. The data in blue is the \tess\ PDCSAP photometry (see Section~\ref{sec:obs}). The dashed purple line is the best fit transit model with 1 and 2 standard deviations either side.}
    \label{fig:lightcurves4}
\end{figure}

\clearpage
\section{Author affiliations} \label{sec:affiliations}
$^{1}$ Department of Physics, University of Warwick, Gibbet Hill Road, Coventry CV4 7AL, UK \\
$^{2}$ Centre for Exoplanets and Habitability, University of Warwick, Gibbet Hill Road, Coventry CV4 7AL, UK \\
$^{3}$ Physikalisches Institut, University of Bern, Gesellsschaftstrasse 6, 3012 Bern, Switzerland \\
$^{4}$ Astrophysics Research Centre, School of Mathematics and Physics, Queen’s University Belfast, Belfast BT7 1NN, UK \\
$^{5}$ Observatoire de Genève, Université de Genève, 51 Chemin Pegasi, 1290 Versoix, Switzerland \\
$^{6}$ Centre for Exoplanet Research, School of Physics and Astronomy, University of Leicester, University Road, Leicester, LE7 7RH, UK \\
$^{7}$ Institute of Planetary Research, German Aerospace Center, Rutherfordstrasse 2, 12489 Berlin, Germany \\
$^{8}$ European Space Agency (ESA), European Space Research and Technology Centre (ESTEC), Keplerlaan 1, 2201 AZ Noordwijk, The Netherlands \\
$^{9}$ Instituto de Estudios Astrof\'isicos, Facultad de Ingenier\'ia y Ciencias, Universidad Diego Portales, Av. Ej\'ercito 441, Santiago, Chile \\
$^{10}$ Centro de Astrof\'isica y Tecnolog\'ias Afines (CATA), Casilla 36-D, Santiago, Chile \\
$^{11}$ Instituto de Astronom\'ia, Universidad Cat\'olica del Norte, Angamos 0610, 1270709, Antofagasta, Chile \\
$^{12}$ Armagh Observatory and Planetarium, College Hill, Armagh, BT61 9DG, N Ireland, UK \\


\bsp	
\label{lastpage}
\end{document}